\documentclass[%
 reprint,
superscriptaddress,
nofootinbib,
 amsmath,amssymb,
 onecolumn,
prd,
]{revtex4-2}

\usepackage{graphicx}
\usepackage{dcolumn}
\usepackage{bm}
\usepackage[colorlinks]{hyperref}

\usepackage{color}
\usepackage[caption=false]{subfig}

\newcommand{\bv}[1]{\boldsymbol{#1}}

\newcommand{\hMpc}{h\,\mathrm{Mpc}^{-1}}

\newcommand{\delD}[1]{(2\pi)^3\delta_\mathrm{D}\left({#1}\right)}

\newcommand{\av}[1]{\left\langle{#1}\right\rangle} 

\newcommand{\vk}{\bv k}

\newcommand{\vm}{\bv m}

\newcommand{\F}{\mathcal{F}}
\newcommand{\G}{\mathcal{G}}

\renewcommand{\P}{\mathcal{P}}

\newcommand{\hn}{\hat{\bv n}}
\newcommand{\hz}{\hat{\bv z}}
\newcommand{\hk}{\hat{\bv k}}

\usepackage{color}
\def\beq{\begin{eqnarray}}
\def\eeq{\end{eqnarray}}

\DeclareSymbolFont{toneletters}{T1}{\familydefault}{m}{it}
\SetSymbolFont{toneletters}{bold}{T1}{\familydefault}{bx}{it}

\DeclareMathSymbol\edth{\mathord}{toneletters}{"F0}

\usepackage{empheq}

\definecolor{darkgreen}{RGB}{0,120,0}

\newcommand{\resub}[1]{#1}

\begin{document}


\title{What can galaxy shapes tell us about physics beyond the standard model?}

\author{Oliver H.\,E. Philcox}
\email{ohep2@cantab.ac.uk}
\affiliation{Simons Society of Fellows, Simons Foundation, 160 5th Avenue, New York, NY 10010, USA}
\affiliation{Center for Theoretical Physics,
Columbia University, New York, NY 10027, USA}
\author{Morgane J. K\"onig}
\affiliation{Department of Physics, Massachusetts Institute of Technology, 77 Massachusetts Avenue, Cambridge, MA 02139, USA}
\author{Stephon Alexander}
\affiliation{Brown Theoretical Physics Center and Department of Physics, Brown University, \\ 
182 Hope Street, Providence, Rhode Island, 02903, USA}
\affiliation{Center for Computational Astrophysics, Flatiron Institute, 162 5th Ave, New York, NY 10010, USA}
\author{David N. Spergel}
\affiliation{Simons Foundation, 160 5th Ave, New York, NY 10010, USA}


\begin{abstract}
\noindent The shapes of galaxies trace scalar physics in the late-Universe through the large-scale gravitational potential. Are they also sensitive to higher-spin physics? We present a general study into the observational consequences of vector and tensor modes in the early and late Universe, through the statistics of cosmic shear and its higher-order generalization, flexion. Higher-spin contributions arise from both gravitational lensing and intrinsic alignments, and we give the leading-order correlators for each (some of which have been previously derived), in addition to their flat-sky limits. In particular, we find non-trivial sourcing of shear $EB$ and $BB$ spectra, depending on the parity properties of the source. We consider two sources of vector and tensor modes: scale-invariant primordial fluctuations and cosmic strings, forecasting the detectability of each for upcoming surveys. Shear is found to be a powerful probe of cosmic strings, primarily through the continual sourcing of vector modes; flexion adds little to the constraining power except on very small scales ($\ell\gtrsim 1000$), though it could be an intriguing probe of as-yet-unknown rank-three tensors or halo-scale physics. Such probes could be used to constrain new physics proposed to explain recent pulsar timing array observations.
\end{abstract}

\maketitle

\section{Introduction}
Through current observational projects such as DES, KiDS, and HSC \citep[e.g.,][]{2021arXiv210513549D,2023arXiv230517173E,2021A&A...645A.104A,2023arXiv230400701D}, and forthcoming experiments such as Rubin and Roman \citep[e.g.,][]{2012arXiv1208.4012G,2009arXiv0912.0201L}, we will measure the apparent shapes of hundreds of millions of galaxies across a wide range of redshifts. The technical barriers associated with such measurements are huge: thanks to decades of \resub{computational}, observational, and theoretical work, we are now in a regime where they can be made robustly, facilitating their use in modern cosmological analyses. 

The principal use of galaxy shape catalogs is as a probe of the underlying matter density of the Universe \citep[e.g.,][]{2001PhR...340..291B,2005PhRvD..72b3516C,2015RPPh...78h6901K,2008ARNPS..58...99H,2003ARA&A..41..645R}. Due to gravitational lensing, photons emitted from some distant galaxy are continuously deflected by matter as they traverse the vast expanses of space before reaching our telescopes; as such, their deflection angle encodes the gravitational potential, $\Phi$, projected along the line-of-sight. For coherent sources such as galaxies, the net effect of lensing is to distort the shape of the galaxies, such that an initially circular projection is distorted to an ellipse (at leading order). Furthermore, the intrinsic shape of the galaxy is itself a probe of $\Phi$ \resub{(which itself can contain novel physics \citep[e.g.,][]{Kogai:2020vzz,Schmidt:2015xka,Chisari:2016xki})}, through tidal interactions as the galaxy forms \citep[e.g.,][]{2004PhRvD..70f3526H,2020JCAP...01..025V,2011JCAP...05..010B,2001ApJ...559..552C,2021MNRAS.501..833K,2021JCAP...03..030S}. Collating the distortions measured from millions of objects (accounting for their significant noise) yields a large-scale map of $\Phi$, or, through the Poisson equation, the matter density, $\rho_m$. 

\vskip 4pt

\paragraph{Vectors \& Tensors}
A full description of gravitational lensing involves not just the scalar part of the metric, $h_{00}\propto \Phi$, but also vector and tensor parts, $h_{0i}$ and $h_{ij}$ \citep[e.g.,][]{2008PhRvD..77j3515S,2012PhRvD..86h3513S,2014PhRvD..89h3507S,2012JCAP...10..030Y}. The same holds also for the intrinsic shapes of galaxies, again through the tidal tensor, $t_{ij}$ \citep{2020JCAP...07..005B}, though these have been little discussed for vector modes. As such, the shapes of galaxies provide a window into the vector and tensor sectors of the Universe, that are hard to observe in other probes (such as galaxy density, at leading order \citep{2013PhRvD..87j3006D,2014JCAP...12..050D}). In the standard cosmological model, we expect such contributions to be trivially small; however, non-standard cosmological models can source vector and tensor modes, and thus be probed using galaxy shapes.

There exist a large number of theoretical models capable of generating vector and tensor modes in the late Universe. Many such scenarios are rooted in inflation, principally from the appearance of vector and tensor fields, for example in vector inflation and axionic models optionally including Chern-Simons interactions \citep{1989PhRvD..40..967F,2008JCAP...08..021K,2016PhRvD..93j3520G,2008JCAP...06..009G,2013JCAP...04..046A,2018JCAP...11..030M,2018JCAP...05..003A,2017PhRvD..96f3506A}. They can also be sourced from single field inflation; whilst primordial vectors decay quickly, primordial gravitational waves (tensor modes) are a signature of many canonical scenarios (see \citep[e.g.,][]{2020A&A...641A..10P} for recent constraints). An array of late-time sources are also possible, driven by old and new fields. Many of these are motivated by our lack of understanding of dark energy; modified gravity scenarios such as quintessence, vector dark energy, and Chern-Simons general relativity, could lead to a spectrum of new perturbations which become relevant at late-times, thereby evading constraints from the cosmic microwave background \citep[e.g.,][]{2004JCAP...07..007A,2008JCAP...08..021K,2009PhR...480....1A,2003PhRvD..68j4012J,2016arXiv160100057A,2008PhRvD..77h3509P}. A network of cosmic strings (arising from some symmetry breaking scenario, such as GUT models), would also lead to continuous production of vector and tensor modes, and thus distortion of galaxy shapes \citep[e.g.,][]{1997PhRvD..55..573V,1994csot.book.....V,2012JCAP...10..030Y}. These are of particular current relevance \citep[e.g.,][]{EPTA:2023xxk} given their plausability as a model for explaining the recent detection of a stochastic gravitational wave background from pulsar timing array experiments \citep{Reardon:2023gzh,Xu:2023wog,EPTA:2023fyk,NANOGrav:2023gor}. Finally, such physics could conceivably break parity symmetry; to some extent, this is expected from primordial phenomena such as baryogenesis \citep[e.g.,][]{2013JCAP...04..046A,2016IJMPD..2540013A,Abedi:2018top}, and is of particular current interest given the potential detections on the CMB and distribution of galaxies \citep{2022PhRvD.106f3501P,2023PhRvD.107b3523C,2023arXiv230304815C,2023arXiv230312106P,2023PhRvL.130t1002C,2023MNRAS.522.5701H,1999PhRvL..83.1506L,2020PhRvL.125v1301M} (though see the no-go theorems in \citep{Liu:2019fag,Cabass:2022rhr}).    

\vskip 4pt

\paragraph{Shear \& Flexion}
To make use of the tranche of cosmological information contained within galaxy shapes, we require some way of distilling this information. Conventionally, this is performed by considering the ``shear'' of galaxies, $\gamma_{ij}$, which measures their ellipticity \citep[e.g.,][]{2015RPPh...78h6901K,2001PhR...340..291B}. Since this is a rank-two tensor, it is sensitive to rank-two components of the metric, including $\partial_i\partial_j h_{00}$, $\partial_i h_{0j}$, $h_{ij}$, \textit{i.e.}\ scalars, vectors, and tensors. In general, the shear is computed for each galaxy of interest, which are combined to make full-sky maps, usually expressed in terms of electric- and magnetic-components, $\gamma^E$ and $\gamma^B$. By measuring the two-point correlation functions, particularly $\av{\gamma^{E}\gamma^{E*}}$, we can probe the power spectrum of matter, and thus constrain parameters such as the matter density and clustering amplitude, \resub{as well as more esoteric physics such as primordial non-Gaussianity from higher-spin particles \citep{Kogai:2020vzz}}. For vectors and tensors, there is information also in the $B$-mode spectra, such as $\av{\gamma^{B}\gamma^{B*}}$, with parity-violating physics appearing in the cross-spectra $\av{\gamma^{E}\gamma^{B*}}$. Further information may also arise in higher-order correlators and cross-spectra, which probe bispectra and beyond (cf.\,\citep{2023arXiv230112890E} for scalar physics, and \citep{2022PhRvD.106h3501P,2023arXiv230202925K,2012JCAP...10..018F,Kogai:2020vzz} for other applications including (anisotropic) primordial non-Gaussianity and higher-spin inflationary phenomena).

The gravitational distortions of galaxies are not fully encapsulated by the shear tensor, $\gamma_{ij}$. At next order, one can define the ``flexion'' components, $\mathcal{F}_{ijk}$, and $\mathcal{G}_{ijk}$, which parametrize the octopole moments of an image \citep[e.g.,][]{2006MNRAS.365..414B,2005ApJ...619..741G,2022PhRvD.105l3521A,2008ApJ...680....1O,2007ApJ...660..995O,2021arXiv210709000G,2011MNRAS.411.2241M}. This is a rank-three tensor, and thus sensitive of scalar, vector, and tensor physics.\footnote{\resub{Whilst one can go further still and define a rank-four tensor, it is not of interest to this work (though can enable constraints on spin-four primordial non-Gaussianity transferred to the scalar sector \citep{Kogai:2020vzz}).}} It may also be a probe of new physics sourced by some irreducible rank-three tensor, such as torsion. In this case, the field can be written in terms of ``gradient'' and ``curl'' modes; only the latter is sourced by standard model physics, e.g., $\av{\mathcal{F}^{g*}\mathcal{F}^g}$. Flexion has been discussed in a range of previous works, with the general conclusion being that it becomes useful in scalar analyses only on very small scales \citep[e.g.,][]{2021arXiv210709000G,2005ApJ...619..741G}; here, we include the flexion in a complete manner, and assess whether the above conclusions remain true for vector and tensor physics.

\vskip 4pt

In the remainder of this work, we will present an in-depth overview of the effects of vector and tensor physics on shear and flexion, considering both intrinsic and extrinsic (lensing) contributions, providing a general dictionary of novel phenomena to galaxy shapes. To this end, we will first set out our conventions for new physics in \S\ref{sec: conventions} via their impact on the metric tensor, and give two motivating examples of beyond-$\Lambda$CDM physics, before presenting a pedagogical overview of galaxy shape observables in \S\ref{sec: observables}. In \S\ref{sec: sources}, we will discuss the sourcing of shear and flexion by scalars, vectors, and tensors, before giving the associated power spectra in \S\ref{sec: power-spectra}, including their small-scale limiting forms. To put our results in context, \S\ref{sec: results} considers the ability of upcoming cosmic shear surveys to detect vector and tensor modes using the two fiducial models discussed in \S\ref{sec: conventions} (sourced by inflation and cosmic strings). Finally, we conclude in \S\ref{sec: conclusions}. Appendix \ref{appen: 2-sphere-math} provides a brief summary of the two-sphere mathematics used in this work, with Appendix \ref{appen: cosmic-string-tensors} outlining our cosmic string tensor model. Finally, Appendices \ref{appen: intrinsic-vector-derivations} and \ref{appen: lensing-spectra} derive full-sky power spectra associated with intrinsic and extrinsic vector modes, and Appendix \ref{appen: kernels} lists the mathematical kernels appearing in the power spectra.

\section{Scalars, Vectors, \& Tensors}\label{sec: conventions}

\subsection{General Formalism}
\noindent Weak gravitational interactions induce a distortion in the background (FRW) metric of the Universe, which takes the form
\beq
    ds^2 = a^2(\eta)\left(\eta_{\mu\nu}+h_{\mu\nu}\right)dx^\mu dx^\nu,
\eeq
where $\eta$ and $x^i$ are conformal time and comoving space, $\eta_{\mu\nu}$ is the flat-space metric, and $a(\eta)$ is the scale factor \citep[e.g.,][]{1995ApJ...455....7M,1984PThPS..78....1K,1992PhR...215..203M}. In the weak field limit, the perturbation $h_{\mu\nu}$ can be decomposed into independent scalar, vector, and tensor modes, with the scalar part taking the form 
\beq\label{eq: scalar-metric}
    h^S_{00} = -2\Psi(\bv{x},\eta), \qquad h^S_{0i} = h^S_{i0} = 0, \qquad h^S_{ij} = -2\Phi(\bv{x},\eta)\delta^{\rm K}_{ij},
\eeq
in the conformal-Newtonian gauge, where $\Phi$ and $\Psi$ are the Bardeen potentials. In the absence of anisotropic stress, $\Phi=\Psi$; this will be assumed below. In the presence of vector modes, we can write
\beq
    h^V_{00} = 0, \qquad h^V_{0i} = h^V_{i0} = B_i(\bv{x},\eta), \qquad  h^V_{ij} = \partial_jH_i+\partial_iH_j,
\eeq
for divergence-free vectors $B_i$ and $H_i$. Again working in the conformal-Newtonian gauge, we can set $H_i=0$ and consider only $B_i$ (equivalent to working with the gauge-invariant potential $-\sigma_{g,i}=B_i-\partial_\eta H_i$ \citep{2012JCAP...10..030Y}). Finally, the tensor degrees of freedom can be written in the form (again working in the conformal-Newtonian gauge)
\beq
    h^T_{00} = 0, \qquad h^T_{0i} = h^T_{i0} = 0, \qquad  h^T_{ij} = h_{ij},
\eeq
where $h_{ij}$ (not to be confused with $h_{\mu\nu}$) is the transverse-traceless rank-two tensor associated with gravitational wave propagation.

In the below, we will find it useful to expand the metric perturbations in Fourier-space via the relations \citep[e.g.,][]{1997PhRvD..56..596H,2012PhRvD..86l5013D}
\beq\label{eq: helicity-expansion}
    \Phi(\bv{x},\eta) = \int_{\vk}e^{i\vk\cdot\bv{x}}\Phi(\vk,\eta), \quad B_i(\bv{x},\eta) = \int_{\vk}e^{i\vk\cdot\bv{x}}\sum_{\lambda=\pm}e^{(\lambda)}_i(\hk)B_\lambda(\vk,\eta), \quad h_{ij}(\bv{x},\eta) = \int_{\vk}e^{i\vk\cdot\bv{x}}\sum_{\lambda=\pm}e^{(\lambda)}_{ij}(\hk)h_\lambda(\vk,\eta),
\eeq
where $\int_{\vk}\equiv \int d\vk/(2\pi)^3$. Here, we have decomposed the vector and tensor modes in \textit{helicity states}, defined as
\beq
    \bv{e}^{(\pm)}(\hk) \equiv (\hat{\bv e}_1\mp i\hat{\bv e}_2)/\sqrt{2}, \qquad e^{(\pm)}_{ij}(\hk) \equiv e^{(\pm)}_i(\hk)e^{(\pm)}_j(\hk)
\eeq
where $\{\bv e_1, \bv e_2, \hk\}$ is a locally orthonormal basis set. As such, the impact of scalar, vector, and tensor modes on the metric, and thus gravitational lensing, is controlled by the (scalar) helicity components $\{\Phi, B_\pm, h_\pm\}$.

Most metric perturbations of interest in cosmology are controlled by stochastic processes such as inflationary quantum mechanical fluctuations, and are thus random fields. As such, their properties can be encapsulated by correlations functions, the simplest of which are power spectra:
\beq\label{eq: Phi,B,h-correlators}
    \av{\Phi(\vk,\eta)\Phi^*(\vk',\eta')} &=& P_{\Phi}(k,\eta,\eta')\,\times\,\delD{\vk-\vk'} \\\nonumber
    \av{B_\lambda(\vk,\eta)B^*_{\lambda'}(\vk',\eta')} &=& \delta^{\rm K}_{\lambda\lambda'}\,\times\,P_{B_\lambda}(k,\eta,\eta')\,\times\,\delD{\vk-\vk'}\\\nonumber
    \av{h_\lambda(\vk,\eta)h^*_{\lambda'}(\vk',\eta')} &=& \delta^{\rm K}_{\lambda\lambda'}\,\times\,P_{h_\lambda}(k,\eta,\eta')\,\times\,\delD{\vk-\vk'},
\eeq
(noting that the mean of each field can be set to zero according to the equivalence principle). Here, we have assumed that the relevant physics is homogeneous (leading to the Dirac delta) and isotropic (such that the correlators depend only on $|\vk|$); furthermore, the different helicity states are uncorrelated. Finally, we may optionally assert parity conservation (\textit{i.e.}\ symmetry under point reflections): this enforces $P_{B_+}=P_{B_-}$ and $P_{h_+}=P_{h_-}$, though this is not generically required.\footnote{In gravitational wave literature, the total gravitational wave power spectrum is often denoted $P_h = P_{h_+}+P_{h_-}$.}

Additional information on the metric perturbations is provided by the Einstein equations, which specify the time evolution of fluctuations. For the scalar sector, their linearization leads to the well-known form $\Phi(\vk,\eta)\equiv T_S(\eta)\Phi(\vk,\eta_0)$, where the transfer function $T_S(\eta)$ is constant in matter domination, and we normalize to the value of the spectrum today (at $\eta=\eta_0$). From the $ij$ part of the perturbed Einstein equation, we obtain the following equations for the evolution of vector and tensor modes \citep[e.g.,][]{Hu:1997hp,2012PhRvD..85d3009I}:
\beq\label{eq: vec-tensor-einstein}
    B_i'(\bv{x},\eta)+2\mathcal{H}(\eta)B_i(\bv{x},\eta) &=& -16\pi Ga^2(\eta)\frac{\partial^j}{\nabla^2}\delta T_{ij}^V(\bv{x},\eta)\\\nonumber
    h_{ij}''(\bv{x},\eta)+2\mathcal{H}(\eta)h_{ij}'(\bv{x},\eta)-\nabla^2h_{ij}(\bv{x},\eta) &=& 16\pi Ga^2(\eta)\delta T_{ij}^T(\bv{x},\eta).
\eeq
Here, primes denote derivatives with respect to conformal time, $\mathcal{H}(\eta)\equiv a'(\eta)/a(\eta)$ is the conformal Hubble parameter, and $\delta T_{ij}^{V,T}$ are the vector and tensor parts of the perturbed stress-energy tensor. From this equation, it is clear that vector and tensor modes can be either (a) primordial, evolving under the homogeneous equations, or (b) dynamically sourced by phenomena contributing to the non-scalar stress-energy tensor (such as cosmic strings and neutrino free-streaming). In the below, we will consider examples of each form.

\subsection{Motivating Examples}
\subsubsection{Inflationary Perturbations}\label{subsec: ex-inflationary}

\noindent In the standard paradigm, quantum fluctuations in inflation source scalar perturbations to the metric, which seed structure formation in the late Universe. A variety of models also predict a primordial spectrum of vector and tensor modes, for example due to gauge fields active in inflation or large excursions of the inflaton. Conventionally, such spectra are parametrized as power laws:
\beq\label{eq: primordial-spectra}
    P_B(k) = r_V(k_0)\,\times\,\frac{2\pi^2}{k^3}\Delta^2_\zeta(k_0)\left(\frac{k}{k_0}\right)^{n_V-1}, \qquad P_h(k) = r_T(k_0)\,\times\,\frac{2\pi^2}{k^2}\Delta^2_\zeta\left(\frac{k}{k_0}\right)^{n_T},
\eeq
where $k_0 = 0.002\mathrm{Mpc}^{-1}$ is a fiducial scale, $n_X$ is a slope, and $r_X$ specifies the amplitude relative to the scalar power spectrum amplitude, $\Delta_\zeta^2$. For $n_V=1, n_T=0$, we find a scale-invariant form. Observations of the CMB place strong constraints on tensors, with $r_T(k_0)\lesssim 0.03$ \citep{2020A&A...641A..10P,2022PhRvD.105h3524T} with a fiducial value $n_T=-r_T/8$. In some scenarios, the primordial perturbations can be violate parity symmetry (as discussed in Appendix B of \citep{2020JCAP...07..005B}); in this case, there is a different amplitude for the two parity states, with
\beq
    P_{B_{\pm}}(k) = (1\pm \epsilon_V)P_{B}(k), \qquad P_{h_{\pm}}(k) = (1\pm \epsilon_T)P_{h}(k),
\eeq
for chirality parameter $\epsilon_X$.

To model vector and tensor modes in the late Universe, we require the evolution of $B_i$ and $h_{ij}$ as well as the primordial form. This is obtained by solving the homogeneous version of \eqref{eq: primordial-spectra}, yielding \citep[e.g.,][]{2012PhRvD..85d3009I}:
\beq
    B_i(\vk,\eta) &=& T_V(\eta)B_i(\vk,\eta_*), \qquad T_V(\eta) = \left(\frac{a(\eta_*)}{a(\eta)}\right)^2\\\nonumber
    h_{ij}(\vk,0) &=& T_T(k,\eta)h_{ij}(\vk), \qquad T_T(k,\eta) \approx \frac{3j_1(k\eta)}{k\eta},
\eeq
where $\eta_*$ is some reference scale and we give the approximate form for the tensor transfer function in matter domination. In this regime, the vector modes decay away quickly, whilst the tensor modes evolve much more slowly, and propagate as a damped wave. For this reason, vector modes are rarely considered in cosmology, since those produced only by inflation are vanishingly small at late times. 

Incorporating these transfer function definitions, we can write the power spectra in separable form:
\beq\label{eq: Phi,B,h-correlators-w-transfer}
    \av{\Phi(\vk,\eta)\Phi^*(\vk',\eta')} &=& T_S(\eta)T_S(\eta')P_{\Phi}(k)\,\times\,\delD{\vk-\vk'} \\\nonumber
    \av{B_\lambda(\vk,\eta)B^*_{\lambda'}(\vk',\eta')} &=& \delta^{\rm K}_{\lambda\lambda'}\,\times\,T_V(\eta)T_V(\eta')P_{B_\lambda}(k)\,\times\,\delD{\vk-\vk'}\\\nonumber
    \av{h_\lambda(\vk,\eta)h^*_{\lambda'}(\vk',\eta')} &=& \delta^{\rm K}_{\lambda\lambda'}\,\times\,T_T(k,\eta)T_T(k,\eta')P_{h_\lambda}(k)\,\times\,\delD{\vk-\vk'};
\eeq
this holds true in the linear regime for any primordial perturbations in the early Universe.

\subsubsection{Cosmic Strings}\label{subsec: ex-strings}
\noindent An example of late-time new physics that could generate detectable vector and tensor modes is cosmic strings \citep[e.g.,][]{1994csot.book.....V,1997PhRvD..55..573V,1994ApJ...431..534H,2012JCAP...10..030Y,1996PhRvD..54.2535M,2009JCAP...10..003T,2008PhRvD..77h3509P}. These are one-dimensional topological defects that are a generic prediction of any grand unification model, sourced by phenomena such as spontaneous symmetry breaking or from superstring theory. A network of strings will contribute to the late-time energy momentum tensor through anisotropic stress, and thus act as a non-decaying source for both vector and tensor metric perturbations $B_i$ and $h_{ij}$. These source photon-baryon vorticity, and thus generate CMB $B$-modes, allowing them to be indirectly constrained \citep{2008PhRvD..77h3509P}. They could also be a possible source of the signal detected in recent pulsar timing array experiments \citep{EPTA:2023xxk}.

Full modelling of the cosmic string correlators is non-trivial, and beyond the scope of this work. For our purposes, we consider only a simple prescription (known as the ``velocity-dependent one-scale'' model), which posits a Poissonian sample of string segments with some tension $G\mu$ and intercommuting (\textit{i.e.}\ reconnecting) probabilty $P$ (following \citep[e.g.,][]{2012JCAP...10..030Y}). Each Nambu-Goto string segment can be described by a position $\sigma$ and a (conformal) time $\eta$ on the string world-sheet, and give the following contribution to the stress-energy tensor in the transverse gauge \citep{1994csot.book.....V}: 
\beq\label{eq: stress-energy-string}
    \delta T^{\mu\nu}(\bv{x},\eta) = \mu \int d\sigma\,\begin{pmatrix} 1 & -\partial_\eta x^i\\ -\partial_\eta x^j & \partial_\eta x^i\partial_\eta x^j-\partial_\sigma x^i\partial_\sigma x^j\end{pmatrix}\delta_{\rm D}(\bv{x}-\bv{x}(\sigma,\eta)).
\eeq
Using the Einstein equations \eqref{eq: vec-tensor-einstein}, we can compute the vector and tensor metric perturbations arising from the strings, and thus the associated power spectrum of $B_i$ and $h_{ij}$. For the vector modes, an explicit calculation can be found in \citep{2012JCAP...10..030Y}; for tensors, we provide a derivation in Appendix \ref{appen: cosmic-string-tensors}. Following a number of simplifying assumptions (detailed in the Appendix), we obtain the following forms for the equal-time power spectrum of vector and tensor modes:
\beq\label{eq: vector-tensor-cosmic-string-spectrum}
    P_{B_{\pm}}(k,\eta,\eta) &=& (16\pi G\mu)^2\frac{2\sqrt{6\pi}v_{\rm rms}^2}{3(1-v_{\rm rms}^2)}
    \frac{4\pi \chi^2a^4}{H}\left(\frac{a}{k\xi}\right)^5\mathrm{erf}\left[\frac{k\xi/a}{2\sqrt 6}\right],\\\nonumber
    P_{h_{\pm}}(k,\eta,\eta) &=& (16\pi G\mu)^2\frac{\sqrt{6\pi}}{9(1-v_{\rm rms}^2)}
    \frac{4\pi \chi^2a^4}{H}\left(\frac{a}{k\xi}\right)^5\mathrm{erf}\left[\frac{k\xi/a}{2\sqrt 6}\right]\left[v_{\rm rms}^4+(1-v_{\rm rms}^2)^2\right],
\eeq
where $v_{\rm rms}$ and $\xi$ are the average string velocity and length, $H$ is the Hubble parameter and $a$ is the scale factor (dropping the $\eta$-dependence for clarity). As in \citep{2012JCAP...10..030Y}, we will assume the fiducial values $\xi=1/(H\gamma_s)$, $v_{\rm rms}^2 = (1-\pi/3\gamma_s)/2$, with correlation length $\gamma_s = (\pi\sqrt{2}/3\tilde c P)^{1/2}$ for $\tilde c\approx 0.23$ and $P\approx 10^{-3}$ \citep{2009JCAP...10..003T}. Notably, the spectra do not depend on the helicity state, thus the effects are parity-conserving. \resub{These forms} will be used in \S\ref{sec: results} to forecast the detectability of the cosmic string tension parameter $G\mu$ from galaxy shape statistics.

\section{Galaxy Shape Observables}\label{sec: observables}

\subsection{Shear and Flexion: Single Sources}\label{subsec: shear+flexion-single}
\noindent A photometric galaxy dataset consists of a large number of two-dimensional galaxy images, sorted into redshift bins. As discussed above, our main interest is not the images itself, but their distortions; this can be parametrized in terms of the image moments (defined in some locally orthogonal basis $\{\hat{\bv e}_1, \hat{\bv e}_2\}$) \citep[e.g.,][]{2015RPPh...78h6901K}: 
\beq\label{eq: holic-def}
    Q_{i_1\ldots i_n} \equiv \frac{\int d\vec\theta\,I(\vec\theta)\Delta\theta_{i_1}\cdots\Delta\theta_{i_n}}{\int d\vec\theta\,I(\vec\theta)}, \qquad i\in\{1,2\}.
\eeq
This depends on the observed brightness profile $I(\vec\theta)$, and the position vector $\Delta\vec\theta$, relative to some suitably defined image center. Noting that the first moment $Q_i$ vanishes (due to the definition of $\Delta\vec\theta$), the simplest way to parametrize the image distortions is via the dimensionless \textit{shear parameters}:
\beq\label{eq: shear-from-quadrupole}
    \gamma_1 \equiv \frac{1}{2}\frac{Q_{11}-Q_{22}}{\zeta_2}, \qquad \gamma_2 \equiv \frac{1}{2}\frac{Q_{12}+Q_{21}}{\zeta_2},
\eeq
for $\zeta_2 \equiv Q_{11}+Q_{22}$. The complex shear, $\gamma_1\pm i\gamma_2$, is a spin-$\pm2$ quantity; under rotation of the $\hat{\bv e}_i$ basis vectors by an angle $\varphi$, it transforms as $(\gamma_1\pm i\gamma_2)\to e^{\pm 2i\varphi}(\gamma_1\pm i\gamma_2)$. Analysis of almost all current photometric surveys proceeds by measuring the shear of each galaxy, combining them to produce redshift-binned maps, then using the statistics of these to place constraints on cosmology, through a theoretical model of lensing and intrinsic effects \citep{2021arXiv210513549D,2020PASJ...72...16H,2023arXiv230400701D,2021A&A...645A.104A,2023arXiv230517173E,2021arXiv210513549D}.

\begin{figure}
    \centering
    \subfloat[Unlensed]{\includegraphics[width=0.243\linewidth]{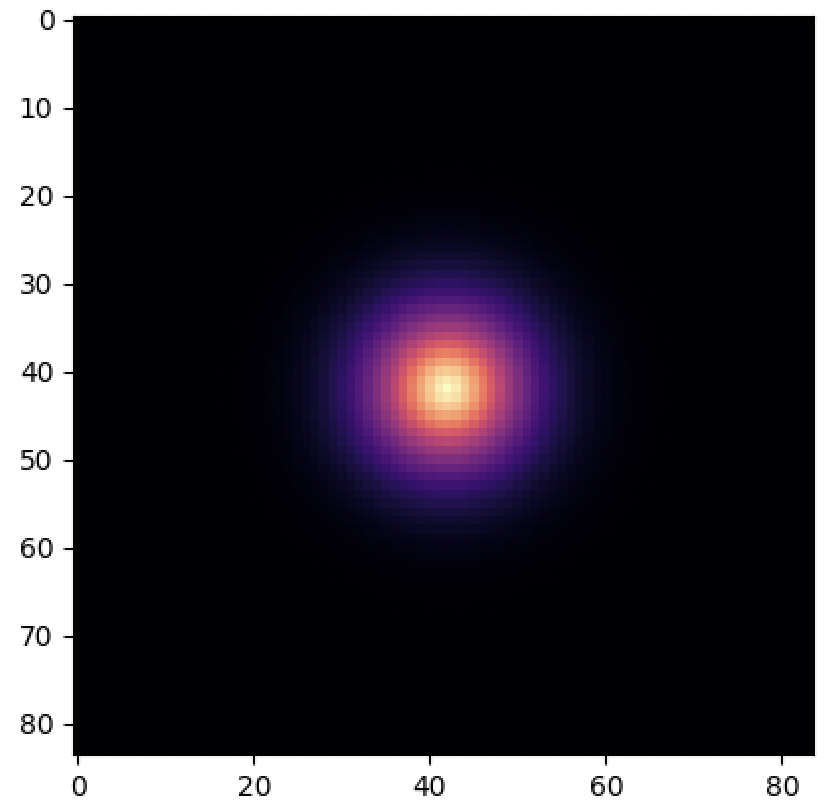}}
    \subfloat[$\gamma_1$]{\includegraphics[width=0.243\linewidth]{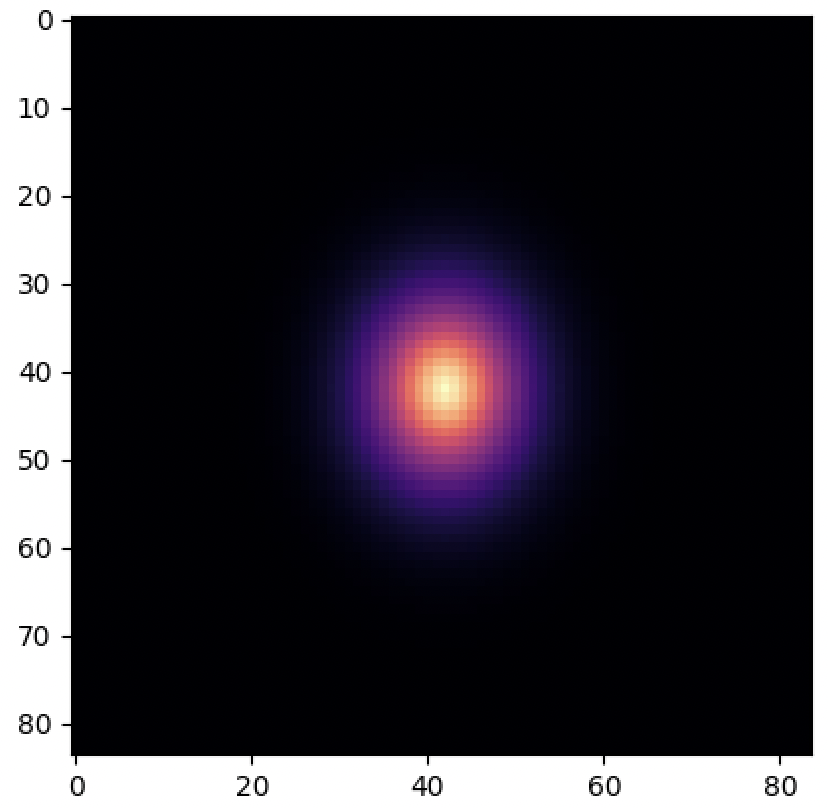}}
    \subfloat[$\F_1$]{\includegraphics[width=0.243\linewidth]{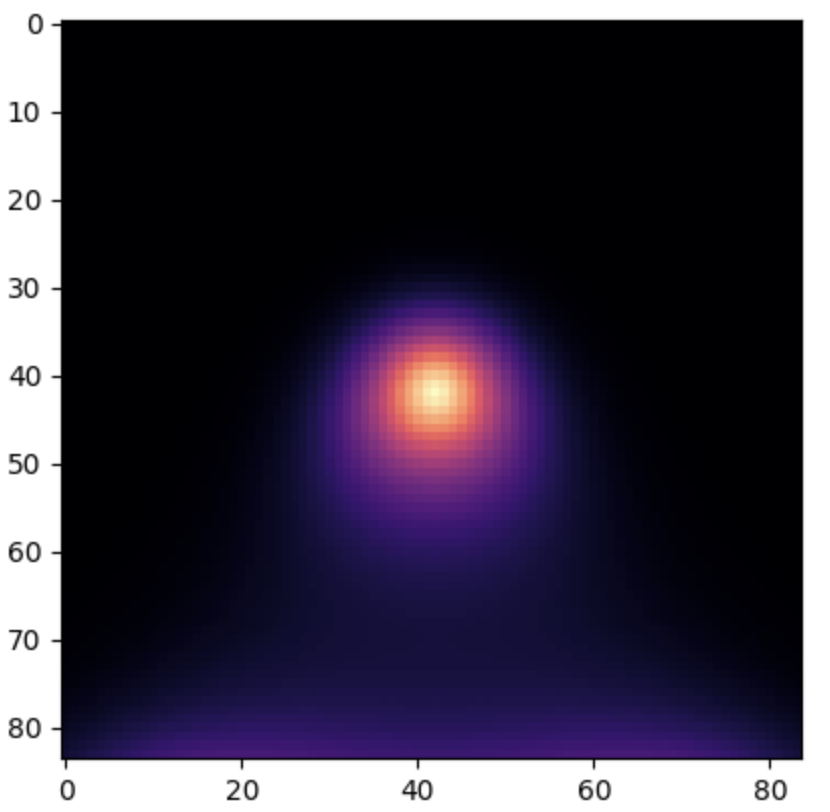}}
    \subfloat[$\G_1$]{\includegraphics[width=0.243\linewidth]{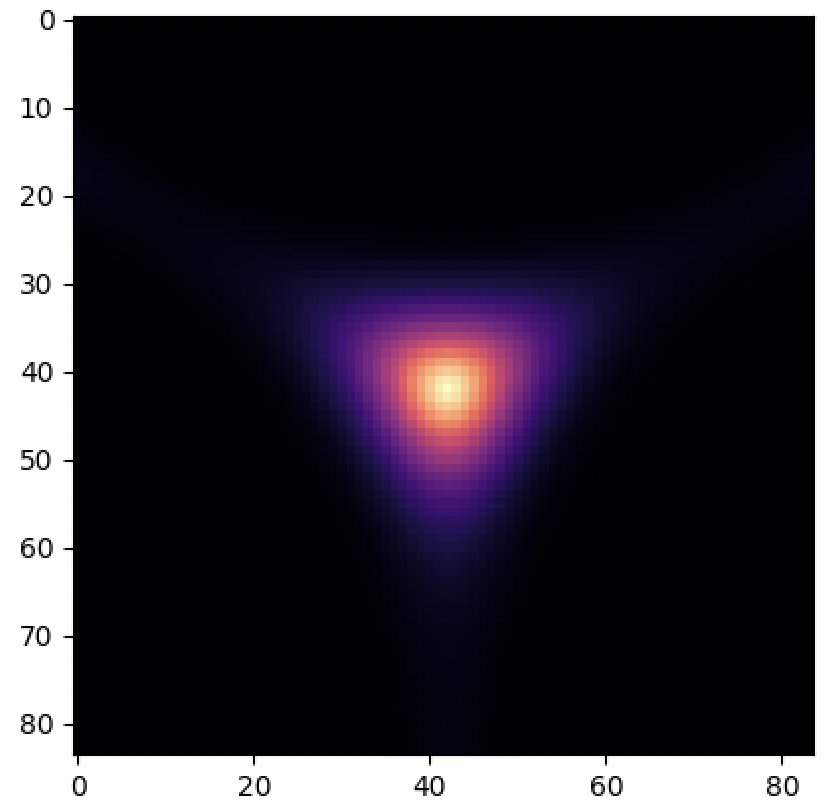}}\\
    \subfloat[$\kappa$]{\includegraphics[width=0.243\linewidth]{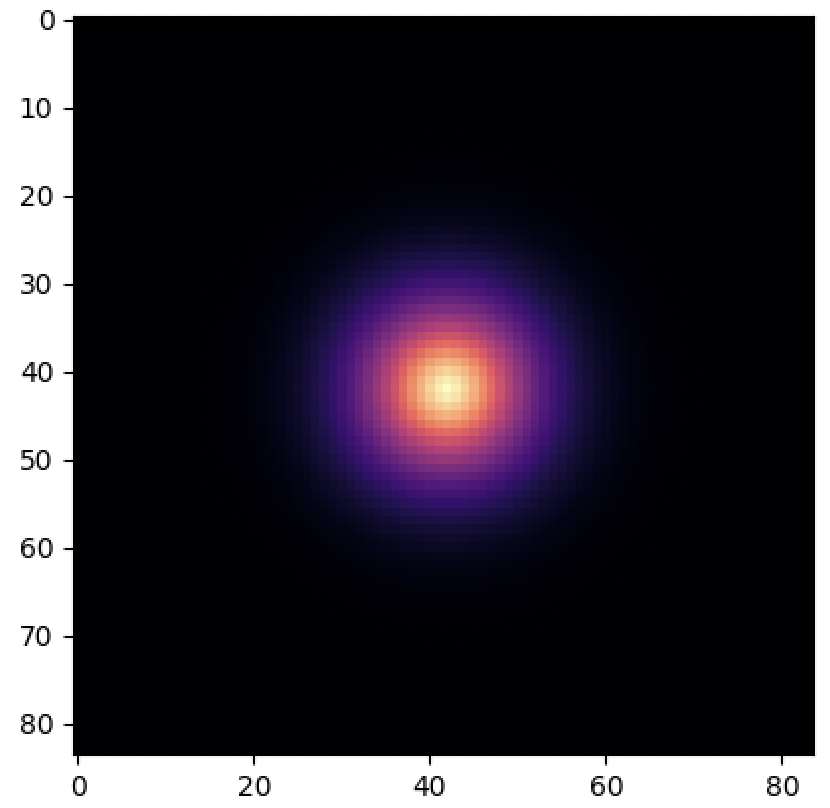}}
    \subfloat[$\gamma_2$]{\includegraphics[width=0.243\linewidth]{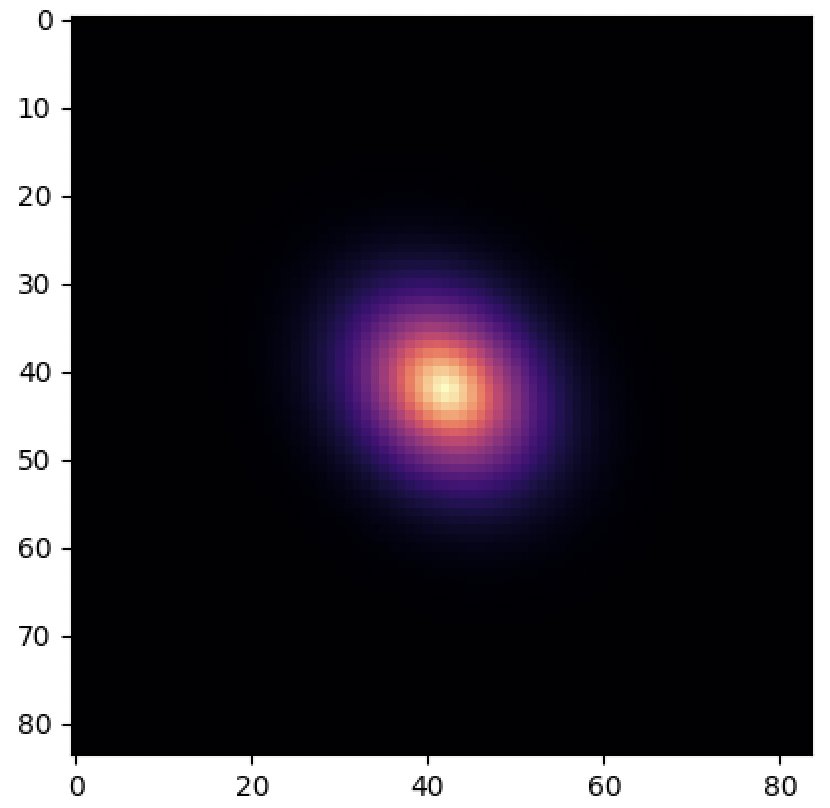}}
    \subfloat[$\F_2$]{\includegraphics[width=0.243\linewidth]{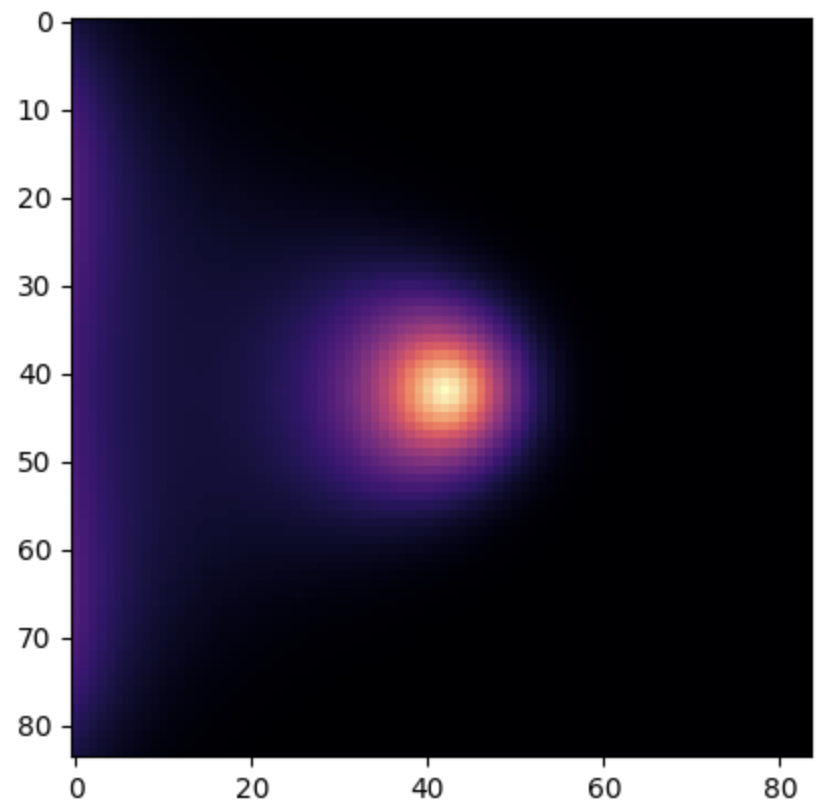}}
    \subfloat[$\G_2$]{\includegraphics[width=0.243\linewidth]{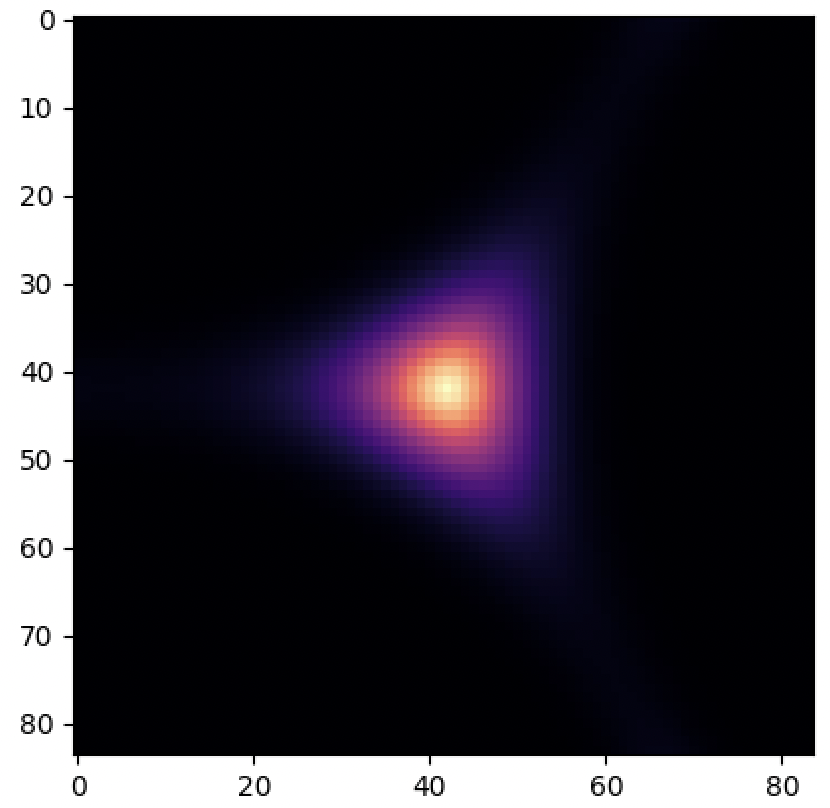}}
    \caption{Visual depiction of weak lensing distortions applied to a simulated galaxy. The unlensed galaxy in the first figure corresponds to a perfect Gaussian circle (neglecting shape noise for the purpose of visualization), and the following figures show the distortion of that galaxy with 10\% convergence/shear ($\gamma_{1,2}$ and $\kappa$), 1\% flexion of the first kind ($\F_{1,2}$) and 10\% flexion of the second kind ($\G_{1,2}$). The differing spins ($0, 2, 1, 3$ in the four columns respectively) show different characteristic distortion shapes that can be constrained from real galaxies. Figure created using the flexion code of \citep{2006MNRAS.365..414B}.}\label{fig: cartoon}
\end{figure}

Whilst the shear fully encapsulates the quadrupolar distortions of the image, there is more information to be found if one looks also to the octopole moments, through the fully symmetric tensor $Q_{ijk}$  \citep{2007ApJ...660..995O,2008ApJ...680....1O,Kogai:2020vzz}. Much as the image quadrupole can be used to form spin-$\pm2$ quantities ($\gamma_1\pm i\gamma_2$), the image octopole can be used to form spin-$\pm1$ and spin-$\pm3$ quantities, known as flexion of the first and second kind:
\beq\label{eq: F,G-from-holics-single-source}
    \F_1 \equiv \frac{4}{9}\frac{Q_{111}+Q_{122}}{\xi}, \quad \F_2 \equiv \frac{4}{9}\frac{Q_{112}+Q_{222}}{\zeta_3}, \quad \G_1 \equiv \frac{4}{3}\frac{Q_{111}-3Q_{122}}{\zeta_3}, \quad \G_2 \equiv \frac{4}{3}\frac{3Q_{112}-Q_{222}}{\zeta_3},
\eeq
where the normalization is given in terms of the image hexadecapole; $\zeta_3 \equiv Q_{1111}+2Q_{1122}+Q_{2222}$. Under rotation by $\varphi$, the complex-valued flexions transform as $(\F_1\pm i\F_2)\to e^{\pm i\varphi}(\F_1\pm i\F_2)$, $(\G_1\pm i\G_2)e^{\pm 3i\varphi}$, thus they are spin-one and spin-three respectively. Unlike a number of previous works \citep[e.g.,][]{2007ApJ...660..995O,2006MNRAS.365..414B}, we will take flexion to be defined by \eqref{eq: F,G-from-holics-single-source}, rather than considering them to be a derived quantity arising in second-order lensing. The effects of shear and flexion on a circular source are depicted in Fig.\,\ref{fig: cartoon}.

\subsection{Shear and Flexion: Full-Sky}\label{subsec: shear+flexion-full-sky}
\noindent By combining the shear and flexion measurements from a large catalog of galaxies, we can produce maps of the large-scale distortion fields binned in comoving distance, $\chi$. These are defined as
\beq\label{eq: redshift-binning-definition}
    X_a(\hn) = \int_0^{\chi_H}d\chi\,n_a(\chi)X(\chi\hn,\chi),\qquad\qquad X\in\{\gamma,\mathcal{F},\G\}
\eeq
where $X(\chi\hn,\chi)$ is the shear or flexion at position $\bv{x} = \chi\hn$ and comoving distance $\chi=\eta_0-\eta$ (which acts as a time coordinate, with horizon $\chi_H$). Here, we have averaged over a bin $n_a(\chi)$ in distance (or redshift, via $n_a(\chi)d\chi\equiv n_a(z)dz$); due to the difficulties in measuring the redshifts of lensed galaxies, these are usually broad. 

In practice, computing maps of shear and flexion over large swathes of the sky is non-trivial. Unless one restricts to very small scales, it is imperative to account for the variation of the basis vectors entering the shear and flexion definitions \eqref{eq: holic-def} across the sky.\footnote{Note that this is ignored in many previous flexion treatments (though see \citep{2011MNRAS.411.2241M} for a counterexample).} To this end, we rely on full-sky mathematics (the basics of which are outlined in Appendix \ref{appen: 2-sphere-math}), first defining the basis vectors
\beq\label{eq: basis-vectors}
    \bv{m}_{\pm}(\hn) = \frac{1}{\sqrt{2}}\left(\hat{\bv e}_\theta \mp i\hat{\bv e}_\varphi\right)(\hn) = \frac{1}{\sqrt{2}}\begin{pmatrix}\cos\theta\cos\varphi\pm i\sin\varphi\\\cos\theta\sin\varphi\mp i\cos\varphi\\ -\sin\theta\end{pmatrix},
\eeq
\citep[e.g.,][]{2012PhRvD..86h3527S,2005PhRvD..72b3516C}
where we henceforth omit the dependence on $\hn\equiv(\theta,\varphi)$ for clarity. These satisfy $m^i_{\pm}m_{\pm\,i} = 0$, $ m^i_{\pm}m_{\mp\,i} = 1$, and $m^i_{\pm}\hat n_i = 0$. Given some locally orthogonal vector set $\{\hat{\bv e}_1,\hat{\bv e}_2,\hn\}$, these asymptote to $\bv m_{\pm} = (\hat{\bv e}_1\mp i\hat{\bv e}_2)/\sqrt{2}$, restoring the definitions of \S\ref{subsec: shear+flexion-single}. 

As discussed in Appendix \ref{appen: 2-sphere-math}, any rank-$n$ tensor, $X_{i_1\ldots i_n}(\hn)$, defined on the two-sphere can be written uniquely in terms of the basis vectors $\bv{m}_{\pm}$ and coefficients with spin-$\pm n$, spin-$\pm(n-2)$, \textit{et cetera}. As such, the (redshift-binned) image quadrupole can be written in a coordinate independent form in terms of spin-$\pm2$ coefficients ${}_{\pm2}\gamma(\hn)$ (which generalize the complex $\gamma_1\pm i\gamma_2$ quantities defined above):
\beq\label{eq: quadrupole-to-shear-full-sky}
    \frac{1}{\zeta_2}Q_{ij}(\hn) = {}_{+2}\gamma(\hn)m_{+\,i}m_{+\,j}+{}_{-2}\gamma(\hn)m_{-\,i}m_{-\,j}.
\eeq
Similarly, the octopole distortions can be decomposed into a spin-$\pm1$ piece ${}_{\pm1}\F(\hn)$ and a spin-$\pm3$ contribution ${}_{\pm3}\G(\hn)$:
\beq\label{eq: octopole-to-flexion-full-sky}
    \frac{1}{\zeta_3}Q_{ijk}(\hn) &=& -\frac{9}{8}\left[\F_{ijk}(\hn) + \frac{1}{3}\G_{ijk}(\hn)\right],\\\nonumber
    \F_{ijk}(\hn) &\equiv& -\frac{1}{2\sqrt 2}\left[{}_{+1}\F(\hn)\left(\eta_{ij}m_{+\,k}+\text{2 perms.}\right)+{}_{-1}\F(\hn)\left(\eta_{ij}m_{-\,k}+\text{2 perms.}\right)\right]\\\nonumber
    \G_{ijk}(\hn) &\equiv&-\frac{1}{\sqrt 2}\left[{}_{+3}\G(\hn)m_{+i}m_{+j}m_{+k}+{}_{-3}\G(\hn)m_{-i}m_{-j}m_{-k}\right],
\eeq
where $\eta_{ij}\equiv \mathrm{diag}\left(1,\sin^2\theta\right)$ is the spatial metric on the two-sphere. Here, ${}_{\pm1}\F$ and ${}_{\pm3}\G$ are the full-sky generalization of the first and second flexion introduced previously. In the complex notation of \citep{2006MNRAS.365..414B,2005ApJ...619..741G}, ${}_{+1}\F = \F$, ${}_{-1}\F = \F^*$, with an analogous result for the spin-$\pm3$ flexion.

Since the basis functions are orthogonal, \eqref{eq: quadrupole-to-shear-full-sky}\,\&\,\eqref{eq: octopole-to-flexion-full-sky} may be used to extract the shear and flexion directly:
\beq\label{eq: g,F,G-from-Q}
    {}_{\pm2}\gamma(\hn) &=& \frac{1}{\zeta_2}m^i_{\mp}m^j_{\mp}Q_{ij}(\hn) \to \frac{1}{2\zeta_2}\left[(Q_{11}-Q_{22})\pm i(Q_{12}+Q_{21})\right](\hn)\\\nonumber 
    {}_{\pm 1}\F(\hn) &=& \frac{8}{9}\frac{\sqrt{2}}{\zeta_3}\,m_{\mp}^im_{\mp}^jm_{\pm}^kQ_{ijk}(\hn) \to \frac{4}{9\zeta_3}\left[(Q_{111}+Q_{112})\pm i(Q_{221}+Q_{222})\right](\hn)\\\nonumber
    {}_{\pm3}\G(\hn) &=& \frac{8}{3}\frac{\sqrt{2}}{\zeta_3}\,m_{\mp}^im_\mp^jm_\mp^kQ_{ijk}(\hn) \to \frac{4}{3\zeta_3}\left[(Q_{111}-3Q_{112})\pm i(3Q_{122}-Q_{222})\right](\hn),
\eeq
where we have taken the local limit on the RHS, finding equivalence with the previous results. For the remainder of this paper, we will work only with the shear and flexion components of definite spin, \textit{i.e.}\ ${}_{\pm2}\gamma(\hn)$, ${}_{\pm1}\F(\hn)$ and ${}_{\pm3}\G(\hn)$ as in previous work using the full-sky shear \citep[e.g.,][]{2005PhRvD..72b3516C,2011MNRAS.411.2241M}.

\subsection{Angular Spectra}\label{subsec: angular-spectra}
Much as Fourier modes are a convenient manner in which to characterize information for 3D observables, spherical harmonic coefficients provide a natural description of quantities on the two-sphere \citep[e.g.,][]{2005PhRvD..72b3516C,2001PhR...340..291B}. In general, a spin-$s$ quantity ${}_sX(\hn)$ can be written (cf.\,\ref{eq: spin-spherical-decomposition})
\beq
    {}_{s}X(\hn) = \sum_{\ell=0}^\infty\sum_{m=-\ell}^\ell {}_{s}X_{\ell m}\,{}_{s}Y_{\ell m}(\hn) \quad \leftrightarrow \quad {}_{s}X_{\ell m}(\hn) = \int d\hn\,{}_{s}X(\hn)\left[{}_{s}Y_{\ell m}(\hn)\right]^*,
\eeq
where ${}_{s}X_{\ell m}$ are the coefficients of ${}_sX(\hn)$, and ${}_sY_{\ell m}(\hn)$ are spin-weighted spherical harmonics, which are discussed in Appendix \ref{appen: 2-sphere-math}. In this work, the quantities of interest are the shear, and the first and second flexion, whose harmonic coefficients are respectively ${}_{\pm2}\gamma_{\ell m}$, ${}_{\pm1}\F_{\ell m}$ and ${}_{\pm3}\G_{\ell m}$. These can be further decomposed into components with definite behavior under parity-transforms:
\beq
    {}_{\pm2}\gamma_{\ell m} = \gamma^E_{\ell m}\pm i \gamma^B_{\ell m}, \quad {}_{\pm1}\F_{\ell m} = \F_{\ell m}^c\pm i\F_{\ell m}^g, \quad {}_{\pm3}\G_{\ell m} = \G^c_{\ell m}\pm i \G^g_{\ell m}; 
\eeq
where $E$ and $c$ (`curl') are parity-even, whilst $B$ and $g$ (`gradient') are parity-odd. 

Using the harmonic coefficients, we can form angular power spectra, which are the primary quantity of interest in photometric analyses. For two fields $X$ and $Y$ (e.g., $\gamma^E$ and $\F^c$), this is defined by
\beq\label{eq: angular-spectra-def}
    C_\ell^{XY,ab} \equiv \av{X^a_{\ell m}Y^{b\,*}_{\ell m}}, \qquad \widehat C_\ell^{XY,ab} \equiv \frac{1}{2\ell+1}\sum_{m=-\ell}^\ell X^a_{\ell m}Y^{b\,*}_{\ell m}
\eeq
for redshift bins $a,b$, where the expectation is taken over the underlying random fields. The RHS gives the standard power spectrum estimator. For a parity-conserving source, $EE$, $BB$, $gg$, $cc$, $Eg$, and $Bc$ spectra can be non-zero; however $EB$, $gc$, $Ec$, and $Bg$ spectra indicate parity-violating physics, since the overall spectrum is parity-odd.

\section{Sources of Shear and Flexion}\label{sec: sources}
\noindent We now ask the following question: what are the possible sources of galaxy shear and flexion, and how do these depend on scalar, vector, and tensor perturbations in the Universe? Three possibilities arise: (1) intrinsic distortions present in the true galaxy shapes induced by tidal fields, (2) extrinsic distortions from weak lensing of the galaxy photons by the intervening potentials, and (3) stochastic contributions, from observational effects and noise. Below, we consider each in turn, before discussing the corresponding power spectra in \S\ref{sec: power-spectra}.

\subsection{Intrinsic Alignments}\label{subsec: intrinsic-source}
\noindent The galaxy brightness profile, $I(\vec\theta)$, probes the projected shape of the galaxy, and is thus sensitive to any physical processes that source non-sphericity, such as large-scale tidal fields. Understanding the intrinsic contributions from scalar degrees-of-freedom has been the subject of a large array of works \citep[e.g.,][]{2001ApJ...559..552C,2004PhRvD..70f3526H,2015JCAP...10..032S,2020JCAP...01..025V,2021MNRAS.501..833K,2021JCAP...03..030S,2021MNRAS.505.2594G,2011JCAP...05..010B,Akitsu:2020jvx,Akitsu:2023eqa}; comparatively less attention has been paid to higher-spin perturbations and flexion (though see \citep{2021arXiv210709000G} for an overview of the scalar sources of intrinsic galaxy flexion, and \citep{2020JCAP...07..005B,2012PhRvD..86h3513S,2014PhRvD..89h3507S,2015JCAP...10..032S,Akitsu:2020jvx,Akitsu:2022lkl} for a detailed discussion of tensorial intrinsic alignments).

As discussed in \S\ref{subsec: shear+flexion-single}, galaxy shear and flexion measure the quadrupole and octopole image moments, $Q_{ij}$ and $Q_{ijk}$. These are themselves projections of the three-dimensional galaxy shape, quantified by the generalized moment of mass tensor $I_{i_1\ldots i_n} \propto \int d\bv{x}\,\rho(\bv{x})\,\Delta x_i\cdots\Delta x_n$ for mass distribution $\rho(\bv{x})$:
\beq
    Q_{ij}(\bv{x},\chi) &\propto& \P_{i}^{\,\,i'}\P_{j}^{\,\,j'}\left(I_{i'j'}(\bv{x},\chi)-\frac{1}{3}\delta^{\rm K}_{i'j'}\,\mathrm{Tr}\,{I_{kl}(\bv{x},\chi)}\right), \qquad Q_{ijk}(\bv{x},\chi) \propto \P_i^{\,\,i'}\P_j^{\,\,j'}\P_k^{\,\,k'}\,I_{i'j'k'}(\bv{x},\chi)
\eeq
\citep[e.g.,][]{2015JCAP...10..032S}, where $\P_{ij} \equiv \delta^{\rm K}_{ij}-\hat n_i\hat n_j$ is a projection operator and $i,j,\ldots\in\{1,2,3\}$. Crucially, the moment of mass tensor traces the metric perturbations, which thus leads to an intrinsic shear and flexion contribution.

In the simplest approximation, we can assume that the three-dimensional galaxy shape responds linearly to the Newtonian potential $\Phi$; invoking symmetry and the equivalence principle, we find $I_{ij}(\bv{x}) \propto \partial_i\partial_j \Phi(\bv{x})$, $I_{ijk}(\bv{x}) \propto \partial_i\partial_j\partial_k\Phi(\bv{x})$ \citep[e.g.,][]{2020JCAP...01..025V}. Notably, the image quadrupole depends on the trace-free combination $t_{ij}(\bv{x})\equiv \left(\partial_{i}\partial_{j}-\frac{1}{3}\delta^{\rm K}_{ij}\nabla^2\right)\Phi(\bv{x},\chi)$, thus we are assuming that $Q_{ij}$ responds linearly to the local tidal field \citep[e.g.,][]{2001ApJ...559..552C,2004PhRvD..70f3526H,2015JCAP...10..032S,2020JCAP...01..025V,2021arXiv210709000G}, \textit{i.e.}
\beq\label{eq: shear-flexion-tidal}
    \left.\frac{1}{\zeta_2}Q_{ij}(\bv{x},\chi)\right|_{\rm int} &\propto& \P_{i}^{\,\,i'}\P_{j}^{\,\,j'}t_{i'j'}(\bv{x},\chi), \qquad \left.\frac{1}{\zeta_3}Q_{ijk}(\bv{x},\chi)\right|_{V, \rm int} \propto \chi\,\P_i^{\,\,i'}\P_j^{\,\,j'}\P_k^{\,\,k'}\,\partial_{(i'}t_{j'k')}(\bv{x},\chi),
\eeq
inserting a factor of $\chi$ to the second equation on dimensional grounds. Using \eqref{eq: g,F,G-from-Q}, we thus obtain the intrinsic contributions to shear and flexion from a scalar metric perturbation at some distance $\chi$:
\beq\label{eq: scalar-g,F,G-source}
    \left.{}_{\pm2}\gamma(\bv{x},\chi)\right|_{S, \rm int} &=& b_S(\chi)\, \left[m^i_{\mp}m^j_{\mp}\partial_{i}\partial_{j}\right]\Phi(\bv{x},\chi),\\\nonumber
    \left.{}_{\pm1}\F(\bv{x},\chi)\right|_{S, \rm int} &=& \left(\frac{8\sqrt 2}{9}\chi\,\tilde b_S(\chi)\right)\,\left[m_{\mp}^im_{\mp}^jm_{\pm}^k\partial_i\partial_j\partial_k\right]\Phi(\bv{x},\chi),\\\nonumber
    \left.{}_{\pm3}\G(\bv{x},\chi)\right|_{S, \rm int} &=& \left(\frac{8\sqrt 2}{3}\chi\,\tilde b_S(\chi)\right)\,\left[m_{\mp}^im_{\mp}^jm_{\mp}^k\partial_i\partial_j\partial_k\right]\Phi(\bv{x},\chi),
\eeq
using that $\P^{i}_{\,\,j}m_{\pm}^j \equiv m_\pm^i$ and $\delta^{\rm K}_{ij}\,m^i_{\pm}m^j_{\pm}=0$. This introduces bias parameters $b_S(\chi)$ and $\tilde b_S(\chi)$: these quantify the linear responses of the shear and flexion to scalar metric perturbations, and depend primarily on the formation epoch of the galaxy. Working in $2a(\chi)/(3H_0^2\Omega_m)$ units (\resub{such that $\nabla^2\Phi\sim \delta$} in the Poisson equation), each bias is expected to be $\mathcal{O}(1)$ \citep{2015JCAP...10..032S,2021arXiv210709000G,2021MNRAS.505.2594G}, \resub{as confirmed experimentally \citep{2011JCAP...05..010B}}. To form the observables, \eqref{eq: scalar-g,F,G-source} must be averaged over a redshift bin, as in \eqref{eq: redshift-binning-definition}. 

By a similar logic, vector and tensor metric perturbation also lead to distortions in the galaxy shape. Whilst these could be modeled using the above symmetry arguments (which imply that $Q_{ij}$ is a projection of $\partial_{(i}B_{j)}$ and $h_{ij}$), this does not properly account for time evolution. Here, we instead adopt the approach of \citep{2012PhRvD..86h3513S,2014PhRvD..89h3507S}, which computes the shape distortion by first estimating the induced tidal field, $t_{ij}$. Using Fermi normal coordinates and assuming the synchronous gauge such that $h_{0i}=0$, we have \citep{2012PhRvD..86h3513S,2012PhRvD..86h3527S}
\beq\label{eq: general-tidal-field}
    t_{ij}(\bv{x},\chi) = \frac{1}{2}\left[a^{-1}(ah'_{ij})'+h_{00,ij}\right](\bv{x},\chi).
\eeq
For scalar perturbations (defined by \eqref{eq: scalar-metric}, with time-invariant $\Phi$), this recovers the results of \eqref{eq: scalar-g,F,G-source} \citep[cf.,][]{2004PhRvD..70f3526H,2011JCAP...05..010B}. For vector perturbations, the synchronous gauge implies $\left.h'_{ij}\right|_{\rm sync} = \left.-2B_{(i,j)}\right|_{\rm cN}$ (for the previously used conformal Newtonian gauge variable $B_i$), thus
\beq\label{eq: calB-def}
    \left.t_{ij}(\bv{x},\chi)\right|_{V} = -\left[a^{-1}(aB_{(i,j)})'\right](\bv{x},\chi) = -\left[\frac{\partial}{\partial\eta}+a(\chi)H(\chi)\right]B_{(i,j)}(\bv{x},\chi)\equiv -\mathcal{I}_VB_{(i,j)}(\bv{x},\chi),
\eeq
defining a time-derivative operator $\mathcal{I}_V$ for simplicity. Similarly, the tidal field induced by a tensor mode is given by
\beq\label{eq: calH-def}
    \left.t_{ij}(\bv{x},\chi)\right|_T = \frac{1}{2}\left[\frac{\partial}{\partial\eta^2}+a(\chi)H(\chi)\frac{\partial}{\partial\eta}\right]h_{ij}(\bv{x},\chi) \equiv \frac{1}{2}\mathcal{I}_Th_{ij}(\bv{x},\chi),
\eeq
defining the $\mathcal{I}_T$ operator. Using \eqref{eq: shear-flexion-tidal}, the leading-order shear and flexion contributions from vector and tensor metric perturbations are given by \citep[cf.,][]{2020JCAP...07..005B}
\beq\label{eq: vector-tensor-g,F,G-source}
    &&\left.{}_{\pm2}\gamma(\bv{x},\chi)\right|_{V, \rm int} = b_V(\chi)\, \left[m^i_{\mp}m^j_{\mp}\partial_{i}\right]\mathcal{I}_VB_{j}(\bv{x},\chi),\qquad \left.{}_{\pm1}\F(\bv{x},\chi)\right|_{V, \rm int} = \left(\frac{8\sqrt 2}{9}\chi\,\tilde b_V(\chi)\right)\,\left[m_{\mp}^{(i}m_{\mp}^jm_{\pm}^{k)}\partial_{i}\partial_j\right]\mathcal{I}_VB_{k}(\bv{x},\chi),\nonumber\\
    &&\qquad\qquad\qquad\qquad\qquad\left.{}_{\pm3}\G(\bv{x},\chi)\right|_{V, \rm int} = \left(\frac{8\sqrt 2}{3}\chi\,\tilde b_V(\chi)\right)\,\left[m_{\mp}^{(i}m_{\mp}^{j}m_{\mp}^{k)}\partial_i\partial_j\right]\mathcal{I}_VB_k(\bv{x},\chi),
\eeq
and
\beq\label{eq: tensor-g,F,G-source}
    &&\left.{}_{\pm2}\gamma(\bv{x},\chi)\right|_{T, \rm int} = b_T(\chi)\, \left[m^i_{\mp}m^j_{\mp}\right]\mathcal{I}_Th_{ij}(\bv{x},\chi),\qquad\left.{}_{\pm1}\F(\bv{x},\chi)\right|_{T, \rm int} = \left(\frac{8\sqrt 2}{9}\chi\,\tilde b_T(\chi)\right)\,\left[m_{\mp}^{(i}m_{\mp}^jm_{\pm}^{k)}\partial_{i}\right]\mathcal{I}_Th_{jk}(\bv{x},\chi),\nonumber\\
    &&\qquad\qquad\qquad\qquad\qquad\left.{}_{\pm3}\G(\bv{x},\chi)\right|_{T, \rm int} = \left(\frac{8\sqrt 2}{3}\chi\,\tilde b_T(\chi)\right)\,\left[m_{\mp}^{(i}m_{\mp}^jm_{\mp}^{k)}\partial_i\right]\mathcal{I}_Th_{jk}(\bv{x},\chi).
\eeq
As before, the bias parameters $b_X(\chi)$ and $\tilde b_X(\chi)$ encode the responses of shear and flexion to metric perturbations \resub{(which we allow to differ between scalars, vectors, and tensors, for full generality)}, and are expected to be $\mathcal{O}(1)$ in $2a(\chi)/(3H_0^2\Omega_m)$ units, \resub{whence we can switch between density and potential fluctuations \citep{2012PhRvD..86h3513S,2021MNRAS.505.2594G,2014PhRvD..89h3507S}, with $b_X\sim 0.1$ found experimentally \citep{2011JCAP...05..010B}}. Given these results, coupled with the power spectra of $\Phi$, $B_i$, and $h_{ij}$ \eqref{eq: Phi,B,h-correlators}, we may compute the statistics of shear and flexion; this is detailed in \S\ref{sec: power-spectra}.

The above derivation makes a number of limiting assumptions. Firstly, we assume that the galaxy at some lookback time $\eta = \eta_0-\chi$ traces the tidal field at the same same time, rather than its full history. Secondly, we neglect dependence of the bias parameters $b_{V,T}$ on the wavelength of vector or tensor mode. A number of works have considered a more nuanced ``fossil'' approach \citep{2014PhRvD..89h3507S,2020JCAP...07..005B,Akitsu:2022lkl,Akitsu:2022lkl}, by explicitly computing the impact of tidal fields on the second-order density field, both in theory and simulations. For primordial gravitational waves, this was found to somewhat enhance the intrinsic alignment signal, and give some scale-dependence \citep[cf.][Fig.\,1]{Akitsu:2020jvx}. For the purposes of forecasting, we neglect these effects since (a) a uniform rescaling is fully degenerate with the unknown bias parameters if the perturbations are separable in time and space, and (b) the scale-dependence is weak for tensor (and vector) modes with $k\gtrsim 10^{-3}\hMpc$. We proceed with a general warning however that the precise values of vector and tensor biases are not well understood.

\subsection{Weak Lensing}\label{subsec: extrinsic-source}
\noindent Next, we consider extrinsic distortions to the observed galaxy shape, which arise from the gravitational deflection of photons as they travel from the source galaxy to the observer \citep[e.g.,][]{1995ApJ...449..460K,2001PhR...340..291B,2005PhRvD..72b3516C}. This is a well-known source of shear, convergence, and flexion \citep{2001PhR...340..291B,2006MNRAS.365..414B,2005ApJ...619..741G,2007ApJ...660..995O}, and encodes information on all types of metric perturbation.

The effects of lensing are encoded in the transformation between unlensed and lensed coordinates, $\theta_i$ and $\theta_i'$;
\beq\label{eq: theta-distortions}
    \theta_i'(\vec\theta) &=& A_{ij}\theta^j + \frac{1}{2}D_{ijk}\theta^j\theta^k+\mathcal{O}(\theta^3), 
\eeq
where the matrices $A_{ij}$ and $D_{ijk}$ are linear transformations of the scalar perturbation $\Phi(\bv{x},\chi)$ (in the weak-field limit):
\beq\label{eq: A-ij-def}
    A^{ij}(\bv{x},\chi) = \delta^{ij}_{\rm K} - \partial^i_\theta\partial^j_\theta\phi(\bv{x},\chi), \qquad D^{ijk}(\bv{x},\chi) = - \partial^i_\theta\partial^j_\theta\partial^k_\theta\phi(\bv{x},\chi)
\eeq
\citep[e.g.,][]{2001PhR...340..291B,2006MNRAS.365..414B,1995ApJ...449..460K}. This uses the projected potential $\phi(\bv{x},\chi)$ for a source at $\bv{x}\equiv\chi\hn$ and redshift $\chi$:
\beq\label{eq: phi-def}
    \phi(\chi\hn,\chi) &=& \frac{2}{c^2}\int_0^\chi d\chi'\,\frac{\chi-\chi'}{\chi\chi'}\Phi(\chi'\hn,\chi').
\eeq
\citep[e.g.,][]{2005PhRvD..72b3516C}, which integrates over the distribution of matter between the source (at $\chi'=\chi$) and the observer (at $\chi'=0$).

Associated with this transformation is a distortion in the galaxy image, $I(\vec\theta)$, which manifests as a quadrupole moment (and beyond). From the definitions in \S\ref{sec: observables}, this sources convergence, shear, and flexion fields of the form:
\beq\label{eq: Aij,Dijk-def}
    -A_{ij}(\bv{x},\chi) &=& \kappa(\bv{x},\chi) \eta_{ij} + {}_{+2}\gamma(\bv{x},\chi)m_{+i}m_{+j}+{}_{-2}\gamma(\bv{x},\chi)m_{-i}m_{-j}\\\nonumber
    D_{ijk}(\bv{x},\chi)&=& \F_{ijk}(\bv{x},\chi)+\G_{ijk}(\bv{x},\chi)
\eeq
\citep[e.g.,][]{2005PhRvD..72b3516C}, where $\F_{ijk}$ and $\G_{ijk}$ can be written in terms of spin-$\pm1$ and $\pm3$ components as in \eqref{eq: octopole-to-flexion-full-sky}.\footnote{We will ignore convergence contributions in this work since they are largely determined by shear (though see also \citep{2014ApJ...780L..16H}).} Unlike for the intrinsic contribution, there is no prefactor of $1/3$ in front of $\G_{ijk}$, thus the spin-$\pm3$ flexion is here enhanced relative to the spin-$\pm1$ form. In the local limit, \eqref{eq: Aij,Dijk-def} reduces to the standard relations \citep[e.g.,][]{2006MNRAS.365..414B,2005ApJ...619..741G}:
\beq\label{eq: A-ij-flat-sky}
    A_{ij} &=& \begin{pmatrix} 1-\kappa-\gamma_1 & -\gamma_2 \\ -\gamma_2 & 1-\kappa+\gamma_1\end{pmatrix},\quad  D_{ij1} = -\frac{1}{2}\begin{pmatrix}3\F_1+\G_1 & \F_2+\G_2 \\ \F_2+\G_2 & \F_1-\G_1\end{pmatrix}, \quad  D_{ij2} = -\frac{1}{2}\begin{pmatrix}\F_2+\G_2 & \F_1-\G_1 \\ \F_1-\G_1 & 3\F_2-\G_2\end{pmatrix},
\eeq
where ${}_{\pm2}\gamma = \gamma_1\pm\gamma_2$ \textit{et cetera}. As in \eqref{eq: g,F,G-from-Q}, orthogonality of the basis vectors allows us to derive expressions for the shear and flexion in terms of $\phi$:
\beq\label{eq: shear,flexion-from-phi}
    \left.{}_{\pm 2}\gamma(\chi\hn,\chi)\right|_{S,\rm ext} &=& \left[m_{\mp\,i} m_{\mp\,j} \partial^i_\theta\partial^j_\theta\right]\phi(\chi\hn,\chi),\\\nonumber
    \left.{}_{\pm 1}\F(\chi\hn,\chi)\right|_{S,\rm ext} &=& \sqrt{2}\,\left[m_{\mp\,i}m_{\mp\,j}m_{\pm\,k}\,\partial^{(i}_\theta\partial^{j_{} }_\theta\partial^{k)}_\theta\right]\phi(\chi\hn,\chi),\\\nonumber \left.{}_{\pm 3}\G(\chi\hn,\chi)\right|_{S,\rm ext} &=& \sqrt{2}\,\left[m_{\mp\,i}m_{\mp\,j}m_{\mp\,k}\,\partial^i_\theta\partial^j_\theta\partial^k_\theta\right]\phi(\chi\hn,\chi).
\eeq
Alternatively, shear and flexion can be written in terms of ``spin-derivatives'', themselves defined in Appendix \ref{appen: 2-sphere-math}. In this form:
\beq\label{eq: lensing-from-spin-derivs}
    \left.{}_{+2}\gamma(\chi\hn,\chi)\right|_{S,\rm ext} &=&  \frac{1}{2}\edth\edth\phi(\chi\hn,\chi), \\\nonumber
    \left.{}_{+1}\F(\chi\hn,\chi)\right|_{S,\rm ext} = -\frac{1}{6}\left[\edth\edth\bar\edth+\edth\bar\edth\edth+\bar\edth\edth\edth\right]\phi(\chi\hn,\chi) &\qquad&
    \left.{}_{+3}\G(\chi\hn,\chi)\right|_{S,\rm ext} = -\frac{1}{2}\edth\edth\edth\phi(\chi\hn,\chi)
\eeq
\citep[e.g.,][]{2005PhRvD..72b3516C}, where $\edth$ ($\bar\edth$) raises (lowers) the spin by one, and components with negative spins are obtained by interchanging $\edth$ and $\bar\edth$. This allows for convenient computation of the statistics of shear and flexion from those of $\phi$.

Vector and tensor perturbations provide an additional source of photon deflections, and thus contributions to shear and flexion. To obtain their form, we use the general gauge-invariant results of \citep[Eq.\,64]{2012PhRvD..86h3527S} (see also \citep{2008PhRvD..77j3515S}), finding:
\beq\label{eq: lensing-schmidt-paper}
    \left.{}_{\pm2}\gamma(\chi\hn,\chi)\right|_{S,\rm ext} &=& 2\int_0^{\chi}d\chi'\frac{\chi'}{\chi}(\chi-\chi')m_\mp^im_\mp^j\partial_i\partial_j\Phi(\chi'\hn,\chi'),\\\nonumber
    \left.{}_{\pm2}\gamma(\chi\hn,\chi)\right|_{V,\rm ext} &=& -\int_0^{\chi}d\chi'\left[\left(1-2\frac{\chi'}{\chi}\right)m_{\mp}^k\partial_{\pm}B_k+\frac{\chi'}{\chi}(\chi-\chi')\left(\hat n^km_\mp^im_\mp^j\partial_i\partial_jB_k\right)\right](\chi'\hn,\chi'),\\\nonumber
    \left.{}_{\pm2}\gamma(\chi\hn,\chi)\right|_{T,\rm ext} &=& -\frac{1}{2}h_{\pm}(\chi\hn,\chi)-\frac{1}{2}h_\pm(\bv 0,0)\\\nonumber
    &&\quad-\int_0^{\chi}d\chi'\left[\left(1-2\frac{\chi'}{\chi}\right)[(\partial_\pm h_{lk})m_\mp^l\hat n^k]-\frac{1}{\chi}h_\pm+\frac{\chi'}{2\chi}(\chi-\chi')m_\mp^im_\mp^j(\partial_i\partial_jh_{kl})\hat n^k\hat n^l\right](\chi'\hn,\chi'),
\eeq
where $h_{ij}(\bv 0,0)$ is the tensor perturbation evaluated at the observers position, $\partial_\pm \equiv m_{\mp}^i\partial_i$, and $h_{\pm}\equiv m_{\mp}^im_{\mp}^jh_{ij}$. Noting that $\partial_\theta^i = \chi^{-1}\P_{j}^{\,\,i}\partial_j$, it is straightforward to show the the scalar piece equals that of \eqref{eq: shear,flexion-from-phi}. For the vector piece (which is equivalent to that of \citep{2012JCAP...10..030Y}), we have two contributions, depending on the first and second derivatives of $B$ respectively; in practice, the second term is expected to dominate on all but the largest scales. The tensor spectrum contains unintegrated terms corresponding to the field at the source and observer positions; the former contributes analogously to the intrinsic piece discussed in \S\ref{subsec: intrinsic-source}, and can be absorbed by redefining $\mathcal{I}_Th_{ij}\to\left[\mathcal{I}_T- 1/(2b_T(\chi))\right]h_{ij}$. The latter piece contributes only to the $\ell=2$ modes \citep{2012PhRvD..86h3527S,2012PhRvD..86h3513S}, and acts to remove the $k\to0$ limit.

The approach of \citep{2012PhRvD..86h3527S} may be further extended to compute the lensing contribution to the flexion parameters $\F$ and $\G$ from vector and tensor modes. However, this is non-trivial, and, foreshadowing the conclusions of \S\ref{sec: results}, unlikely to be of significant practical use. For our purposes, we will make the following assumption for vector and tensor sources:
\beq\label{eq: flexion-lensing-approx}
    \left.{}_{+1}\F(\chi\hn,\chi)\right|_{\rm ext} \approx -\left.\bar\edth{}_{+2}\gamma(\chi\hn,\chi)\right|_{\rm ext}, \qquad  \left.{}_{+3}\G(\chi\hn,\chi)\right|_{\rm ext} \approx -\left.\edth{}_{+2}\gamma(\chi\hn,\chi)\right|_{\rm ext},\\\nonumber
    \left.{}_{-1}\F(\chi\hn,\chi)\right|_{\rm ext} \approx -\left.\edth{}_{-2}\gamma(\chi\hn,\chi)\right|_{\rm ext}, \qquad  \left.{}_{-3}\G(\chi\hn,\chi)\right|_{\rm ext} \approx -\left.\bar\edth{}_{-2}\gamma(\chi\hn,\chi)\right|_{\rm ext}.
\eeq
This is justified by noting that the flexion lensing tensor $D^{ijk} = \partial^i_\theta A^{jk}$ contains one additional spin-derivative compared to $A^{jk}$ (since $m_{-i}\partial_{\theta}^i=-\edth$ and $m_{+i}\partial_\theta^i=-\bar\edth$ locally). On small scales, $\bar\edth$ and $\edth$ permute, thus this recovers the scalar results of \eqref{eq: lensing-from-spin-derivs} \citep[e.g.,][]{2005ApJ...619..741G,2006MNRAS.365..414B}. In practice, the contribution of flexion is strongly subdominant outside the small-scale limit, thus this approximation is generally appropriate.

\subsection{Other Sources}\label{subsec: stochastic-source}
\noindent The last major source of shear and flexion is stochasticity due to both experimental noise and the intrinsic variation in galaxy shapes \citep[e.g.,][]{2021arXiv210513549D}. Here, we assume that the local measurement of a given observable, $X\in\{\gamma,\F,\G\}$, follows a Gaussian distribution with variance $\sigma^2_X$, and (ignoring other sources) zero mean, \textit{i.e.}
\beq
    \hat X_i \sim \mathcal{N}\left(0,\frac{\sigma_{X}^2}{2}\right), \qquad i\in\{1,2\},
\eeq
for the component projected onto the $\hat{\bf e}_i$ axis. We will further assert that the noise is uncorrelated (such that each galaxy is independent), and that the variances are independent of redshift and position. This yields a Poissonian distribution, with a simple power spectrum (cf.\,\S\ref{subsec: angular-spectra}). Typical values for the noise amplitudes are $\sigma_\gamma = 0.4$ \citep{2021MNRAS.505.2594G}, and $\sigma_{\F} = \sigma_\G/3 = 0.009\,\mathrm{sr}^{-1}$ \citep{2007ApJ...660..995O}; the additional factor of three in the second flexion arises from its definition in \eqref{eq: octopole-to-flexion-full-sky}.

Additional contributions to shear and flexion arise from higher-order effects, neglected in the above sections. Under our previous assumption, conventional physics generates only parity-conserving modes ($\gamma^E$, $\F^g$, $\G^g$), thus any robust detection of the parity-breaking modes ($\gamma^B$, $\F^c$, $\G^c$) would imply evidence for new phenomena. In praxis, second-order effects can generate $B$- and $c$-modes even from scalar sources, and thus yield correlators such as $\av{\gamma^B\gamma^{B*}}$, which requires new physics). For example, the intrinsic alignment of galaxies contains a term $\propto t_{ij}\delta$ (for matter overdensity $\delta$) due to the non-uniform distribution of source galaxies; as shown in \citep{2004PhRvD..70f3526H,2020JCAP...07..005B}, this generates both shear $E$ and $B$ components, whose spectra involve the convolution of two $P_\Phi$ spectra (via the Poisson equation). Similarly, post-Born effects in weak lensing lead to a shear $B$-mode signal \citep{2003PhRvD..68h3002H}, with another generated from the relation between shear and ``reduced shear'', $g\equiv \gamma/(1-\kappa)$ \citep{2012PhRvD..86h3513S}. In the forecasts of \S\ref{sec: results}, we will contributions of thse phenomena to the $E$- and $g$-mode spectra (since they are heavily subdominant), but add a rough estimate of their size in the $B$-mode spectra, following previous works. Importantly, such effects cannot contribute to the parity-odd correlators such as $EB$, since their underlying physics is parity-conserving. Ignoring systematic contamination, this makes $EB$-spectra a bountiful place in which to search for new physics.

\section{Power Spectra}\label{sec: power-spectra}
\noindent Armed with the statistics of scalar, vector, and tensor perturbations and their relation to shear and flexion, we may now proceed to compute the observed angular power spectra, $C_\ell^{XY}$. We will first compute the general full-sky form for both intrinsic and extrinsic sourcing, before considering the flat-sky simplifications in \S\ref{sec: flat-spectra}.

\subsection{Full-Sky Spectra}\label{subsec: full-spectra}
\noindent In the weak-field regime, the observable quantities $\gamma^{E,B}$, $\F^{g,c}$ and $\G^{g,c}$ are linear transformations of the underlying metric perturbations (\S\ref{sec: observables}). As such, the power spectra, $C_\ell^{XY}$, depend on two copies of the perturbation in question and can be written in the general form
\beq\label{eq: general-spectra}
    \left.C_{\ell}^{XY}(\chi,\chi')\right|_S &=&  \frac{2}{\pi}\int_0^\infty k^2dk\,F_{\ell}^{X,S}(k,\chi)F_{\ell}^{Y,S*}(k,\chi')P_\Phi(k,\chi,\chi')\\\nonumber
    \left.C_{\ell}^{XY}(\chi,\chi')\right|_V &=&  \frac{2}{\pi}\int_0^\infty k^2dk\,F_{\ell}^{X,V}(k,\chi)F_{\ell}^{Y,V*}(k,\chi')\left[P_{B_+}(k,\chi,\chi')\pm P_{B_-}(k,\chi,\chi')\right]\\\nonumber
    \left.C_{\ell}^{XY}(\chi,\chi')\right|_T &=&  \frac{2}{\pi}\int_0^\infty k^2dk\,F_{\ell}^{X,T}(k,\chi)F_{\ell}^{Y,T*}(k,\chi')\left[P_{h_+}(k,\chi,\chi')\pm P_{h_-}(k,\chi,\chi')\right],
\eeq
where the transfer function $F_\ell^{X,P}$ encodes the (local) response of the field $X$ to the perturbation $P\in\{S,V,T\}$, and the second and third lines have a negative sign only for parity-odd spectra. In practice, the observables are binned in redshift via \eqref{eq: redshift-binning-definition}, which modifies the kernels
\beq\label{eq: redshift-binned-kernels}
    F_\ell^{X,S}(k,\chi) \to F^X_{\ell,a}(k) \equiv \int_0^{\chi_H} d\chi\,n_a(\chi)F_\ell^{X,S}(k,\chi)
\eeq
and $C_\ell^{XY}(\chi,\chi')\to C_{\ell,ab}^{XY}$ (absorbing the redshift-dependent parts of $P_\Phi,P_B,P_h$ into the transfer functions). In the below, we will present the relevant transfer functions, which fully specify the observable power spectra.

\subsubsection{Intrinsic Contributions}\label{subsec: intrinsic-spectra}
\noindent As shown in \citep{2012PhRvD..86h3527S}, the angular power spectrum of a general spin-$\pm s$ quantity ${}_{\pm s}\mathcal{W}(\chi\hn,\chi)$ (with $s>0$), can be obtained as follows:
\begin{enumerate}
    \item Compute the contributions to ${}_{\pm s}\mathcal{W}(\chi\hn,\chi)$ arising from a single Fourier mode parallel to the $\hz$ axis, \textit{i.e.}\ $\vk = k\hz$, and a single helicity state $\lambda$.
    \item Apply spin-raising and spin-lowering operators (see Appendix \ref{appen: 2-sphere-math}) to ${}_{\pm s}\mathcal{W}(\chi\hn,\chi;k\hz,\lambda)$ to obtain the spin-zero quantities $\bar\edth^s{}_{+s}\mathcal{W}(\chi\hn,\chi;k\hz,\lambda)$ and $\edth^s{}_{-s}\mathcal{W}(\chi\hn,\chi;k\hz,\lambda)$.
    \item Extract the spherical harmonic coefficients, ${}_{\pm s}\mathcal W_{\ell m}(\chi;k\hz,\lambda)$, by integrating the scalars against $Y_{\ell m}^*(\hn)$ \eqref{eq: spin-sph-coeffs2}. These may be split into $E$ and $B$ (or $g$ and $c$) contributions, as in \S\ref{subsec: angular-spectra}.
    \item Compute the angular power spectra by taking the expectation of two $\mathcal W^X_{\ell m}(\chi;k\hz,\lambda)$ functions, integrating over Fourier modes $\vk$ and averaging over $m$. Via isotropy, this gives the full power spectrum from all $\vk$ modes, not just $\vk = k\hz$, and allows extraction of the associated transfer functions via \eqref{eq: general-spectra}.
\end{enumerate}

In Appendix \ref{appen: intrinsic-vector-derivations}, we utilize this scheme to derive the shear and flexion power spectra sourced by intrinsic vector alignments (a novel feature of this work). Since the computations for scalars and tensors are analogous (and the shear results can be found in previous work, e.g., \citep{2012PhRvD..86h3513S,2015JCAP...10..032S} for scalars, though with less attention paid to parity-breaking contributions), we present only the final expressions here. Before redshift integration, the shear kernels are given by (dropping factors of $i^\ell$)
\beq\label{eq: shear-kernels}
    \left.F^{\gamma^E,P}_\ell(k,\chi)\right|_{\rm int} &=& \sqrt{\frac{(\ell-2)!(\ell+s_P)!}{(\ell+2)!(\ell-s_P)!}}\frac{b_P(\chi)}{2\chi^2}\left(\frac{\chi}{\sqrt 2}\right)^{s_P}\mathrm{Re}\left[\hat Q^{(+1)}_{\gamma,P}(x)\right]\frac{j_\ell(x)}{x^{s_P}}\mathcal{I}_P(\chi)\\\nonumber
    \left.F^{\gamma^B,P}_\ell(k,\chi)\right|_{\rm int} &=& \sqrt{\frac{(\ell-2)!(\ell+s_P)!}{(\ell+2)!(\ell-s_P)!}}\frac{b_P(\chi)}{2\chi^2}\left(\frac{\chi}{\sqrt 2}\right)^{s_P}\mathrm{Im}\left[\hat Q^{(+1)}_{\gamma,P}(x)\right]\frac{j_\ell(x)}{x^{s_P}}\mathcal{I}_P(\chi),
\eeq
where $P\in\{S,V,T\}$ labels the source-type, with corresponding source spin $s_P = \{0,1,2\}$. This involves two sets of operators: (a) the time-derivative operators $\mathcal{I}_P$ of \S\ref{subsec: intrinsic-source} (with $\mathcal{I}_S(\chi) = 1$), acting on the source power spectrum; (b) the spatial $\hat Q_{\gamma,P}(x)$ operators acting on the spherical Bessel functions, $j_\ell(x)$, with $x = k\chi$. These are low-order polynomials in $x$ and $\partial_x$, and their form is given explicitly in \eqref{eq: Q-gamma} of Appendix \ref{appen: kernels}.

A similar calculation yields the flexion kernels:
\beq\label{eq: flexion-kernels}
    \left.F^{\F^g,P}_\ell(k,\chi)\right|_{\rm int} &=& \sqrt{\frac{(\ell-1)!(\ell+s_P)!}{(\ell+1)!(\ell-s_P)!}}\frac{4\tilde b_P(\chi)}{27\chi^2}\left(\frac{\chi}{\sqrt{2}}\right)^{s_P}\mathrm{Re}\left[\hat Q_{\F,P}^{(+1)}(x)\right]\frac{j_\ell(x)}{x^{s_P}}\mathcal{I}_P(\chi)\\\nonumber
    \left.F^{\F^c,P}_\ell(k,\chi)\right|_{\rm int} &=& -\sqrt{\frac{(\ell-1)!(\ell+s_P)!}{(\ell+1)!(\ell-s_P)!}}\frac{4\tilde b_P(\chi)}{27\chi^2}\left(\frac{\chi}{\sqrt{2}}\right)^{s_P}\mathrm{Im}\left[\hat Q_{\F,P}^{(+1)}(x)\right]\frac{j_\ell(x)}{x^{s_P}}\mathcal{I}_P(\chi)\\\nonumber
    \left.F^{\G^g,P}_\ell(k,\chi)\right|_{\rm int} &=& \sqrt{\frac{(\ell-3)!(\ell+s_P)!}{(\ell+3)!(\ell-s_P)!}}\frac{4\tilde b_P(\chi)}{3\chi^2}\left(\frac{\chi}{\sqrt{2}}\right)^{s_P}\mathrm{Re}\left[\hat Q_{\G,P}^{(+1)}(x)\right]\frac{j_\ell(x)}{x^{s_P}}\mathcal{I}_P(\chi)\\\nonumber
    \left.F^{\G^c,P}_\ell(k,\chi)\right|_{\rm int} &=& -\sqrt{\frac{(\ell-3)!(\ell+s_P)!}{(\ell+3)!(\ell-s_P)!}}\frac{4\tilde b_P(\chi)}{3\chi^2}\left(\frac{\chi}{\sqrt{2}}\right)^{s_P}\mathrm{Im}\left[\hat Q_{\G,P}^{(+1)}(x)\right]\frac{j_\ell(x)}{x^{s_P}}\mathcal{I}_P(\chi);
\eeq
these depend on a new set of derivative operators $\hat Q_{\F,P}$ and $\hat Q_{\G,P}$ given in \eqref{eq: Q-F} and \eqref{eq: Q-G} respectively. Each of these kernels can be binned in redshift via \eqref{eq: redshift-binned-kernels}.

Notably, the above scalar operators $\hat{Q}_{X,S}(x)$ are explicitly real, thus the scalar $B$- and $c$-mode transfer functions vanish. This is as expected; a scalar contains insufficient degrees-of-freedom to generate parity-breaking distortions. As such, if one is interested only in scalar physics (which is usually the case), we need analyze only $E$ and $g$ mode spectra. In contrast, vector and tensor modes generically source both $E/g$ and $B/c$ distortions, making the latter observable a smoking gun for new physics (once noise and higher-order effects are taken into account). Furthermore, we note that the vector and tensor spectra entering \eqref{eq: general-spectra} involve either the sum or the difference of the two helicity spectra; the enters in the $EB$, $gc$, $Ec$, and $Bg$ correlators, and implies that such spectra provide access to the parity-violating physical sector \citep[e.g.,][]{2020JCAP...07..005B}. 

\subsubsection{Extrinsic Contributions}\label{subsec: extrinsic-spectra}
\noindent For scalar perturbations, computation of the weak lensing power spectrum is straightforward \citep{2005PhRvD..72b3516C}. First, we write the definition of shear and flexion in terms of the lensing potential $\phi$ \eqref{eq: lensing-from-spin-derivs}, expanding the latter in spherical harmonics:
\beq
    \left.{}_{+2}\gamma(\chi\hn,\chi)\right|_{S,\rm ext} =& & \frac{1}{2}\sum_{\ell m}\phi_{\ell m}(\chi)\edth\edth Y_{\ell m}(\hn), \\\nonumber
    \left.{}_{+1}\F(\chi\hn,\chi)\right|_{S,\rm ext} = -\frac{1}{6}\sum_{\ell m}\phi_{\ell m}(\chi)\left[\edth\bar\edth\bar\edth+\bar\edth\edth\bar\edth+\bar\edth\bar\edth\edth\right]&&Y_{\ell m}(\hn), \quad \left.{}_{+3}\G(\chi\hn,\chi)\right|_{S,\rm ext} = -\frac{1}{2}\sum_{\ell m}\phi_{\ell m}(\chi)\edth\edth\edth Y_{\ell m}(\hn).
\eeq
Using the spin-derivative relations given in \eqref{eq: spin-weighted-sph-def}, this can be rewritten in terms of spin-weighted spherical harmonics, allowing the relevant $E$ and $B$ mode coefficients to be extracted via orthogonality \eqref{eq: spin-sph-orthonormality}: 
\beq
    &&\qquad\qquad\qquad\qquad\qquad\qquad\qquad\qquad\left.\gamma^E_{\ell m}(\chi)\right|_{S,\rm ext} =  \frac{1}{2}\sqrt{\frac{(\ell+2)!}{(\ell-2)!}}\phi_{\ell m}(\chi), \\\nonumber
    &&\left.i\F^g_{\ell m}(\chi)\right|_{S,\rm ext} = \frac{1}{6}\sum_{\ell m}\phi_{\ell m}(\chi)\sqrt{\frac{(\ell+1)!}{(\ell-1)!}}\left[(\ell-1)(\ell+2)+2\ell(\ell+1)\right], \quad \left.i\G^g_{\ell m}(\chi)\right|_{S,\rm ext} = \frac{1}{2}\sqrt{\frac{(\ell+3)!}{(\ell-3)!}}\phi_{\ell m}(\chi).
\eeq
As in \S\ref{subsec: intrinsic-spectra}, scalar physics sources only $E$ and $g$ modes at leading order. To form power spectra we proceed to expand the scalar perturbation in Fourier space and evaluate its two-point correlator yielding:
\beq\label{eq: lensing-harmonics}
    \phi_{\ell m}(\chi) &=& \frac{2}{c^2}\int_0^\chi d\chi'\,\frac{\chi-\chi'}{\chi\chi'}\int d\hn\,\Phi(\chi'\hn,\chi')Y_{\ell m}^*(\hn) = \frac{2}{c^2}\int_0^\chi d\chi'\,\frac{\chi-\chi'}{\chi\chi'}\int_{\vk}4\pi i^\ell j_\ell(k\chi')\Phi(\vk,\chi')Y_{\ell m}^*(\hk),
\eeq
using \eqref{eq: phi-def} \citep[e.g.,][]{2005PhRvD..72b3516C}. Integrating over redshift, we find the extrinsic scalar transfer functions:
\beq\label{eq: extrinsic-transfer-scalar}
    \left.F_{\ell,a}^{\gamma^E,S}(k)\right|_{\rm ext} &=& \frac{1}{c^2}\sqrt{\frac{(\ell+2)!}{(\ell-2)!}}\int_0^{\chi_H} \frac{d\chi}{\chi}\,q_a(\chi)j_\ell(k\chi)\\\nonumber
    \left.F_\ell^{\F^g,S}(k,\chi)\right|_{\rm ext} &=& \frac{1}{3c^2}\sqrt{\frac{(\ell+1)!}{(\ell-1)!}}\left[(\ell-1)(\ell+2)+2\ell(\ell+1)\right]\int_0^{\chi_H} \frac{d\chi}{\chi}q_a(\chi)j_\ell(k\chi)\\\nonumber
    \left.F_\ell^{\G^g,S}(k,\chi)\right|_{\rm ext} &=& \frac{1}{c^2}\sqrt{\frac{(\ell+3)!}{(\ell-3)!}}\int_0^{\chi_H}\frac{d\chi}{\chi}q_a(\chi)j_\ell(k\chi).\\\nonumber
\eeq
where $q_a(\chi) \equiv \int_\chi^{\chi_H}d\chi'\,n_a(\chi')(\chi'-\chi)/\chi'$ is the lensing efficiency. We note that any redshift-dependent contribution to the power spectrum (\textit{i.e.}\ the potential growth factor) should be included in the $\chi$ integrals. Finally, we note that the full scalar transfer functions are the sum of the intrinsic and extrinsic contributions (which are non-trivially correlated).

The power spectra arising from vector and tensor lensing can be computed analogously to the intrinsic contributions (\S\ref{subsec: intrinsic-spectra}); this is presented in Appendix \ref{appen: lensing-spectra}, with the tensor case following \citep{2012PhRvD..86h3513S} (see also \citep{2012PhRvD..86h3527S,2014PhRvD..89h3507S,2015JCAP...10..032S}). The main result is the following set of shear transfer functions
\beq\label{eq: ext-vector-tensor-shear-spectra}
    \left.F_{\ell,a}^{\gamma^E,V}(k)\right|_{\rm ext} &=& \frac{1}{2\sqrt{2}}\sqrt{\frac{(\ell-2)!(\ell+1)!}{(\ell+2)!(\ell-1)!}}\int_0^{\chi_H} \frac{d\chi}{\chi}\left(m_a(\chi)\mathrm{Re}\left[\hat Q_{\gamma,V,1}^{(+)}(x)\right]+(m_a(\chi)-q_a(\chi))\,\mathrm{Re}\left[\hat Q_{\gamma,V,2}^{(+)}(x)\right]\right)\frac{j_\ell(x)}{x}\nonumber\\   
    \left.F_{\ell,a}^{\gamma^B,V}(k)\right|_{\rm ext} &=&\frac{1}{2\sqrt{2}}\sqrt{\frac{(\ell-2)!(\ell+1)!}{(\ell+2)!(\ell-1)!}}\int_0^{\chi_H}\frac{d\chi}{\chi}m_a(\chi)\,\mathrm{Im}\left[\hat Q_{\gamma,V,1}^{(+)}(x)\right]\frac{j_\ell(x)}{x}\nonumber\\
    \left.F_{\ell,a}^{\gamma^E,T}(k)\right|_{\rm ext} &=& \frac{1}{4}\int_0^{\chi_H}\frac{d\chi}{\chi}\left(m_a(\chi)\,\mathrm{Re}\left[\hat Q_{\gamma,T,1}^{(+)}(x)\right]+(m_a(\chi)-q_a(\chi))\mathrm{Re}\left[\hat Q_{\gamma,T,2}^{(+)}(x)\right]\right)\frac{j_\ell(x)}{x^2}\nonumber\\
    \left.F_{\ell,a}^{\gamma^B,T}(k)\right|_{\rm ext} &=& \frac{1}{4}\int_0^{\chi_H}\frac{d\chi}{\chi}m_a(\chi)\,\mathrm{Im}\left[\hat Q_{\gamma,T,1}^{(+)}(x)\right]\frac{j_\ell(x)}{x^2},
\eeq
defining the redshift-integrated kernel $m_a(\chi) \equiv\int_\chi^{\chi_H}d\chi'\,n_a(\chi')$. This depends on the $\hat Q$ derivative operators whose forms are given in \eqref{eq: Q-gamma-ext}. As before, the vector and tensor modes source both $E$ and $B$ mode contributions, and, if the underlying physics is parity-violating, leads to an $EB$ spectrum.

Under the approximation \eqref{eq: flexion-lensing-approx}, we may compute the flexion correlators directly from those for the shear. In particular, expanding ${}_{\pm2}\gamma$ in spin-weighted spherical harmonics and using relation \eqref{eq: spin-sph-relations} leads to:
\begin{align}\label{eq: flexion-from-shear-kernels}
    &\left.iF^{\F^g,V/T}_{\ell,a}(k)\right|_{\rm ext} = \sqrt{\frac{(\ell+2)!(\ell-1)!}{(\ell-2)!(\ell+1)!}}\left.F^{\gamma^E,V/T}_{\ell,a}(k) \right|_{\rm ext}, \quad &&\left.iF^{\F^c,V/T}_{\ell,a}(k)\right|_{\rm ext}  = -\sqrt{\frac{(\ell+2)!(\ell-1)!}{(\ell-2)!(\ell+1)!}}\left.F^{\gamma^B,V/T}_{\ell,a}(k)\right|_{\rm ext} \\\nonumber
    &\left.iF^{\G^g,V/T}_{\ell,a}(k)\right|_{\rm ext} = -\sqrt{\frac{(\ell+3)!(\ell-2)!}{(\ell-3)!(\ell+2)!}}\left.F^{\gamma^E,V/T}_{\ell,a}(k) \right|_{\rm ext}, \quad &&\left.iF^{\G^c,V/T}_{\ell,a}(k)\right|_{\rm ext} = \sqrt{\frac{(\ell+3)!(\ell-2)!}{(\ell-3)!(\ell+2)!}}\left.F^{\gamma^B,V/T}_{\ell,a}(k)\right|_{\rm ext}.
\end{align}
The flexion power spectra thus take the forms of \eqref{eq: ext-vector-tensor-shear-spectra}, but with the additional $\ell$-dependent prefactors given above, which asymptote to $\pm\ell$ in the large-$\ell$ limit.

\subsubsection{Other Contributions}\label{subsec: stoch-spectra}
\noindent As discussed in \S\ref{subsec: stochastic-source}, stochastic contributions to shear and flexion are modelled as a random fluctuations in each of two orthogonal directions \citep[e.g.,][]{2001PhR...340..291B}. Averaging over sources within some bin $a$, the contribution to the spin-$\pm s$ piece in harmonic space is 
\beq
    {}_{\pm s}\hat X_{a,\ell m} = \int d\chi\,n_a(\chi)\int d\hn\,{}_{\pm s}Y^*_{\ell m}(\hn)\left[\hat{X}_1\pm i\hat{X}_2\right](\chi\hn,\chi),
\eeq
where the LHS can be decomposed into $E$ and $B$ (or $g$ and $c$) modes, both of which are non-vanishing. Assuming each galaxy to be uncorrelated, implies that $\av{\hat{X}_i(\bv{x})\hat{X}_i^*(\bv{x}')} = \sigma^2_X \delta_{\rm D}(\bv{x}-\bv{x'})$; after imposing spin-weighted spherical harmonic orthonormality \eqref{eq: spin-sph-orthonormality}, this leads to the shear spectra
\beq
    \left.C_\ell^{\gamma^E\gamma^E,ab}\right|_{\rm noise} = \left.C_\ell^{\gamma^B\gamma^B,ab}\right|_{\rm noise} = \int_0^{\chi_H} d\chi\,n_a(\chi)n_b(\chi)\,\sigma^2_{\gamma} \approx \delta^{\rm K}_{ab}\frac{\sigma_\gamma^2}{\bar n_a},
\eeq
with analogous forms for flexion. The RHS is obtained in the limit of non-overlapping bins containing $\bar{n}_a$ discrete sources per unit steradian. As expected, the Poisson noise is scale-independent, and contributes only to auto-spectra.

Higher-order contributions to the $B$ and $g$ modes are more difficult to compute since the underlying integrals are convolutional. \citep{2003PhRvD..68h3002H,2009ApJ...702..593S} present a careful treatment of second-order lensing effects; here, we use a rough estimate of their magnitude only, since our focus is on non-$\Lambda$CDM physics. This is obtained by assuming that the spectra are dominated by reduced shear contributions (\textit{i.e.} those $\propto \kappa \gamma_{ij}$), which lead to the approximate form $\left.C_\ell^{\gamma^B\gamma^B}\right|_{S^2,\rm ext}\approx 10^{-3}\times (1+q)^2\left.C_\ell^{\gamma^E\gamma^E}\right|_{S,\rm ext}$, where $q\approx 1$ is the lensing bias (cf.\,Fig.\,7 of \citep{2012PhRvD..86h3513S}). We assume a similar form for the flexion $c$-mode spectra; $\left.C_\ell^{\F^c\F^c}\right|_{S^2,\rm ext}\approx 10^{-3}\times (1+q)^2\left.C_\ell^{\F^g\F^g}\right|_{S,\rm ext}$, and similarly for $\G$.

\subsection{Flat-Sky Limits}\label{sec: flat-spectra}
\noindent The results presented in \S\ref{subsec: full-spectra} give the complete descriptions of the angular power spectra of shear and flexion on all scales; however, they are expensive to implement, and difficult to interpret. At large $\ell$, one can make use of the Limber approximation, which states \citep{1953ApJ...117..134L} (see \citep[e.g.,][]{2017JCAP...05..014L,2008PhRvD..78l3506L,2020JCAP...05..010F} for associated works):
\beq\label{eq: limber-approx}
    j_\ell(x) \approx \sqrt{\frac{\pi}{2\nu}}\delta_{\rm D}(x-\nu), \qquad \ell\gg1
\eeq
where $\nu \equiv \ell+1/2$. This can be used to remove the Bessel function integrals present in each transfer function, allowing the full spectrum to be computed via a single integral.\footnote{This assumption is not always appropriate: primordially-sourced tensors evolve under an oscillatory transfer function $T_T\propto j_1(k\eta)/(k\eta)$, thus the integrals in question contain two Bessel functions, and are thus not well approximated by \eqref{eq: limber-approx}.} Below, we discuss the limiting forms for both the intrinsic and extrinsic spectra, paying close attention to their scale dependence.

\subsubsection{Intrinsic Contributions}\label{subsec: intrinsic-flatsky}
\noindent In the large-$\ell$ limit, the scalar intrinsic alignment kernels \eqref{eq: shear-kernels}\,\&\,\eqref{eq: flexion-kernels} have the limiting forms:
\beq
    \left.F_\ell^{\gamma^E,S}(k,\chi)\right|_{\rm int} &\approx& \ell^2\frac{b_S(\chi)}{2\chi^2}j_\ell(x), \qquad \left.F_\ell^{\F^g,S}(k,\chi)\right|_{\rm int} \approx - \frac{1}{3}\left.F_\ell^{\G_g,S}(k,\chi)\right|_{\rm int} \approx \ell^3\frac{4\tilde b_S(\chi)}{9\chi^2}j_\ell(x),
\eeq
using the small-scale limit of the $\hat Q_{X,S}^{(+1)}$ operators given in Appendix \ref{appen: kernels}. Integrating over redshift and using \eqref{eq: limber-approx} leads to the simplified forms:
\beq
    \left.F_\ell^{\gamma^E,S,a}(k)\right|_{\rm int} &=& \frac{k}{2\nu^{1/2}}\sqrt{\frac{\pi}{2\nu}}n_a(\nu/k)b_S(\nu/k), \\\nonumber
    \left.F_\ell^{\F^g,S,a}(k)\right|_{\rm int} &\approx& -\frac{1}{3}\left.F_\ell^{\G^g,S,a}(k)\right|_{\rm int} \approx \frac{4k\nu}{9}\sqrt{\frac{\pi}{2\nu}}n_a(\nu/k)\tilde b_S(\nu/k),
\eeq
where omit the $\chi$-dependent part of the power spectrum for clarity. The utility of such expressions is clear if we consider the power spectra themselves, for example:
\beq\label{eq: intrinsic-scalar-limit}
    \left.C_\ell^{\gamma^E\gamma^E,ab}\right|_{S,\rm int} &\approx& \frac{\nu^4}{4}\int_0^\infty \frac{d\chi}{\chi^6}n_a(\chi)n_b(\chi)b_S^2(\chi)P_\Phi(\nu/\chi,\chi,\chi)\\\nonumber
    \left.C_\ell^{\F^g\F^g,ab}\right|_{S,\rm int}&\approx&\frac{1}{9}\left.C_\ell^{\G^g\G^g,ab}\right|_{S,\rm int} \approx \frac{16\nu^6}{81}\int \frac{d\chi}{\chi^6}n_a(\chi)n_b(\chi)\tilde b_S^2(\chi)P_\Phi(\nu/\chi,\chi,\chi)
\eeq
\citep{2017MNRAS.472.2126K,2017JCAP...05..014L}; this can now be computed as a single redshift integral, which is highly accurate for $\ell\gtrsim 100$. Notably, the spectra are diagonal in redshift if bins $a$ and $b$ do not overlap.

The limiting forms for vector and tensor sources are a little more complex. Using the results of Appendix \ref{appen: kernels}, the limiting forms of the vector unbinned kernels are given by
\beq
    &&\left.F_\ell^{\gamma^E,V}(k,\chi)\right|_{\rm int} \approx \ell\frac{b_V(\chi)}{2\sqrt 2\chi}j_\ell'(x)\mathcal{I}_V(\chi), \qquad \left.F_\ell^{\gamma^B,V}(k,\chi)\right|_{\rm int} \approx \ell\frac{b_V(\chi)}{2\sqrt 2\chi}j_\ell(x)\mathcal{I}_V(\chi), \\\nonumber 
    &&\left.F_\ell^{\F^g,V}(k,\chi)\right|_{\rm int} \approx -\frac{1}{3}\left.F_\ell^{\G^g,V}(k,\chi)\right|_{\rm int} \approx \ell^2\frac{2\sqrt 2\tilde b_V(\chi)}{9\chi}j_\ell'(x)\mathcal{I}_V(\chi), \\\nonumber
    &&\left.F_\ell^{\F^c,V}(k,\chi)\right|_{\rm int} \approx -\frac{1}{9}\left.F_\ell^{\G^c,V}(k,\chi)\right|_{\rm int} \approx -\ell^2\frac{2\sqrt 2\tilde b_V(\chi)}{27\chi}j_\ell(x)\mathcal{I}_V(\chi).
\eeq
Notably, the $B$- and $c$-modes are proportional to $j_\ell(x)$, whilst the $E$ and $g$-modes depend on the Bessel function derivatives $j_\ell'(x)$, and are thus suppressed. This occurs due to projection: the 3D vector field is parity-odd, but can induce a small parity-even component when projected onto the two-sphere \citep[cf.][]{2017PhRvL.118v1301M}. 

Integrals involving Bessel function derivatives are harder to simplified using the Limber approximation (though see Appendix E of \citep{2020JCAP...07..005B} for a number of useful simplifications).\footnote{Whilst one can rewrite $j_\ell'(x)$ as the difference between two spherical Bessel functions then apply the Limber approximation, this is numerically unstable.} Instead, one can obtain the redshift-binned kernels using integration-by-parts, for example:
\beq\label{eq: integ-by-part-kernels}
    \left.F_\ell^{\gamma^E,V,a}(k)\right|_{\rm int} &\approx& \ell\int_{\chi_1}^{\chi_2}d\chi\,n_a(\chi)\frac{b_V(\chi)}{2\sqrt 2\chi}j_\ell'(x) = \left[n_a(\chi)\frac{b_V(\chi)}{2\sqrt 2k\chi}j_\ell(x)\right]_{\chi_1}^{\chi_2} -\int_{\chi_1}^{\chi_2}d\chi\,\partial_\chi\left[n_a(\chi)\frac{b_V(\chi)}{2\sqrt 2k\chi}\right]j_\ell(x),\\\nonumber
    &\approx&\left[n_a(\chi)\frac{b_V(\chi)}{2\sqrt 2k\chi}j_\ell(x)\right]_{\chi_1}^{\chi_2} -\sqrt{\frac{\pi}{2\nu}}\partial_\chi\left[n_a(\chi)\frac{b_V(\chi)}{2\sqrt 2k^2\chi}\right]_{\chi=\nu/k},
\eeq
assuming the redshift bin to have finite support over $[\chi_1,\chi_2]$, and inserting \eqref{eq: limber-approx} in the second expression.\footnote{For clarity, we have dropped the $\chi$ dependence arising from $\mathcal{I}_V(\chi)\mathcal{I}_V(\chi')P_{B_\pm}(k,\chi,\chi')$. In practice, this can be included by enforcing a separable correlator via $P_X(k,\chi,\chi')\approx \sqrt{P_X(k,\chi,\chi)P_X(k,\chi',\chi')}$ for $X\in\{\Phi,B_\pm,h_\pm\}$.} If the number density distribution falls smoothly to zero at the boundaries, then the first term is negligible; however, it can be important in the case of (largely unphysical) sharp bin boundaries. Using these relations, one can derive the flat-sky limits of power spectra, for example, the auto-spectra of $B$- and $c$-modes:
\beq
    \left.C_\ell^{\gamma^B\gamma^B,ab}\right|_{V,\rm int} &\approx& \frac{\nu^2}{8}\int_0^\infty \frac{d\chi}{\chi^4}n_a(\chi)n_b(\chi)b_V^2(\chi)\left[P_{\mathcal{I}_VB_+}(\nu/\chi,\chi,\chi)+P_{\mathcal{I}_VB_-}(\nu/\chi,\chi,\chi)\right]\\\nonumber
    \left.C_\ell^{\F^c\F^c,ab}\right|_{V,\rm int} &\approx& \frac{1}{81}\left.C_\ell^{\G^c\G^c,ab}\right|_{V,\rm int} \approx \left(\frac{2\sqrt 2\nu^2}{27}\right)^2\int_0^\infty \frac{d\chi}{\chi^4}n_a(\chi)n_b(\chi)\tilde b_V^2(\chi)\left[P_{\mathcal{I}_VB_+}(\nu/\chi,\chi,\chi)+P_{\mathcal{I}_VB_-}(\nu/\chi,\chi,\chi)\right],
\eeq
absorbing the $\mathcal{I}_V$ operators into the power spectrum. 

The flat-sky limits of tensor correlators can be derived similarly. The unbinned kernels can be written:
\beq
    &&\left.F_\ell^{\gamma^E,T}(k,\chi)\right|_{\rm int} \approx \left(\frac{\ell^2}{x^2}-2\right)\frac{b_T(\chi)}{4}j_\ell(x)\mathcal{I}_T(\chi), \qquad \left.F_\ell^{\gamma^B,T}(k,\chi) \right|_{\rm int} \approx \frac{b_T(\chi)}{2}j_\ell'(x)\mathcal{I}_T(\chi),\\\nonumber
    &&\left.F_\ell^{\F^g,T}(k,\chi)\right|_{\rm int}\approx \ell\left(3\frac{\ell^2}{x^2}-2\right)\frac{2\tilde b_T(\chi)}{27}j_\ell(x)\mathcal{I}_T(\chi),\qquad \left.F_\ell^{\G^g,T}(k,\chi)\right|_{\rm int}\approx -\ell\left(\frac{\ell^2}{x^2}-2\right)\frac{2\tilde b_T(\chi)}{3}j_\ell(x)\mathcal{I}_T(\chi),\\\noindent
    && \left.F_\ell^{\F^c,T}(k,\chi)\right|_{\rm int} \approx -\frac{1}{9}\left.F_\ell^{\G^c,T}(k,\chi)\right|_{\rm int}\approx -\ell \frac{4\tilde b_T(\chi)}{27}j_\ell'(x)\mathcal{I}_T(\chi).
\eeq
We find the opposite conclusion to the vector case: the $B$- and $c$-modes are derivative-suppressed compared to the $E$- and $g$-modes \citep{2017PhRvL.118v1301M}. Their binned forms can be obtained using integration by parts, as in \eqref{eq: integ-by-part-kernels}. The unsuppressed flat-sky auto-power spectra take the form:
\beq
    \left.C_\ell^{\gamma^E\gamma^E,ab}\right|_{T,\rm int} &\approx& \frac{1}{16}\int_0^\infty \frac{d\chi}{\chi^2}n_a(\chi)n_b(\chi)b_T^2(\chi)\left[P_{\mathcal{I}_Th_+}(\nu/\chi,\chi,\chi)+P_{\mathcal{I}_Th_-}(\nu/\chi,\chi,\chi)\right]\\\nonumber
    \left.C_\ell^{\F^g\F^g,ab}\right|_{T,\rm int} &\approx& 
    \frac{1}{81}\left.C_\ell^{\G^g\G^g,ab}\right|_{T,\rm int} \approx 
    \left(\frac{2\ell}{9}\right)^2\int_0^\infty \frac{d\chi}{\chi^2}n_a(\chi)n_b(\chi)\tilde b_T^2(\chi)\left[P_{\mathcal{I}_Th_+}(\nu/\chi,\chi,\chi)+P_{\mathcal{I}_Th_-}(\nu/\chi,\chi,\chi)\right].
\eeq

The utility of the above simplifications is twofold: firstly, they allow for efficient computation of the high-$\ell$ spectra (which would otherwise require a dense three-dimensional numerical integration grid), and secondly, one can easily assess the $\ell$-scalings of the correlators. These are given explicitly in Tab.~\ref{tab: Cl-scalings} for each intrinsic auto-spectrum sourced by scalar, vector, and tensor physics. This clearly demonstrates three points, noted above:
\begin{enumerate}
    \item All flexion spectra are suppressed by $\ell^2$ with respect to shear, and thus contribute significantly only on small scales.
    \item Scalar perturbations generate only $E$- and $g$-modes at leading-order, whilst vectors and tensors generate also $B$- and $c$-modes.
    \item Intrinsic vector $E$- and $g$-mode spectra are derivative-suppressed compared to $B$- and $c$-modes; the reverse is true for tensors.
\end{enumerate}
The first property implies that flexion will be useful only on small scales, matching previous conclusions \citep[e.g.][]{2006MNRAS.365..414B}). From the third, we conclude that the often-ignored $B$- and $c$-mode spectra are an excellent place in which to search for vector perturbations; in contrast, tensors require $E$- and $g$-modes, which are contain large contributions from scalar physics, thus hampering their detectability. If the sources are parity-violating, the cross-spectra (particularly $EB$) can be of significant use, though the projection effects partially suppress such signals \citep{2017PhRvL.118v1301M,2020JCAP...07..005B}.

\begin{table}
    \centering
    \begin{tabular}{c|cccc}
    \textbf{Intrinsic} & $\gamma^E\gamma^E$ & $\gamma^B\gamma^B$ & $\F^g\F^g$ & $\F^c\F^c$\\\hline
     Scalar & 1 & 0 & $\ell^2$ & 0\\
     Vector & 1 & $\ell^{2}$ & $\ell^2$ & $\ell^4$\\
     Tensor & 1 & $\ell^{-2}$ & $\ell^2$ & $\ell^0$\\
    \end{tabular}
    \hskip 20 pt
    \begin{tabular}{c|cccc}
    \textbf{Extrinsic} & $\gamma^E\gamma^E$ & $\gamma^B\gamma^B$ & $\F^g\F^g$ & $\F^c\F^c$\\\hline
     Scalar & 1 & 0 & $\ell^2$ & 0\\
     Vector & 1 & $\ell^{-2}$ & $\ell^2$ & $\ell^0$\\
     Tensor & 1 & $\ell^{-2}$ & $\ell^2$ & $\ell^0$\\
    \end{tabular}
    \caption{Approximate scalings of the intrinsic alignment and lensing power spectra in the large-$\ell$ limit, as derived from \S\ref{sec: flat-spectra}. In each case, we give the $\ell$-dependence of $C_\ell^{XX}/C_\ell^{\gamma^E\gamma^E}$, considering both shear and flexion, and scalar, vector, and tensor physics. We display results only for the first flexion, noting that $\F$ and $\G$ have the same scalings at large-$\ell$. The scalings of cross-spectra may be found by taking the harmonic mean of auto-spectra scalings, \textit{i.e.}\ the $C_\ell^{\gamma^B\F^c}$ scaling for vector intrinsic alignments is $\sqrt{\ell^2\times \ell^4}=\ell^3$ that of $C_\ell^{\gamma^E\gamma^E}$. Note that the amplitudes of intrinsic vector $E$/$g$-modes and tensor intrinsic $B$/$c$-modes are suppressed due to the presence of Bessel function derivatives.}
    \label{tab: Cl-scalings}
\end{table}

\subsubsection{Extrinsic Contributions}\label{subsec: extrinsic-flatsky}
\noindent The large-$\ell$ limit of scalar lensing spectra is straightforward. From \eqref{eq: lensing-harmonics}, the scalar transfer functions have the asymptotic form
\beq
    \left.F_{\ell,a}^{\gamma^E,S}(k)\right|_{\rm ext} &\approx& \frac{1}{\ell}\left.F_\ell^{\F^g,S}(k,\chi)\right|_{\rm ext} \approx \frac{1}{\ell}\left.F_\ell^{\G^g,S}(k,\chi)\right|_{\rm ext} \approx \frac{\ell^2}{c^2}\int_0^{\chi_H} \frac{d\chi}{\chi}\,q_a(\chi)j_\ell(k\chi),
\eeq
leading to auto-spectra of the form
\beq
    C_\ell^{\gamma^E\gamma^E,ab} &\approx& \frac{\nu^4}{c^4}\int_0^{\infty} \frac{d\chi}{\chi^4}q_a(\chi)q_b(\chi)P_\Phi(\nu/\chi,\chi,\chi)
\eeq
\citep[e.g.,][]{2017JCAP...05..014L}, invoking the Limber approximation \eqref{eq: limber-approx}. Unlike the intrinsic spectra, this is not diagonal in redshift. 

For the vector and tensor kernels, computation proceeds analogously to \S\ref{subsec: intrinsic-flatsky}. From \eqref{eq: ext-vector-tensor-shear-spectra}, and using the limiting forms of the $\hat{Q}$ operators given in Appendix \ref{appen: kernels}, we find
\beq
    &&\left.F_{\ell,a}^{\gamma^E,V}(k)\right|_{\rm ext} \approx -\frac{\ell^3}{2\sqrt{2}}\int_0^{\chi_H} \frac{d\chi}{\chi}q_a(\chi)\frac{j_\ell(x)}{x}, \qquad \left.F_{\ell,a}^{\gamma^B,V}(k)\right|_{\rm ext} \approx\frac{\ell}{2\sqrt{2}}\int_0^{\chi_H}\frac{d\chi}{\chi}m_a(\chi)j_\ell(x)\\\nonumber
    &&\left.F_{\ell,a}^{\gamma^E,T}(k)\right|_{\rm ext} \approx -\frac{\ell^4}{8}\int_0^{\chi_H}\frac{d\chi}{\chi}q_a(\chi)\frac{j_\ell(x)}{x^2}, \qquad \quad\left.F_{\ell,a}^{\gamma^B,T}(k)\right|_{\rm ext} \approx \frac{\ell^2}{4}\int_0^{\chi_H}\frac{d\chi}{\chi}m_a(\chi)\frac{j_\ell(x)}{x}.
\eeq
This leads to power spectra of the form
\beq
    \left.C_\ell^{\gamma^E\gamma^E,ab}\right|_{V,\rm ext} &\approx& \frac{\nu^4}{8}\int_0^\infty\frac{d\chi}{\chi^4}q_a(\chi)q_b(\chi)\left[P_{B_+}(\nu/\chi,\chi,\chi)+P_{B_-}(\nu/\chi,\chi,\chi)\right]\\\nonumber
    \left.C_\ell^{\gamma^B\gamma^B,ab}\right|_{V,\rm ext} &\approx& \frac{\nu^2}{8}\int_0^\infty\frac{d\chi}{\chi^4}m_a(\chi)m_b(\chi)\left[P_{B_+}(\nu/\chi,\chi,\chi)+P_{B_-}(\nu/\chi,\chi,\chi)\right]\\\nonumber
    \left.C_\ell^{\gamma^E\gamma^E,ab}\right|_{T,\rm ext} &\approx& \frac{\nu^4}{64}\int_0^\infty\frac{d\chi}{\chi^4}q_a(\chi)q_b(\chi)\left[P_{h_+}(\nu/\chi,\chi,\chi)+P_{h_-}(\nu/\chi,\chi,\chi)\right]\\\nonumber
    \left.C_\ell^{\gamma^B\gamma^B,ab}\right|_{T,\rm ext} &\approx& \frac{\nu^2}{16}\int_0^\infty\frac{d\chi}{\chi^4}m_a(\chi)m_b(\chi)\left[P_{h_+}(\nu/\chi,\chi,\chi)+P_{h_-}(\nu/\chi,\chi,\chi)\right],
\eeq
under the Limber approximation. For the flexion spectra, we employ the simplifying \textit{ansatz} of \eqref{eq: flexion-from-shear-kernels}, which implies 
\beq
    &&\left.iF^{\F^g,V/T}_{\ell,a}(k)\right|_{\rm ext} \approx -\left.iF^{\G^g,V/T}_{\ell,a}(k)\right|_{\rm ext} \approx \ell\times\left.F^{\gamma^E,V/T}_{\ell,a}(k) \right|_{\rm ext},\\\nonumber
    &&\left.iF^{\F^c,V/T}_{\ell,a}(k)\right|_{\rm ext}  \approx - \left.iF^{\G^c,V/T}_{\ell,a}(k)\right|_{\rm ext} \approx -\ell\times\left.F^{\gamma^B,V/T}_{\ell,a}(k)\right|_{\rm ext},
\eeq
allowing simple computation of the flexion spectra \citep[e.g.,][]{2006MNRAS.365..414B,2005ApJ...619..741G}.

A summary of the extrinsic auto-spectrum scalings is given in Table \ref{tab: Cl-scalings}, alongside the intrinsic results. As before, the flexion spectra are suppressed by $\ell^2$ compared to the shear spectra, and the tensor $B$- and $c$-modes are suppressed compared to the $E$- and $g$-modes, though there are no Bessel function derivatives in this case. Here, we find the opposite scalings of vector modes (with $E$- and $c$-modes enhanced on large scales rather than $B$- and $g$-modes). This can be rationalized by looking at how vector modes impact intrinsic and extrinsic shear, \eqref{eq: vector-tensor-g,F,G-source} \& \eqref{eq: lensing-schmidt-paper}. In the former case, we have dependence only through the projected derivatives $\partial_\pm B_\pm$, whilst the latter also contains also a term involving $\partial_\pm\partial_\pm B_\parallel$ (proportional to $q(\chi)$) for $B_\parallel \equiv B_i\hat{n}^i$, which dominates in the large-$\ell$ regime. Since this is a radial projection instead of an angular projection, $E$- and $B$-modes are interchanged, leading to the different scalings.

\section{Numerical Results}\label{sec: results}
\noindent Having discussed the physical sources of galaxy shape distortions and their associated power spectra, we now proceed to perform a numerical study of new physics in shear and flexion. To this end, we will consider two beyond-$\Lambda$CDM physics models, described in \S\ref{sec: conventions}: (1) non-standard inflation, with independent primordial sourcing of vector and tensor degrees of freedom; (2) a late-Universe network of cosmic strings, which jointly sources vector and tensor modes. By combining the power spectra of the metric fluctuations with the above results, we can consider the heuristic forms of the generated shear and flexion spectra, and forecast the possible constraints on the model amplitudes. 

\subsection{Set-Up}\label{subsec: forecast-setup}
\noindent \paragraph{Perturbations} 
As discussed in \S\ref{subsec: ex-inflationary}, we model primordial power spectra using power laws, relative to a fiducial scale $k_0 = 0.002\mathrm{Mpc}^{-1}$ \citep[e.g.,][]{2020A&A...641A..10P}. Here, we assume a tensor exponent $n_T = -r_T/8$ (motivated by simple inflationary models), with \resub{an independent scale-invariant vector exponent, $n_V = 1$}.\footnote{\resub{Strictly, it is inconsistent to simultaneously assume \textit{almost} scale-invariant tensors ($n_T=-r_T/8$) and \textit{exactly} scale-invariant vectors ($n_V=1$), if sourced by a common inflationary origin. Here, we consider the two perturbations separately, and note that small deviations of either from scale-invariance will have very minor impact on our constraints.}} For plotting, we will assume the fiducial values $r_T = r_V = 1$, with the vector transfer functions normalized to their values at the last scattering surface (noting the strong decay as $a^{-2}(z)$). To test constraints on parity-violating physics, we will allow for maximum fiducial chirality $\epsilon_{V,T} = 1$, noting that the set of spectra sourced by parity-odd and parity-even physics are disjoint at leading order.

The power spectra of the cosmic shear metric perturbations is modeled as described in \S\ref{subsec: ex-strings} (using the results derived in Appendix \ref{appen: cosmic-string-tensors}), using the fiducial parameter values given therein, with a baseline (squared) string tension $(G\mu)^2 = 10^{-16}$ and an intercommuting probability of $P=10^{-3}$, which sets the amplitude of both vector and tensor contributions \citep[e.g.,][]{2012JCAP...10..030Y}. In the standard string model, only parity-even correlators are sourced; here, we will allow for parity-breaking contributions via a global chirality parameter $\epsilon=1$, analogous to the primordial case. Finally, the intrinsic alignment spectra require the time-differentiated vector and tensor spectra: following a similar derivation to that given in Appendix \ref{appen: cosmic-string-tensors}, the equal-time correlators can be shown to equal
\beq
    P_{\mathcal{I}_VB_{\pm}}(k,\eta,\eta) &=& \left\{(3aH)^2+\frac{1}{3}k^2v^2_{\rm rms}\right\}P_{B_{\pm}}(k,\eta,\eta),\\\nonumber
    P_{\mathcal{I}_Th_{\pm}}(k,\eta,\eta) &=& \left\{\left[6(aH)^2+2\partial_\eta(aH)\right]^2+\frac{1}{3}(5aH)^2k^2v_{\rm rms}^2+\frac{2}{9}k^4v_{\rm rms}^4\right\}P_{h_{\pm}}(k,\eta,\eta)
\eeq
assuming time-independence for $v^2_{\rm rms}$ and the comoving string density.

\vskip 4pt

\paragraph{Experimental Parameters}
To compute the shear and flexion power spectra we must also specify the number-density of source galaxies and their noise properties. Following \citep{2021arXiv210709000G}, we will assume a projected galaxy density of $40\,\mathrm{arcmin}^{-2}$, with the normalized distribution 
\beq
    n(z) \propto \left(\frac{z}{z_0}\right)^\alpha e^{-(z/z_0)^\beta},
\eeq
where $\alpha = 2$, $\beta = 3/2$ and $z_0 = 0.64$, as relevant for a \textit{Euclid}-like survey \citep{2011arXiv1110.3193L}. This is split into three equally populated bins, each following a smooth (overlapping) distribution. We will assume a sky fraction of $f_{\rm sky} = 0.3636$, as relevant for \textit{Euclid} \citep{2019JCAP...11..034C}.

To model the intrinsic contributions, we require the shear and flexion alignment biases, $b_X(\chi)$ and $\tilde b_X(\chi)$. Due to hydrodynamic effects, these are difficult to predict from theory or simulations, even for the scalar case. Here, we follow \cite{2015JCAP...10..032S}, and fix the scalar shear bias to $b_S(\chi)=b_1^I(\chi)\frac{2a(\chi)}{3H_0^2\Omega_{m0}}$, with $b_1^I(\chi) = -0.1\Omega_{m0}D(0)/D(\chi)$, consistent with \citep{2011JCAP...05..010B}.  Assuming all components to impact the shear only through the tidal tensor yields the relations $b_T=b_S/2$, $b_V=-b_S$, reinserting the previously dropped constants of proportionality \citep[cf.,][]{2012PhRvD..86h3513S}. We additionally assume the same bias for shear and flexion, \textit{i.e.} $b_X=\tilde b_X$, motivated by the analytic results of \citep[Eq.\,18]{2021arXiv210709000G}.

\vskip 4pt

\paragraph{Numerical Set-Up}
We numerically compute angular power spectra for all auto- and cross-spectra of shear and spin-$\pm1$ flexion for $39$ $\ell$-bins in the range $[2,2000]$, using linear spacing for $\ell\leq 10$ and logarithmic else. We ignore the spin-$\pm3$ flexion for clarity, noting that this will not change our overall conclusions. Spectra are computed in \textsc{python} using numerical quadrature, with the Limber approximation (\S\ref{sec: flat-spectra}) used for $\ell>100$ (except for pathological situations, such as primordially sourced tensors). Throughout, we assume the fiducial cosmology $\{h=0.7,A_s=1.95\times 10^{-9}, n_s=0.96,\Omega_b=0.049,\Omega_m=0.3\}$, and evaluate scalar spectra using the ``halofit'' formalism to approximate non-linear effects \citep{2003MNRAS.341.1311S}.

\subsection{Power Spectra}
\noindent Under null assumptions (\textit{i.e.} without vector and tensor modes), the following spectra are sourced at leading order:
\beq
    &C_\ell^{\gamma^E\gamma^E},\,C_\ell^{\F^g\F^g},\, C_\ell^{\gamma^E\F^g} \qquad &\text{(from scalars)}\\\nonumber
    &C_\ell^{\gamma^E\gamma^E},\,C_\ell^{\F^g\F^g},\, C_\ell^{\gamma^B\gamma^B},\,C_\ell^{\F^c\F^c}\qquad &\text{(from noise)}.
\eeq
For noise and intrinsic-only contributions, only auto-spectra $a=b$ are sourced (for redshift bins $a,b$, in the large-$\ell$) limit, whilst lensing sources also $a\neq b$ bin pairs and cross-correlations with intrinsic effects. At second order, scalars also source contributions to $C_\ell^{\gamma^B\gamma^B}$, $C_\ell^{\F^c\F^c}$, and $C_\ell^{\gamma^B\F^c}$, which are modeled as discussed in \S\ref{subsec: stoch-spectra}. In the presence of vector or tensor perturbations, we source the following spectra: 
\beq
    &C_\ell^{\gamma^E\gamma^E},\,C_\ell^{\gamma^B\gamma^B},\,C_\ell^{\F^g\F^g},\,C_\ell^{\F^c\F^c},\,C_\ell^{\gamma^E\F^g},\,C_\ell^{\gamma^B\F^c} \qquad &\text{(parity-conserving)}\\\nonumber
    &C_\ell^{\gamma^E\gamma^B},\,C_\ell^{\F^g\F^c},\, C_\ell^{\gamma^E\F^c},\,C_\ell^{\gamma^B\F^g}\qquad &\text{(parity-breaking)},
\eeq
depending on the parity-properties of the underlying metric correlators.

\begin{figure}
    \centering
    \includegraphics[width=0.9\textwidth]{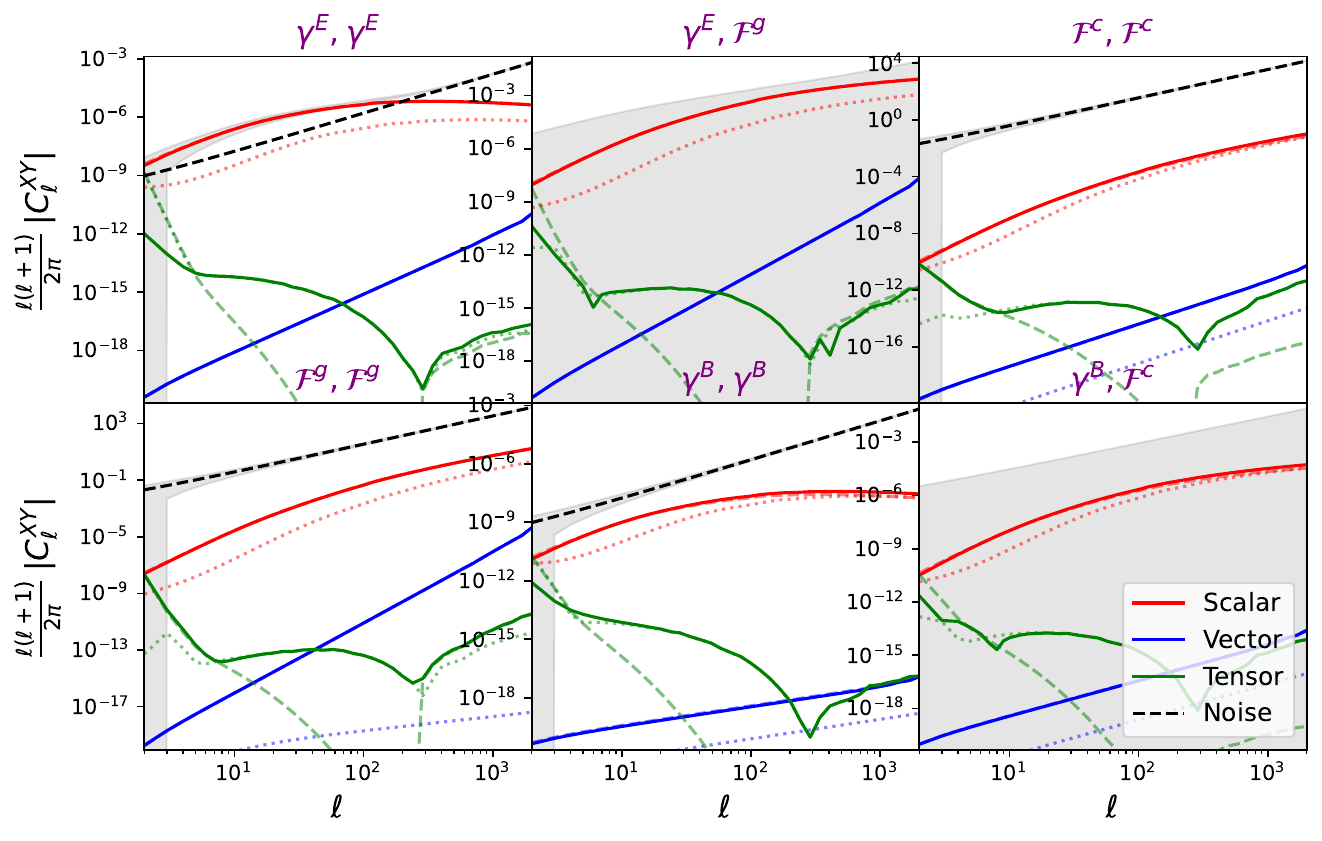}
    \includegraphics[width=0.9\textwidth]{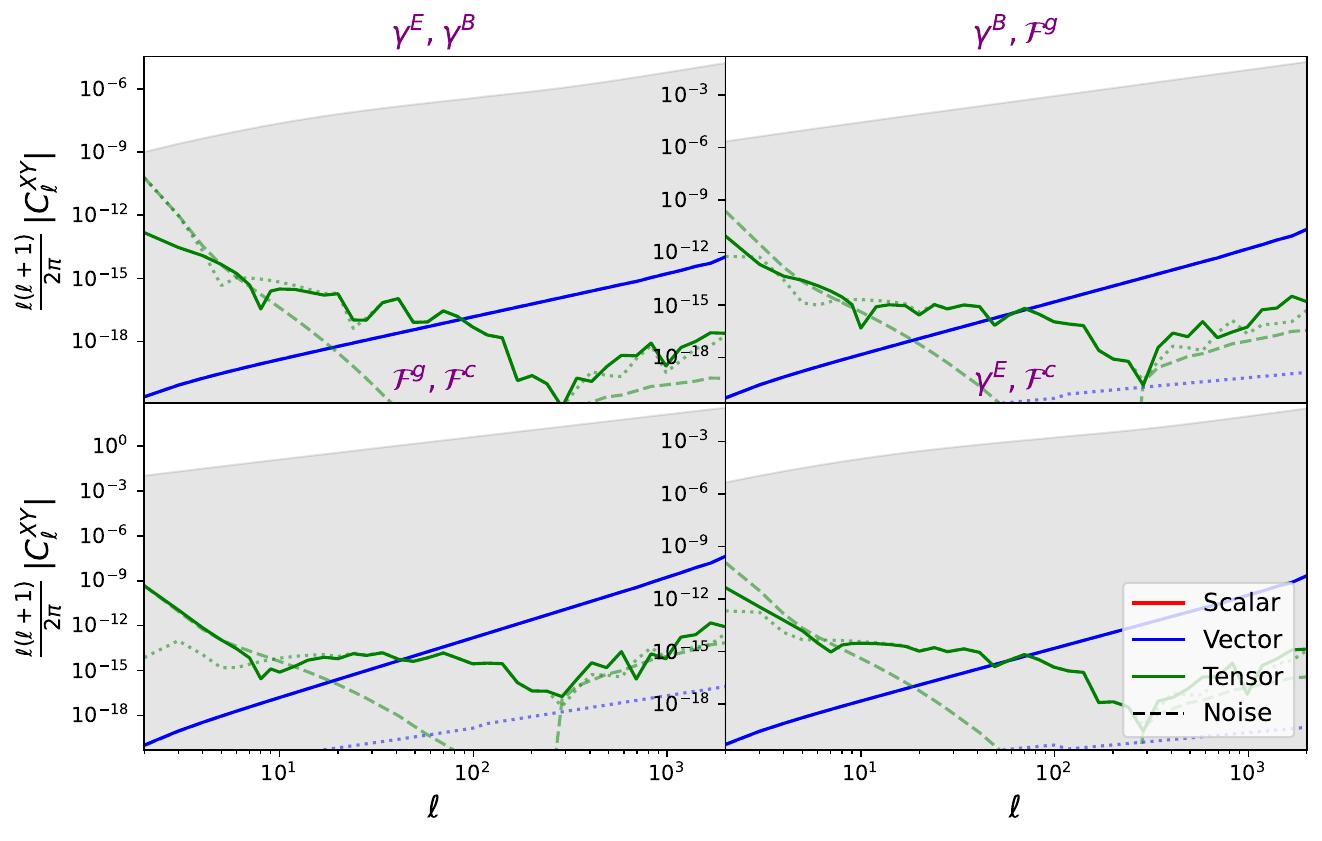}
    \caption{Shear and flexion spectra sourced by scalar physics as well as primordially-generated vector and tensor modes. Each panel shows a separate cross-spectrum, as marked in the title, and the two plots show parity-conserving (top) and parity-breaking (bottom) correlators. The grey bands indicate $1\sigma$ errorbars, with dashed (dotted) lines showing spectra containing only lensing (intrinsic alignment) contributions. The experimental set-up is as described in \S\ref{subsec: forecast-setup} for a \textit{Euclid}-like experiment, and we plot only results from the central redshift bin with $\bar{z}\approx 0.95$. The fiducial spectra have amplitudes $r_T = 1$ for tensors, and $r_V = 1$ for vectors, with the latter spectra defined relative to the surface of last scattering. For parity-violating correlators, we assume a chirality parameter $\epsilon_{V,T} = 1$. Gravitational wave contributions (green) are difficult to detect except on ultra-large scales in shear. Vector modes of this type contribute significantly only on small scales, though this depends on the primordial tilt. The scatter in the tensor curves is not physical; this arises from numerical inaccuracies in the oscillatory integrals.}
    \label{fig: primordial-fig}
\end{figure}

\vskip 4pt
\paragraph{Inflationary Perturbations}
In Fig.\,\ref{fig: primordial-fig}, we plot the primordially-sourced vector and tensor correlators of shear and flexion, alongside the fiducial power spectra arising from scalar physics and noise. Notably, detecting such spectra will be challenging. Whilst we forecast a very strong shear $E$-mode signal (principally arising from lensing), the only other scalar cross-spectrum with a realistic hope of being detected is that between $\gamma^E$ and $\F^g$ (or between $\gamma^E$ and $\G^g$), which requires a large $\ell_{\rm max}$ \citep[cf.][]{2008ApJ...680....1O,2022PhRvD.105l3521A,2021arXiv210709000G}. As previously noted, the parity-odd scalar modes are strongly suppressed compared to the parity-even forms, and do not contribute to parity-violating cross-spectra.

Spectra sourced by primordial vector perturbations show a strong scale-dependence, with largest contributions seen on smallest scales. In this scenario, the correlators are dominated by weak lensing contributions, and, as predicted in Tab.\,\ref{tab: Cl-scalings}, $E$-modes exhibit steeper scalings with $\ell$ than $B$-modes. If one considers the intrinsic contributions alone, the opposite is true, matching \S\ref{sec: flat-spectra}, and analogous conclusions can be drawn for flexion (though at much lower signal-to-noise). In our fiducial parametrization, we find extremely small amplitudes for the vector spectra; we will discuss the resulting constraints on $r_V$ and $\epsilon_Vr_V$ in \S\ref{subsec: fisher}.

The primordial tensor contributions are very different from those of vectors. As in \citep{2012PhRvD..86h3513S}, the spectra peak on ultra-large scales but quickly decay away to small values (whose inference is complicated by numerical error intrinsic to the Bessel function integrals). In this case, the intrinsic alignment $E$-mode contribution dominates at large scales, but much of this power is cancelled by an analogous extrinsic contribution due to the equivalence principle. The gravitational waves contribute non-trivially to an array of correlators with different $\ell$-dependencies; however, their power is always small, even at the fiducial amplitude $r_T = 1$. As shown in \citep{2012PhRvD..86h3513S} and discussed below, the signal can be boosted somewhat by using a higher-redshift galaxy sample.

\begin{figure}
    \centering
    \includegraphics[width=0.9\textwidth]{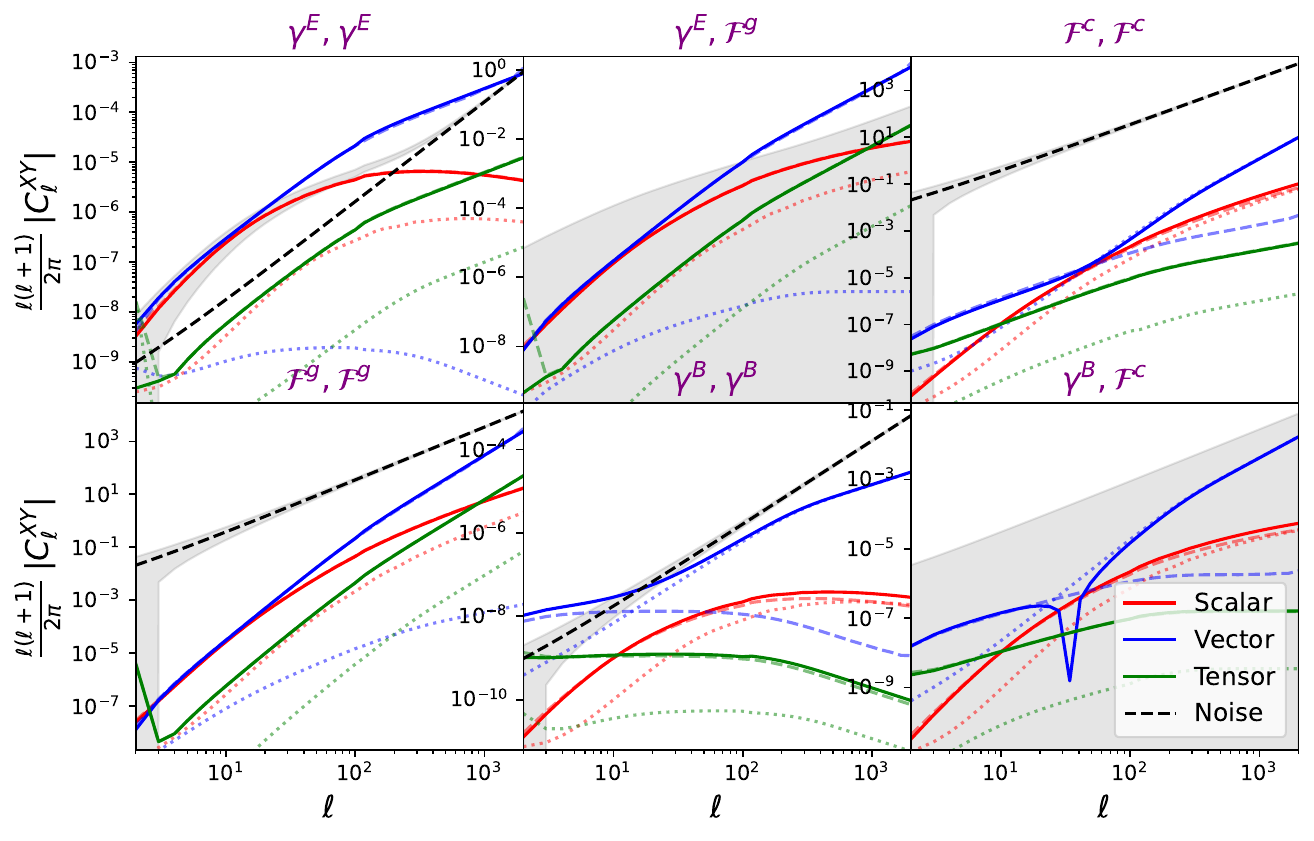}
    \includegraphics[width=0.9\textwidth]{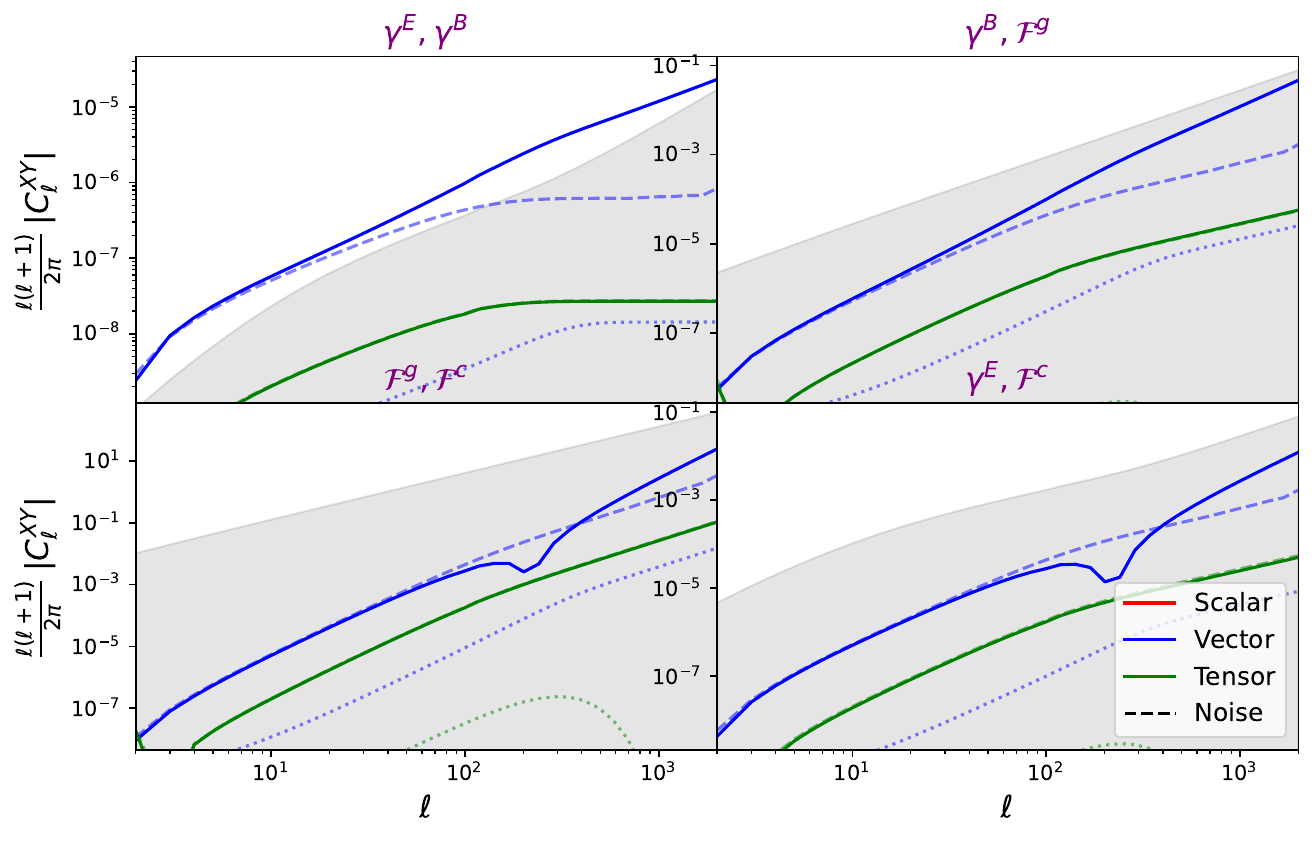}
    \caption{As Fig.\,\ref{fig: primordial-fig}, but for a late-time network of cosmic strings, which sources both vector and tensor correlators. Here, we set the fiducial string tension to $G\mu=10^{-8}$, and optionally allow for parity-odd contributions with chirality $\epsilon=1$. For this choice of amplitude, we find large contributions from vector lensing modes, particularly in shear $E$- and $B$-mode correlators on small scales.}
    \label{fig: string-fig}
\end{figure}

\vskip 4pt
\paragraph{Cosmic Strings}
Fig.\,\ref{fig: string-fig} shows the analogous results for the cosmic string power spectrum, which are much more promising. With the fiducial string tension of $G\mu = 10^{-8}$, vector contributions to galaxy shapes are significant, and are usually larger than the scalar pieces, particularly for low-$\ell$ $B$-modes and high-$\ell$ $E$-modes. Many of spectra are dominated by lensing contributions \citep[cf.,][]{2012JCAP...10..030Y}, though intrinsic alignment contributions are found to be important for $B$-modes, and the cross-correlation of lensing and intrinsic effects dominates in the $\gamma^E-\gamma^B$ spectrum. Flexion spectra are also of interest, with a clear signal appearing in the $\gamma^E-\F^g$ correlator for $\ell\gtrsim 100$. The parity-odd signal may also be detectable through the $\gamma^E-\gamma^B$ cross-spectrum though we note that this is not sourced in the standard string paradigm.

The tensorial contributions induced by cosmic strings are heavily suppressed compared to the vector pieces, with correlators one-to-two orders of magnitude smaller. The overall scalings with $\ell$ also differ slightly from vectors, due to differing absolute importance of intrinsic and extrinsic contributions (cf.\,Tab.\,\ref{tab: Cl-scalings}). For the string scenario, we conclude that the vector contributions will dominate the constraining power on the model amplitude; however, this statement is strongly model-dependent.

\subsection{Fisher Forecasts}\label{subsec: fisher}
\noindent We now elucidate the constraining power of shear and flexion on the physical models themselves. For this, we utilize a Fisher forecast \citep[e.g.,][]{1997ApJ...480...22T}, computing the optimal constraints on some set of parameters $\{p_\alpha\}$, via $\sigma^2(p_\alpha)\geq F^{-1}_{\alpha\alpha}$, for Fisher matrix
\beq
    F_{\alpha\beta} \equiv \sum_\ell\frac{\partial\mathbf{D}_\ell}{\partial p_\alpha}\mathbb{C}_\ell^{-1}\frac{\partial\mathbf{D}_\ell}{\partial p_\beta}.
\eeq
This depends on a data-vector, $\mathbf{D}_\ell$, and covariance $\mathbb{C}_\ell$, where the former contains all independent auto- and cross-spectra. Using three tomographic bins, we find a total of 42 (12) parity-even and 36 (9) parity-odd spectra when flexion is (is not) included. Assuming Gaussian statistics, the covariance between any pair of spectra is given by:
\beq\label{eq: signal-cov}
    \mathrm{Cov}\left(C_{\ell}^{XY,ab},C_{\ell'}^{ZW,cd}\right) = \frac{2\delta^K_{\ell\ell'}}{(2\ell+1)f_{\rm sky}}\left[C_\ell^{XZ,ac}C_\ell^{YW,bd}+C_\ell^{XW,ad}C_\ell^{YZ,bc}\right],
\eeq
which is diagonal in $\ell$. Under null hypotheses, the spectra on the RHS contain only signals from scalar perturbations and stochasticity; furthermore, parity-odd and parity-even spectra are uncorrelated, thus we may compute their contributions to parameter constraints separately. 

\renewcommand{\arraystretch}{1.25}
\begin{table}
    \centering
    \begin{tabular}{l|c|c|c||c|c}
    \textbf{Observable} &\quad\textbf{Parity}\quad\quad& \quad$\sigma[r_{V}]$\quad\quad & \quad$\sigma[r_{T}]$\quad\quad & \quad$\sigma_{\rm V}[(G\mu)^2]$\quad\quad & \quad$\sigma_{\rm T}[(G\mu)^2]$\quad\quad \\\hline\hline
    Shear & Even & $4.5\times 10^4$ & $260$ & $2.2\times 10^{-20}$ & $1.2\times 10^{-19}$\\
    Shear & Odd & $2.6\times 10^7$ & $600$ & $1.1\times 10^{-19}$ & $5.0\times 10^{-17}$\\  
    Shear \& Flexion & Even & $3.9\times 10^4$ & $260$ & $2.0\times 10^{-20}$ & $1.1\times 10^{-19}$\\  
    Shear \& Flexion & Odd & $2.3\times 10^7$ & $600$ & $1.0\times 10^{-19}$ & $5.0\times 10^{-17}$\\  
    \end{tabular}
    \caption{Constraints on the amplitudes of primordial and string-induced vector and tensor modes, obtained from Fisher forecasts appropriate for a \textit{Euclid}-like survey, measuring shear and flexion for $\ell\in[2,2000]$. In each case, we give the $1\sigma$ constraint on the model amplitude ($r_X$ for primordial sources, $(G\mu)^2$ for strings), quoting primordial vector results relative to the amplitude at last scattering. Parity-odd constraints assume maximal chirality ($\epsilon = 1$), and we additionally separate vector and tensor contributions to $(G\mu)^2$. In general, we find that the addition of flexion somewhat improves constraining power for spectra contributing at high-$\ell$, and parity-odd spectra are less constraining than parity-even spectra. The constraints on primordial tensors (\textit{i.e.}\ gravitational waves) are generally weak, whilst those on strings are relatively tight. The $\ell$-dependence of these constraints is explored in Figs.\,\ref{fig: primordial-fish}\,\&\,\ref{fig: string-fish}.}
    \label{tab: constraints-tab}
\end{table}

\begin{figure}
    \centering
    \includegraphics[width=0.49\textwidth]{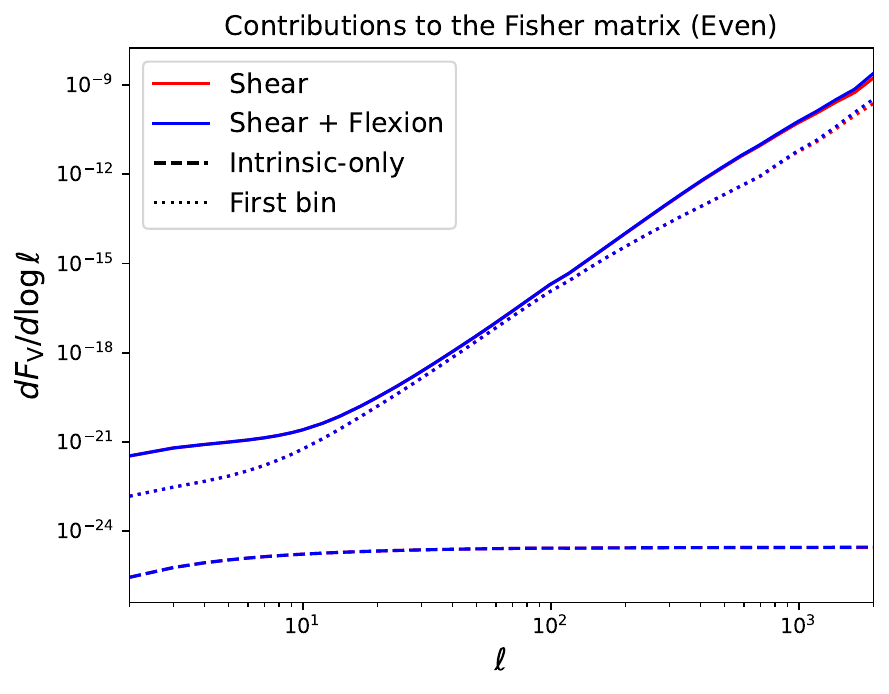}
    \includegraphics[width=0.49\textwidth]{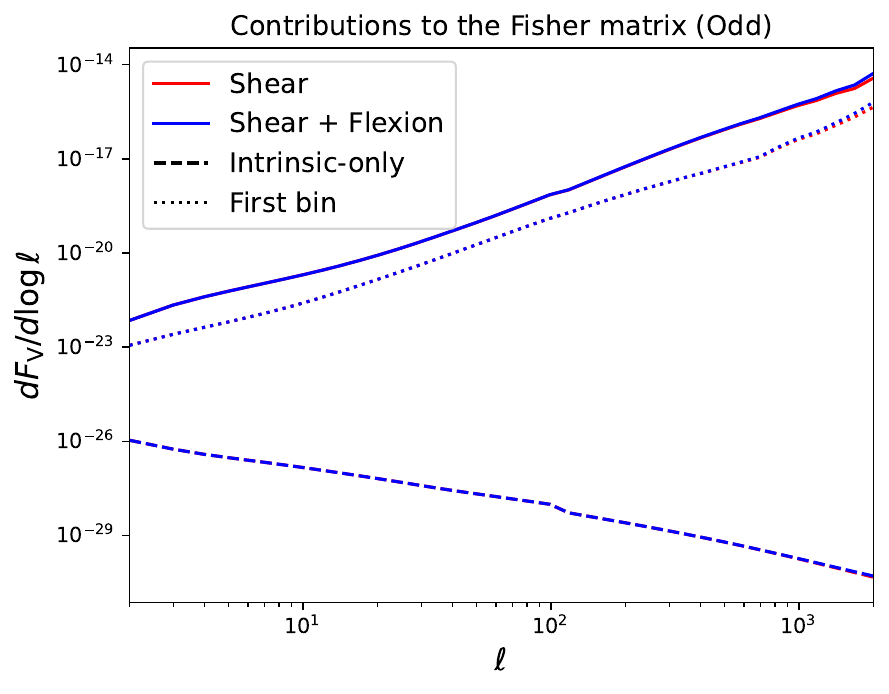}
    \includegraphics[width=0.49\textwidth]{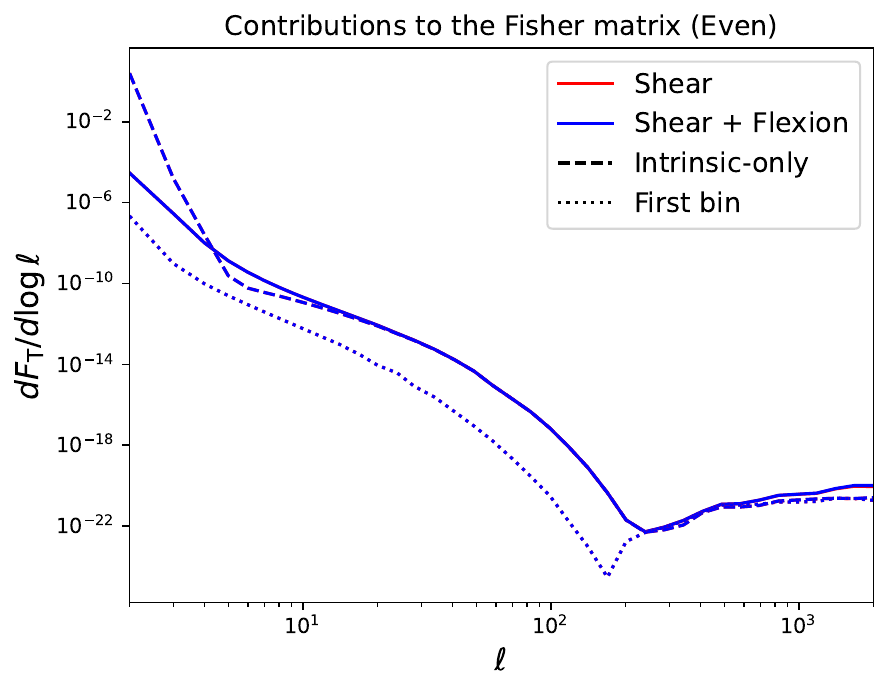}
    \includegraphics[width=0.49\textwidth]{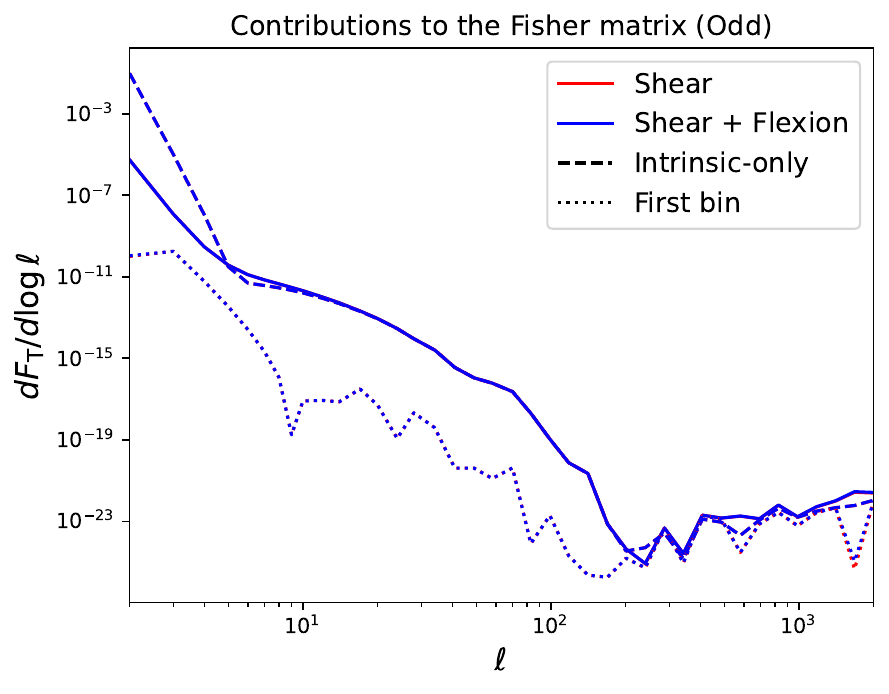}
    \caption{Constraints on primordial vector (top) and tensor (bottom) amplitudes from parity-even (left) and parity-odd (right) shear and flexion power spectra. In each case, we plot the per-$\ell$ contribution to the Fisher matrix on the amplitude parameter $r_X$, motivated by the Cram\'er-Rao bound $\sigma_X\geq F_X^{-1/2}$. Red lines show results using only the shear correlators, whilst blue lines additionally include flexion (and cross-correlations). Dashed lines show the constraints in the absence of weak lensing, whilst the dotted lines show the constraints from only the first redshift bin. We observe that constraints on vector amplitudes are dominated by high-$\ell$, whilst those on tensor amplitudes depend principally on the largest scales measured.}
    \label{fig: primordial-fish}
\end{figure}

\begin{figure}
    \centering
    \includegraphics[width=0.49\textwidth]{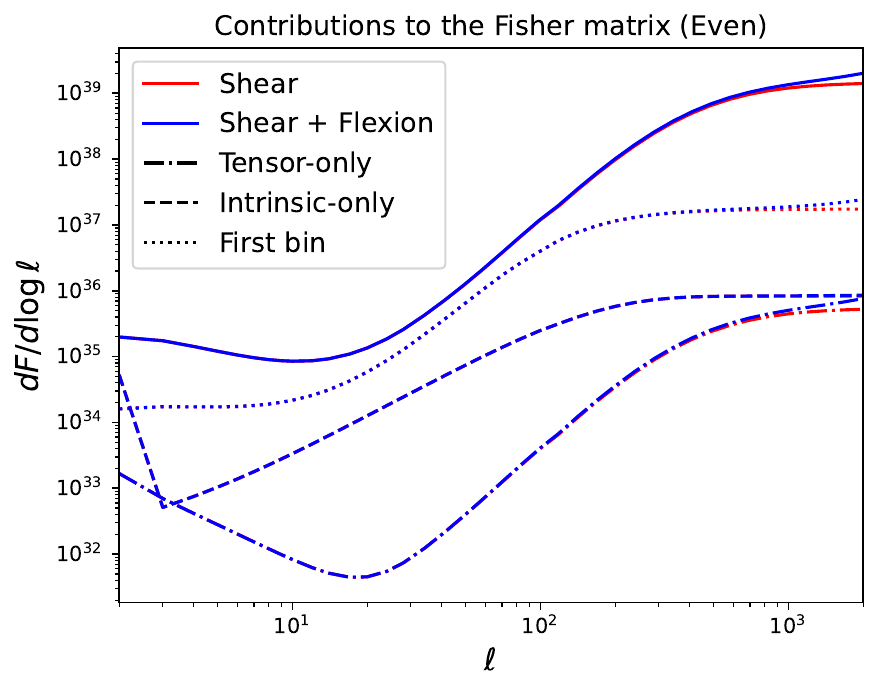}
    \includegraphics[width=0.49\textwidth]{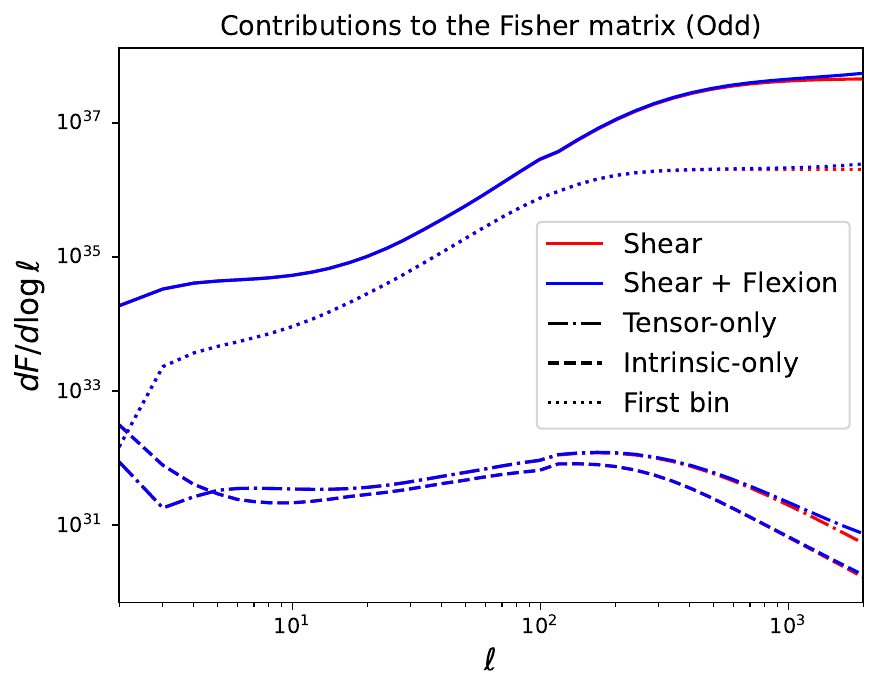}
    \caption{Constraints on the cosmic string tension parameter $(G\mu)^2$ from parity-even (left) and parity-odd (right) shear and flexion power spectra.  This is analogous to Fig.\,\ref{fig: primordial-fish}, but displays the joint constraint from vector and tensor modes (with the tensor-only results shown as dot-dashed lines). In this scenario, information is principally sourced by vector lensing correlations and is more equally distributed across scales than for the primordial constraints (Fig.\,\ref{fig: primordial-fish}). At large $\ell$ ($\gtrsim1000$), we find that the adding flexion into the analysis can lead to improved constraints.}
    \label{fig: string-fish}
\end{figure}

Tab.\,\ref{tab: constraints-tab} lists the forecasted constraints on the model amplitudes of the two sources of vector and tensor physics discussed above: the vector/tensor-to-scalar amplitude $r_X$, and the squared string tension parameter $(G\mu)^2$. We consider constraints from both parity-even and parity-odd correlators, and optionally include the flexion auto- and cross-spectra. To ascertain the scale-dependence of the constraints, we plot the derivatives of the Fisher matrix with $\ell$ in Figs.\,\ref{fig: primordial-fish}\,\&\,\ref{fig: string-fish}; this additionally shows whether the intrinsic or lensing contributions are dominant.

\vskip 4pt
\paragraph{Inflationary Perturbations}

Considering first the primordial sources, we find that the $1\sigma$ constraints on $r_V$ are weak, with the tightest bound of $\sigma(r_V) = \mathcal{O}(10^4)$. This constraint is dominated by the smallest scales and lensing, matching the conclusions drawn from Fig.\,\ref{fig: primordial-fig}; the former property implies that flexion can (and does) add significant information, tightening the bounds by some $10$-$20\%$ at $\ell\leq 2000$. Noting that the amplitude parameter has been normalized relative to the CMB, the implication of this study is that galaxy shapes are highly unlikely to yield useful constraints on primordially sourced vector modes. This occurs due to the $a^{-2}(z)$ transfer function intrinsic to vectorial physics.

The constraints on $r_T$ are considerably tighter than those on $r_V$, yet still unlikely to be competitive in the near future. Here, we find $\sigma(r_T) \approx 260$ from a \textit{Euclid}-like experiment, which is dominated by the lowest $\ell$-modes, and highly sensitive to the sample redshift; restricting to the first redshift bin reduces the constraining power by roughly a factor of ten, matching the conclusions of \citep{2012PhRvD..86h3513S}. Furthermore, intrinsic alignments dominate the information content; this matches the results of \citep[Fig.\,6]{2012PhRvD..86h3513S}, and imparts significant model-dependence since the intrinsic alignment biases are not well-known. The situation is similar for parity-odd tensor modes, with a somewhat weaker bound of $\sigma(\epsilon_Tr_T)\approx 600$; such modes are less commonly constrained in other probes, though our bound is still unlikely to be competitive \citep[cf.,][]{2020JCAP...07..005B}. In this scenario, the addition of flexion does not improve the constraints due to the strong scale-dependence.

\vskip 4pt
\paragraph{Cosmic Strings}

Turning to the cosmic string constraints, we find $1\sigma$ bounds on the squared string tension $(G\mu)^2$ of $\mathcal{O}(10^{-20})$ from parity-even correlators or $\mathcal{O}(10^{-19})$ from the parity-odd sector (if present). These agree with the lensing shear forecast of \citep{2012JCAP...10..030Y}, though we note that the results have strong dependence on the assumed intercommuting probability $P$, with the degeneracy direction $(G\mu)^2P^{-3/2}$. Furthermore, the constraints are of a similar amplitude to the limits from pulsar timing array experments, with $G\mu\lesssim 10^{-10}$ (though this statement is inherently model-dependent) \citep{EPTA:2023xxk}. Both vector- and tensor-derived constraints show strong scale-dependence, with maximal detection significance at $\ell\sim 1000$. As before, this implies that flexion can lead to significantly improved constraints, though all spectra will saturate at high-$\ell$ due to Poisson noise. Here, we find that the overall constraints are dominated by lensing at high-redshift, which is rationalized by noting that vector modes decay quickly after their sourcing. As in Fig.\,\ref{fig: string-fig}, we note that string-induced tensors contain significantly less information than vectors, with roughly $500\times$ weaker constraints. Overall, the prospect of constraining cosmic strings with galaxy shapes is promising, particularly when we include all sources of information available.

\section{Summary \& Conclusions}\label{sec: conclusions}
\noindent Perturbations in the Universe can be decomposed into scalars, vectors, and tensors. To date, only scalar physics has been observed; however, there exist a wealth of cosmological models that could source vector and tensor contributions at early- or late-times, thus their study can yield strong constraints on the Universe's initial conditions and evolution. In this work, we provide an in-depth study of the impact of such modes on galaxy shape statistics. These are a natural place in which to look for new physcis since they are tensorial (and thus respond directly to metric perturbations), and will be observed in great depth in the coming decade.

New physics can distort galaxy shapes both intrinsically (through correlations with the local tidal field \citep[e.g.,][]{2004PhRvD..70f3526H,2011JCAP...05..010B,Kogai:2020vzz}) and extrinsically (through weak gravitational lensing \citep[e.g.,][]{2001PhR...340..291B,2008ARNPS..58...99H}). Both effects can be straightforwardly written in terms of derivatives of the metric perturbations, and thus mapped to galaxy shape observables, such as shear and its higher-order generalization, flexion \citep[e.g.,][]{2006MNRAS.365..414B,2005ApJ...619..741G} (though the scale-dependence of such effects may complicate analyses slightly). For scalar perturbations, only shear-$E$ and flexion-$g$ modes are sourced; vector and tensor physics yields signals also in shear-$B$ and flexion-$c$ observables. Spectra of the latter could thus be a natural place to search for new physics without contamination from scalar modes (at leading order); these are additionally free from cosmic variance limitations (though their scalings with $\ell$ usually differ from the $E$- and $g$-modes) \citep[e.g.,][]{2012PhRvD..86h3513S}. In addition, if the Universe contains parity-violating physics, such as chiral gravitational waves, $EB$ cross-correlations (and beyond) will be sourced \citep[e.g.,][]{2020JCAP...07..005B}, which again do not suffer from cosmic variance at leading order. We caution, however, that $B$ modes are often used as a consistency check in weak lensing analyses (see \citep{2003A&A...408..829S} for a detailed discussion); if a survey finds a non-trivial shear-$B$ power spectrum (in excess of noise), this is usually attributed to systematics effects, and the null tests are said to have failed. Although this could well be the case, new physics could also be hiding in such regimes!

To place these results in quantitative context, we have considered test cases for vector and tensor physics, sourced by primordial and late-time physics, with the latter involving a network of cosmic strings. For a \textit{Euclid}-like lensing experiment, we concluded that primordial gravitational waves would be very difficult to detect (unless one worked at much higher redshift, as discussed in \citep{2012PhRvD..86h3513S,2020JCAP...07..005B}, though without flexion), though it competitive constraints on cosmic string physics could be possible (as in \citep{2012JCAP...10..030Y}, though without flexion, intrinsic alignments, or tensors). Drawing general conclusions from specific physical models, is, in general, difficult; however, we may make the following broad statements:
\begin{itemize}
    \item Shear and flexion spectra induced by new physics models can have a wide range of scale and redshift-dependencies, due to the large array of possible underlying perturbation power spectra and transfer functions. For example, cosmic strings and primordial vectors source spectra dominant at high-$\ell$, whilst gravitational waves contribute only on very large scales.
    \item Due to their strong scaling with redshift, vector modes sourced in the early Universe will be essentially impossible to detect using galaxy shapes (or any other late-Universe probe). Vector modes sourced at late-times could be detectable however, as seen from the cosmic string example.
    \item If the new physics contributes significantly on small scales, the addition of flexion correlators can boost signal-to-noise, and may allow various models to be differentiated. This may be of particular relevance for dark matter studies, and may yield significantly tighter constraints on phenomena such as fuzzy dark matter \citep[e.g.,][]{Ferreira:2020fam,Hui:2016ltb,Hu:2000ke,Mocz:2023adf,Dome:2022eaw,Amin:2022pzv}. If irreducible rank-three tensors also exist in the Universe (e.g., from some modified gravity phenomena, such as torsion), one might also expect them to show up in flexion spectra.
    \item Depending on the type of spectrum and physical model of interest, the shear and flexion correlators could be dominated either by lensing, intrinsic alignments, or their cross-correlation; as such, both effects should always be included in analysis pipelines.
    \item $BB$ and $EB$ spectra could be smoking guns of new physics. Whilst their utility depends on noise parameters (since they are not cosmic-shear limited) and systematic contamination, these spectra are not sourced by scalars at leading order (or, for the cross-spectrum, at any order). They could thus be a robust cosmological probe, particularly for parity-violating physics.
\end{itemize}
Based on the above, the overarching conclusion of this work is clear: if we wish to make full use of the treasure trove of cosmological information provided by upcoming galaxy shape surveys, we must look not just to vanilla $\Lambda$CDM physics and simple observables, but also to non-standard physics and higher-order statistics, such as vectors, tensors, $B$-modes, and flexion.

\begin{acknowledgments}
\footnotesize
We thank Kazuyuki Akitsu, Jo Dunkley, Bhuv Jain, \resub{Fabian Schmidt}, Uros Seljak, and Risa Wechsler for useful discussions across the years this paper took to write. \resub{We are additionally grateful to the anonymous referee for insightful feedback}. OHEP is a Junior Fellow of the Simons Society of Fellows. This work was inspired by the creative output of Funkmaster Flex. This project has been partially completed at the Laboratory for Nuclear Science and the Center for Theoretical Physics at the Massachusetts Institute of Technology (MIT-CTP/5613).
\end{acknowledgments}

\appendix

\section{Tensors on the 2-Sphere}\label{appen: 2-sphere-math}
A general rank-$s$ tensor, $\mathcal{T}$, on the 2-sphere may be written in terms of the $\vm_{\pm}$ basis vectors defined in \eqref{eq: basis-vectors} as
\beq
    \mathcal{T}_{i_1\ldots i_s}(\hn) = {}_{+s}\mathcal{T}(\hn)\left(\vm_+\otimes\cdots\otimes\vm_+\right)_{i_1\ldots i_s}+ {}_{-s}\mathcal{T}(\hn)\left(\vm_-\otimes\cdots\otimes\vm_-\right)_{i_1\ldots i_s}
\eeq
\citep[e.g.,][]{2005PhRvD..72b3516C,2012PhRvD..86h3527S}. Here, ${}_{\pm s}\mathcal{T}(\hn)$ is the spin-$\pm s$ component of $\mathcal{T}$, which transforms as ${}_{\pm s}\mathcal{T}\to e^{\pm is\varphi}{}_{\pm s}\mathcal{T}$ under rotation by $\varphi$. Using the basis vector properties outlined in \S\ref{subsec: shear+flexion-full-sky}, we may extract the spin-$\pm s$ components from the components of $\mathcal{T}$ in some arbitrary coordinate chart as
\beq
    {}_{\pm s}\mathcal{T}(\hn) &=& \left(m_{\mp}^{i_1}\cdots m_{\mp}^{i_s}\right)\mathcal{T}_{i_1\ldots i_s}(\hn),
\eeq
assuming the Einstein summation convention. In a local orthogonal basis $\{\hat{\bv e}_1,\hat{\bv e}_2,\hn\}$, the basis vectors can be written $\bv m_{\pm} = (\hat{\bv e}_1\mp i\hat{\bv e}_2)/\sqrt{2}$, such that the spin components take a simple form. As an example, a vector field has ${}_{\pm 1}v(\hn) = (v_1\pm i v_2)/\sqrt{2}$ locally.

To alter the spin of objects on the 2-sphere, we can use the spin-raising and spin-lowering operators, $\edth$ and $\bar\edth$ (``edth''). These are defined by their action on spin-$s$ functions (for both $s<0$ and $s>0$):
\beq\label{eq: edth-def}
    \edth {}_s\mathcal{T}(\hn) = -(\sin\theta)^s\left[\partial_\theta + i \csc\theta\partial_\varphi\right](\sin\theta)^{-s}\,{}_s\mathcal{T}(\hn), \qquad 
    \bar\edth {}_s\mathcal{T}(\hn) = -(\sin\theta)^{-s}\left[\partial_\theta - i \csc\theta\partial_\varphi\right](\sin\theta)^{s}\,{}_s\mathcal{T}(\hn)
\eeq
\citep[e.g.,][]{1967JMP.....8.2155G,2000PhRvD..62d3007H,2005PhRvD..72b3516C}, assuming a spherical coordinate chart $\{\theta,\varphi\}$. It is straightforward to show that $\edth{}_s\mathcal{T}$ has spin $(s+1)$ and $\bar\edth{}_s\mathcal{T}$ has spin $(s-1)$. In the flat-sky limit, the operators reduce to
\beq\label{eq: edth-flat-sky}
    \edth {}_s\mathcal{T}(\hn) \to -(\partial_1 + i \partial_2){}_s\mathcal{T}(\hn), \qquad \bar\edth {}_s\mathcal{T}(\hn)\to -(\partial_1 -i\partial_2){}_s\mathcal{T}(\hn),
\eeq
\textit{i.e.}\ $-\partial$ and $-\partial^*$ in the complex-valued notation of \citep{2005ApJ...619..741G,2006MNRAS.365..414B}. Furthermore, the derivatives can be written in terms of the covariant derivative $\nabla$ on the 2-sphere:
\beq\label{eq: spin-deriv-applied}
    \edth\left[{}_s\mathcal{T}(\hn)\right] &=& -\sqrt{2}\left(\nabla_k\left[{}_sf(\hn)\right]m_-^k + s\tau{}_s\mathcal{T}(\hn)\right), \qquad 
    \bar\edth\left[{}_s\mathcal{T}(\hn)\right] = -\sqrt{2}\left(\nabla_k\left[{}_s\mathcal{T}(\hn)\right]m_+^k - s\bar\tau{}_s\mathcal{T}(\hn)\right)
\eeq
\citep{2005PhRvD..72b3516C}, where $\tau = \nabla_k m_{+\,i}m_{-}^im_{-}^k$ accounts for the spatial variation of the basis functions. A particularly useful result is the action of $s$ $\bar\edth$ operators on a spin-$s$ function (with $s>0$), or $s$ $\edth$ operators on a spin-$-s$ function. This yields
\beq\label{eq: spin-deriv-s-times}
    \bar\edth^s{}_{s}\mathcal{T}(\hn) = \left\{\partial_\mu+\frac{i\partial_\varphi}{1-\mu^2}\right\}^s(1-\mu^2)^{s/2}{}_s\mathcal{T}(\hn) \qquad \edth^{s}{}_{-s}\mathcal{T}(\hn) = \left\{\partial_\mu-\frac{i\partial_\varphi}{1-\mu^2}\right\}^{s}(1-\mu^2)^{s/2}{}_{-s}\mathcal{T}(\hn),
\eeq
where $\mu\equiv\cos\theta$.

Much like a scalar on the 2-sphere can be written in terms of spherical harmonics, a rank-$s$ tensor can be written in terms of \textit{spin-weighted} spherical harmonics. Explicitly,
\beq\label{eq: spin-spherical-decomposition}
    {}_{s}\mathcal{T}(\hn) = \sum_{\ell m}{}_{s}\mathcal{T}_{\ell m} \,{}_{s}Y_{\ell m}(\hn),
\eeq
where $\ell$ and $m$ are integers satisfying $\ell\geq 0$, $|m|\leq \ell$. The spin-weighted spherical harmonics may be obtained by applying the spin-raising and spin-lowering operators to the standard spherical harmonics, $Y_{\ell m}$:
\beq\label{eq: spin-weighted-sph-def}
    \sqrt{\frac{(\ell+s)!}{(\ell-s)!}}{}_sY_{\ell m}(\hn) = \edth^s Y_{\ell m}(\hn) \qquad \sqrt{\frac{(\ell+s)!}{(\ell-s)!}}{}_{-s}Y_{\ell m}(\hn) = (-1)^s\bar\edth^s Y_{\ell m}(\hn)
\eeq
(for $s\geq 0$). These are orthonormal (for arbitrary $s$):
\beq\label{eq: spin-sph-orthonormality}
    \int d\hn\,{}_{s}Y_{\ell m}(\hn)\left[{}_{s}Y_{\ell'm'}(\hn)\right]^* = \delta^{\rm K}_{\ell\ell'}\delta^{\rm K}_{mm'},
\eeq
and obey the relations \citep[e.g.,][]{2005PhRvD..72b3516C}
\beq\label{eq: spin-sph-relations}
    \edth\,{}_sY_{\ell m}(\hn) &=& \sqrt{(\ell-s)(\ell+s+1)}\,{}_{s+1}Y_{\ell m}(\hn), \qquad  \bar\edth\,{}_sY_{\ell m}(\hn) = -\sqrt{(\ell+s)(\ell-s+1)}\,{}_{s-1}Y_{\ell m}(\hn).
\eeq
Via orthogonality, the coefficients appearing in \eqref{eq: spin-weighted-sph-def} can be extracted straightforwardly:
\beq\label{eq: spin-sph-coeffs}
    {}_s \mathcal{T}_{\ell m} = \int d\hn\,{}_s\mathcal{T}(\hn)\left[{}_{s}Y_{\ell m}(\hn)\right]^*.
\eeq
An additional relation of use is obtained by performing integration by parts on the above expression, and using \eqref{eq: spin-sph-relations}:
\beq\label{eq: spin-sph-coeffs2}
    {}_s \mathcal{T}_{\ell m} = \sqrt{\frac{(\ell-|s|)!}{(\ell+|s|)!}}\int d\hn\,(-\bar\edth)^s{}_s\mathcal{T}(\hn)\left[Y_{\ell m}(\hn)\right]^* = \sqrt{\frac{(\ell-|s|)!}{(\ell+|s|)!}}\int d\hn\,\edth^s{}_{-s}\mathcal{T}(\hn)\left[Y_{\ell m}(\hn)\right]^*
\eeq
for $s>0$. To extract the basis functions, one can thus apply spin-lowering or raising operations to the spin components ${}_{\pm s}\mathcal{T}(\hn)$, and integrate with respect to a spherical harmonic.

\section{Cosmic String Tensor Modes}\label{appen: cosmic-string-tensors}
Below, we provide a brief derivation of the tensor power spectrum arising from cosmic string sources. This follows \citep{2012JCAP...10..030Y}, whereupon vector perturbations were considered. Starting from \eqref{eq: stress-energy-string}, the $ij$ component of the stress-energy tensor from a single string segment can be written
\beq
    \delta T_{ij}(\bv{x},\eta) = \mu \int d\sigma\,e^{i\vk\cdot\bv{x}(\sigma,\eta)}\left(\dot{x}^i\dot{x}^j-x^i{}'x^j{}'\right)(\sigma,\eta).
\eeq
Via the Einstein equation \eqref{eq: vec-tensor-einstein}, this can be related to the metric perturbation $h_{ij}$. Projecting onto the tensor helicity basis (which removes the trace term), we find:
\beq
    \ddot{h}_{\pm}(\vk,\eta)+2\frac{\dot{a}}{a}\dot{h}_{\pm}(\vk,\eta)+k^2h_{\pm}(\vk,\eta) = 16\pi G\mu\,a^2 \int d\sigma \,e^{i\vk\cdot\bv{x}(\sigma,\eta)}e_{ij}^{(\mp2)}(\hk)\left(\dot{x}^i\dot{x}^j-x^i{}'x^j{}'\right)(\sigma,\eta),
\eeq
where, in contrast to the above, dots and primes denote derivatives with respect to $\eta$ and $\sigma$ in this section. To proceed we make the simplifying assumption that the source can be considered time-independent (\textit{i.e.}\ we assume steady-state behavior), which permits the solution:
\beq\label{eq: h-pm-string}
    h_{\pm}(\vk,\eta) = \frac{16\pi G\mu\,a^2}{k^2} \int d\sigma e^{i\vk\cdot\bv{x}(\sigma,\eta)}e_{ij}^{(\mp2)}(\hk)\left(\dot{x}^i\dot{x}^j-x^i{}'x^j{}'\right)(\sigma,\eta),
\eeq
henceforth disregarding the propagating modes (\textit{i.e.}\ gravitational waves). To obtain the shear and flexion predictions, one should properly consider the unequal-time power spectrum of $h_{ij}$; here we follow \citep{2012JCAP...10..030Y} and make the assumption that $P_{h_\pm}(k,\eta,\eta') \approx \sqrt{P_{h_\pm}(k,\eta,\eta)P_{h_\pm}(k,\eta',\eta')}$. This is not exact, but is appropriate for this rough forecast. The equal-time correlation function can be obtained from \eqref{eq: h-pm-string}:
\beq
    P_{h_{\pm}}(k,\eta,\eta) &=& \frac{(16\pi G\mu)^2\,a^4}{k^4}n_sdV\frac{1}{\mathcal V} e_{ij}^{(\mp2)}(\hk)e_{kl}^{(\pm2)}(\hk)\\\nonumber
    &&\,\times\,\int d\sigma_1 d\sigma_2 \av{e^{i\vk\cdot(\bv{x}(\sigma_1,\eta)-\bv{x}(\sigma_2,\eta)}\left(\dot{r}^i\dot{r}^j-r^i{}'r^j{}'\right)(\sigma_1,\eta)\left(\dot{r}^k\dot{r}^l-r^k{}'r^l{}'\right)(\sigma_2,\eta)},
\eeq
for comoving volume element $dV = 4\pi\chi^2/H$, string segment number density $n_s = a^3/\xi^3$, survey volume $\mathcal{V}$ and characteristic string length $\xi$. Asserting the isotropic and time-independent correlators $\av{\dot{r}^i(\sigma_1,\eta)\dot{r}^j(\sigma_2,\eta)} = \frac{1}{3}\delta^{ij}_{\rm K}V_s(\sigma_1-\sigma_2)$, $\av{r^i{}'(\sigma_1,\eta)r^j{}'(\sigma_2,\eta)} = \frac{1}{3}\delta^{ij}_{\rm K}T_s(\sigma_1-\sigma_2)$, with $\av{e^{i\vk\cdot(\bv{x}(\sigma_1,\eta)-\bv{x}(\sigma_2,\eta)}} = e^{-k^2\Gamma_s(\sigma_1-\sigma_2)/6}$ with $\Gamma_s(\sigma) \approx \sigma^2T_s(\sigma)$, we can write
\beq
    P_{h_{\pm}}(k,\eta,\eta) &=& \frac{2}{9}\frac{(16\pi G\mu)^2\,a^4}{k^4}
    \frac{4\pi \chi^2a^3}{H\xi^3}\frac{1}{\mathcal V}\int d\sigma_- d\sigma_+ e^{-(1/6)k^2\sigma_-^2T_s(\sigma_-)}\left[V_s^2(\sigma_-)+T_s^2(\sigma_-)\right],
\eeq
defining $\sigma_{\pm} = \sigma_1\pm \sigma_2$. As in \citep{2012JCAP...10..030Y}, we assume the correlators to have the scale invariant forms $V_s\approx v_{\rm rms}^2$, $T_s\approx (1-v_{\rm rms}^2)$ on scales $\sigma<\xi/a$, and zero else. Noting that $\int d\sigma_+/\mathcal{V}$ is the length of the string segment per unit volume, $\approx a^2/\xi^2\sqrt{1-v_{\rm rms}^2}$, and that $|\sigma_-|\lesssim \xi/2a\sqrt{1-v_{\rm rms}^2}$, we find
\beq\label{eq: tensor-cosmic-string-spectrum}
    P_{h_{\pm}}(k,\eta,\eta) &=& (16\pi G\mu)^2\frac{\sqrt{6\pi}}{9(1-v_{\rm rms}^2)}
    \frac{4\pi \chi^2a^4}{H}\left(\frac{a}{k\xi}\right)^5\mathrm{erf}\left[\frac{k\xi/a}{2\sqrt 6}\right]\left[v_{\rm rms}^4+(1-v_{\rm rms}^2)^2\right].
\eeq

\section{Intrinsic Vector Spectra}\label{appen: intrinsic-vector-derivations}
\noindent Here, we derive the shear kernels sourced by vector intrinsic alignments, following the methodology of \citep{2015JCAP...10..032S}. Starting from \eqref{eq: vector-tensor-g,F,G-source}, and expanding in helicity states using \eqref{eq: helicity-expansion}, we find
\beq
    \left.{}_{\pm2}\gamma(\bv{x},\chi)\right|_{V,\rm int} &=& b_V(\chi)\int_{\vk}e^{i\vk\cdot\bv{x}}(ik_\pm)\sum_\lambda \mathcal{I}_VB_\lambda(\vk,\chi)e^{(\lambda)}_{\pm}(\hk),
\eeq
where $k_\pm\equiv m_{\mp}^ik_i$ and $e^{(\lambda)}_\pm \equiv m_{\mp}^ie^{(\lambda)}_i$. For a single Fourier mode along $\vk=k\hz$, the basis definition of \eqref{eq: basis-vectors} implies $k_{\pm}=-(k/\sqrt{2})\sqrt{1-\mu^2}$, and $e_{\pm}^{(\lambda)} = (1/2)(\mu\pm\lambda)e^{-i\lambda\varphi}$ for $\mu\equiv\cos\theta=\hz\cdot\hn$. Additionally restricting to a single polarization state $\lambda$, we find the contribution
\beq
    \left.{}_{\pm2}\gamma(\chi\hn,\chi;k\hz,\lambda)\right|_{V,\rm int} &=& \frac{-ik}{2\sqrt{2}}b_V(\chi)\sqrt{1-\mu^2}(\mu\pm\lambda)e^{-i\lambda\varphi} \mathcal{I}_VB_\lambda(\vk,\chi)e^{ik\chi\mu}.
\eeq
To obtain the shear harmonic coefficients, it is convenient to first compute the spin-zero quantities $\bar\edth^2{}_{+2}\gamma$ and $\edth^2{}_{-2}\gamma$. Using relation \eqref{eq: spin-deriv-s-times}, these can be written
\beq
    \left.{}\bar\edth^2{}_{+2}\gamma(\chi\hn,\chi;k\hz,\lambda)\right|_{V,\rm int} &=& \frac{-ik}{2\sqrt{2}}b_V(\chi)\mathcal{I}_VB_\lambda(\vk,\chi)e^{-i\lambda\varphi}\left\{\partial_\mu+\frac{\lambda}{1-\mu^2}\right\}^2\left[(1-\mu^2)^{3/2}(\mu+\lambda)e^{ik\chi\mu}\right]\\\nonumber
    &\equiv& \frac{b_V(\chi)}{2\sqrt{2}\chi}\mathcal{I}_VB_\lambda(\vk,\chi)\sqrt{1-\mu^2}e^{-i\lambda\varphi}\hat Q_{\gamma,V}^{(\lambda)}(x)e^{ix\mu}\\\nonumber
    \left.{}\edth^2{}_{-2}\gamma(\chi\hn,\chi;k\hz,\lambda)\right|_{V,\rm int} &=& \frac{-ik}{2\sqrt{2}}b_V(\chi)\mathcal{I}_VB_\lambda(\vk,\chi)e^{-i\lambda\varphi}\left\{\partial_\mu-\frac{\lambda}{1-\mu^2}\right\}^2\left[(1-\mu^2)^{3/2}(\mu-\lambda)e^{ik\chi\mu}\right]\\\nonumber
    &\equiv& \frac{b_V(\chi)}{2\sqrt{2}\chi}\mathcal{I}_VB_\lambda(\vk,\chi)\sqrt{1-\mu^2}e^{-i\lambda\varphi}\hat Q_{\gamma,V}^{(\lambda)\,*}(x)e^{ix\mu}
\eeq
where $x\equiv k\chi$, and we have defined the operator $\hat{Q}^{(\lambda)}_{\gamma,V}(x) = x\left[4x+(12+x^2)\partial_x + 8x\partial_x^2 + x^2\partial_x^3\right]+i\lambda x^2\left[x+4\partial_x + x\partial_x^2\right]$, which satisfies $\hat Q^{(\lambda)\,*}_{\gamma,V}(x) = \hat Q^{(-\lambda)}_{\gamma,V}(x)$. The spin-weighted spherical harmonic coefficients ${}_{\pm2}\gamma_{\ell m}\equiv \gamma^E_{\ell m}\pm i\gamma^B_{\ell m}$ can be obtained via the relation
\beq
    {}_{+2}\gamma_{\ell m} = \sqrt{\frac{(\ell-2)!}{(\ell+2)!}}\int d\hn\,\bar\edth^2{}_{+2}\gamma(\hn)Y^*_{\ell m}(\hn), \qquad {}_{-2}\gamma_{\ell m} = \sqrt{\frac{(\ell-2)!}{(\ell+2)!}}\int d\hn\,\edth^2{}_{-2}\gamma(\hn)Y^*_{\ell m}(\hn)
\eeq
\eqref{eq: spin-sph-coeffs2}. To perform the integral over $\hn$, we use the result proved in \citep[Appendix A]{2012PhRvD..86h3527S}:
\beq\label{eq: sph-integral-useful-result}
    \int d\hn\,Y^*_{\ell m}(\hn)(1-\mu^2)^{|r|/2}e^{ir\varphi}e^{ix\mu} = \sqrt{4\pi(2\ell+1)}\sqrt{\frac{(\ell+|r|)!}{(\ell-|r|)!}}i^ri^\ell \frac{j_\ell(x)}{x^{|r|}}\delta^{\rm K}_{mr}.
\eeq
for integer $r$, $\ell$, and $m$, and spherical Bessel function $j_\ell(x)$. This leads to the following shear coefficients:
\beq
    \left.{}{}_{+2}\gamma_{\ell m}(\chi;k\hz,\lambda)\right|_{V,\rm int} &=& i^{\ell-\lambda}\sqrt{4\pi(2\ell+1)}\sqrt{\frac{(\ell-2)!(\ell+1)!}{(\ell+2)!(\ell-1)!}}\frac{b_V(\chi)}{2\sqrt{2}\chi}\mathcal{I}_VB_\lambda(\vk,\chi)\hat Q_{\gamma,V}^{(\lambda)}(x)\frac{j_\ell(x)}{x}\times\delta^{\rm K}_{m(-\lambda)}\\\nonumber
\eeq
The expression for ${}_{-2}\gamma_{\ell m}$ is analogous, but with $\hat{Q}_{\gamma,V}^{(\lambda)}(x)$ replaced with its conjugate. To form the shear power spectrum arising from vector perturbations, we need simply take the expectation of two copies of $\gamma_{\ell m}^X$ (for $X\in\{E,B\}$), summing over $\lambda$, averaging over $m$, and integrating over $\vk$:
\beq
    \left.C_\ell^{\gamma^X\gamma^Y}(\chi,\chi')\right|_{V,\rm int} = \frac{1}{2\ell+1}\sum_{m=-\ell}^\ell\sum_{\lambda=\pm1} \int_{\vk}\av{\left.\gamma^X_{\ell m}(\chi;k\hz,\lambda)\right|_{V,\rm int}\left.\gamma^{Y*}_{\ell m}(\chi';k\hz,\lambda)\right|_{V,\rm int}}.
\eeq
By isotropy, this recovers the full spectrum, even though we have only considered $\vk = k\hz$. In full, we obtain the spectra
\beq\label{eq: general-spectrum-appendix}
    \left.C_\ell^{\gamma^X\gamma^Y}(\chi,\chi')\right|_{V,\rm int} = \frac{2}{\pi}\int_0^\infty k^2dk\,F^{\gamma^X,V}_\ell(k,\chi)F^{\gamma^Y,V*}_\ell(k,\chi')\left[P_{B_+}(k,\chi,\chi')\pm P_{B_-}(k,\chi,\chi')\right],
\eeq
using the correlation properties of $B_\lambda(\vk,\chi)$ \eqref{eq: Phi,B,h-correlators}, where we take the difference of two helicity states if $X\neq Y$.\footnote{This arises since $\mathrm{Re}\,\hat Q^{(-1)}(x) = \mathrm{Re}\,\hat Q^{(+1)}(x)$, but $\mathrm{Im}\,\hat Q^{(-1)}(x) = -\mathrm{Im}\,\hat Q^{(+1)}(x)$.} From the spectrum, we can read off the kernels (dropping a trivial $i^\ell$ factor)
\beq
    \left.F^{\gamma^E,V}_\ell(k,\chi)\right|_{\rm int} &=& \sqrt{\frac{(\ell-2)!(\ell+1)!}{(\ell+2)!(\ell-1)!}}\frac{b_V(\chi)}{2\sqrt 2\chi}\mathrm{Re}\left[\hat Q^{(+1)}_{\gamma,V}(x)\right]\frac{j_\ell(x)}{x}\mathcal{I}_V\\\nonumber
    \left.F^{\gamma^B,V}_\ell(k,\chi)\right|_{\rm int} &=& \sqrt{\frac{(\ell-2)!(\ell+1)!}{(\ell+2)!(\ell-1)!}}\frac{b_V(\chi)}{2\sqrt 2\chi}\mathrm{Im}\left[\hat Q^{(+1)}_{\gamma,V}(x)\right]\frac{j_\ell(x)}{x}\mathcal{I}_V.
\eeq
As discussed in the main text, vectors source both $E$- and $B$ modes (unlike scalar perturbations); moreover, if the spectrum does not conserve parity $P_{B_+}\neq P_{B_-}$, and an $EB$ cross-spectrum will be sourced. These spectra can be promoted to binned form by simply integrating over $\chi$ and $\chi'$ via \eqref{eq: redshift-binned-kernels}.

The flexion kernels can be similarly computed. Restricting our attention to the spin-$\pm1$ flexion (for brevity), the Fourier-mode expansion of \eqref{eq: vector-tensor-g,F,G-source} gives
\beq
    \left.{}_{\pm1}\F(\bv{x},\chi)\right|_{V,\rm int} = \frac{8\sqrt 2\chi\tilde b_V(\chi)}{27}\int_{\vk}e^{i\vk\cdot\bv{x}}\sum_\lambda \left[(ik_{\pm})^2e_{\mp}^{(\lambda)}(\hk)+2(ik_\mp)(ik_\pm)e_{\pm}^{(\lambda)}(\hk)\right]\mathcal{I}_VB_\lambda(\vk,\chi).
\eeq
The contribution to $\bar\edth{}_{+1}\F$ from a Fourier mode $\vk = k\hz$ and a single helicity state is equal to
\beq
    \left.\bar\edth{}_{+1}\F(\chi\hn,\chi;k\hz,\lambda)\right|_{V,\rm int} &=& -\frac{2\sqrt 2\chi\tilde b_V(\chi)}{27}k^2e^{-i\lambda\varphi}\left\{\partial_\mu+\frac{\lambda}{1-\mu^2}\right\}\left[(1-\mu^2)^{3/2}e^{ix\mu}(3\mu+\lambda)\right]\mathcal{I}_VB_\lambda(\vk,\chi) \\\nonumber
    &=& -\frac{2\sqrt 2\tilde b_V(\chi)}{27\chi}\mathcal{I}_VB_\lambda(\vk,\chi)e^{-i\lambda\varphi}\sqrt{1-\mu^2}\hat{Q}_{\gamma,V}^{(\lambda)}(x)e^{ix\mu},
\eeq
where $\hat Q_{\F,V}^{(\lambda)}(x) \equiv x^2\left[4+12\partial_x^2+3x\partial_x(1+\partial_x^2) + i\lambda x(1+\partial_x^2)\right]$. $\edth{}_{-1}\F$ takes a similar form but replacing $\hat Q_{\F,V}^{(\lambda)}(x)$ with its conjugate. As before, the spin-zero coefficients can be used to obtain the spherical harmonic decomposition of $\F$:
\beq
    {}_{+1}\F_{\ell m}(\chi;k\hz,\lambda) &=& -\sqrt{\frac{(\ell-1)!}{(\ell+1)!}}\int d\hn\,\bar\edth{}_{+1}\F(\chi\hn,\chi;k\hz,\lambda)Y^*_{\ell m}(\hn)\\\nonumber
    &=&i^{\ell-\lambda}\sqrt{4\pi(2\ell+1)}\frac{2\sqrt{2}\tilde b_V(\chi)}{27\chi}\mathcal{I}_VB_\lambda(\vk,\chi)\hat Q_{\F,V}^{(\lambda)}(x)\frac{j_\ell(x)}{x}\times\delta^{\rm K}_{m(-\lambda)}\\\nonumber
\eeq
again using \eqref{eq: sph-integral-useful-result}. The resulting spectrum takes the same form as \eqref{eq: general-spectrum-appendix}, but with the modified kernels
\beq
    \left.F^{\F^g,V}(k,\chi)\right|_{\rm int} &=& \frac{2\sqrt 2\tilde b_V(\chi)}{27\chi}\mathrm{Re}\left[\hat Q_{\F,V}(x)\right]\frac{j_\ell(x)}{x}\mathcal{I}_V, \\\nonumber 
    \left.F^{\F^c,V}(k,\chi)\right|_{\rm int} &=& -\frac{2\sqrt 2\tilde b_V(\chi)}{27\chi}\mathrm{Im}\left[\hat Q_{\F,V}(x)\right]\frac{j_\ell(x)}{x}\mathcal{I}_V.
\eeq
Derivations for other intrinsic alignment spectra, such as those sourced by scalars and tensors, follow similarly, and make use of the kernels given in Appendix \ref{appen: kernels}.

\section{Extrinsic Vector and Tensor Spectra}\label{appen: lensing-spectra}
\noindent Here, we derive the full-sky power spectrum of shear sourced by vector and tensor lensing, following an analogous procedure to Appendix \ref{appen: intrinsic-vector-derivations} (from \citep{2012PhRvD..86h3513S}). The vector contribution can be written
\beq
    \left.{}_{\pm2}\gamma(\chi\hn,\chi)\right|_{V,\rm ext} &=& -\int_0^{\chi}d\chi'\int_{\vk}e^{ik\chi'\hk\cdot\hn}\sum_\lambda B_\lambda(\vk,\chi')\left[\left(1-2\frac{\chi'}{\chi}\right)(ik_\pm)e^{(\lambda)}_{\pm}(\hk)+\frac{\chi'}{\chi}(\chi-\chi')(ik_\pm)^2e^{(\lambda)}_{\parallel}(\hk)\right],\\\nonumber
\eeq
where $e^{(\lambda)}_\parallel \equiv \hat n^ke^{(\lambda)}_k$. Restricting to a single helicity state and Fourier mode with $\vk=k\hz$, we find
\beq
    \left.{}_{+2}\gamma(\chi\hn,\chi;k\hz,\lambda)\right|_{V,\rm ext} &=& \int_0^{\chi}d\chi' B_\lambda(\vk,\chi')\frac{\sqrt{1-\mu^2}}{2\sqrt2}\left[ix'(\mu+\lambda)\left(\frac{1}{\chi'}-\frac{2}{\chi}\right)+x'^2(1-\mu^2)\left(\frac{1}{\chi'}-\frac{1}{\chi}\right)\right]e^{-i\lambda\varphi}e^{ix'\mu},
\eeq
using that $e^{(\lambda)}_\parallel = (\sqrt{1-\mu^2}/\sqrt{2})e^{-i\lambda\varphi}$ and setting $x'\equiv k\chi'$. The spin-zero part, $\bar\edth^2{}_{+2}\gamma$, can be obtained with \eqref{eq: spin-deriv-s-times}:
\beq\label{eq: gamma-spin-zero-ext-tmp}
    \left.\bar\edth^2{}_{+2}\gamma(\chi\hn,\chi;k\hz,\lambda)\right|_{V,\rm ext} &=& \frac{1}{2\sqrt 2}\int_0^{\chi}d\chi' B_\lambda(\vk,\chi')e^{-i\lambda\varphi}\left\{\partial_\mu+\frac{\lambda}{1-\mu^2}\right\}^2\\\nonumber
    &&\,\qquad\,\times\,\left[ix'(\mu+\lambda)(1-\mu^2)^{3/2}\left(\frac{1}{\chi'}-\frac{2}{\chi}\right)+x'^2(1-\mu^2)^{5/2}\left(\frac{1}{\chi'}-\frac{1}{\chi}\right)\right]e^{ix'\mu},\\\nonumber
    &\equiv& \frac{1}{2\sqrt{2}}\int_0^{\chi}\frac{d\chi'}{\chi'} B_\lambda(\vk,\chi')e^{-i\lambda\varphi}\sqrt{1-\mu^2}\left[\hat Q_{\gamma,V,1}^{(\lambda)}(x')+\frac{\chi'}{\chi}Q_{\gamma,V,2}^{(\lambda)}(x')\right]e^{ix'\mu},
\eeq
defining the new operators $Q_{\gamma,V,1/2}^{(\lambda)}(x)$, whose forms are given in \eqref{eq: Q-gamma-ext}. The expression for $\edth^2{}_{-2}\gamma$ is analogous but with all $\hat Q$ operators replaced by their conjugates. As in Appendix \ref{appen: intrinsic-vector-derivations}, we proceed by integrating \eqref{eq: gamma-spin-zero-ext-tmp} against a spherical harmonic using \eqref{eq: sph-integral-useful-result} to yield
\beq
    \left.{}_{+2}\gamma_{\ell m}(\chi;k\hz,\lambda)\right|_{V,\rm ext} &=& \frac{\sqrt{4\pi(2\ell+1)}}{2\sqrt{2}}\sqrt{\frac{(\ell-2)!(\ell+1)!}{(\ell+2)!(\ell-1)!}}i^{\ell-\lambda} \int_0^{\chi}\frac{d\chi'}{\chi'} B_\lambda(\vk,\chi')\left[\hat Q_{\gamma,V,1}^{(\lambda)}(x')+\frac{\chi'}{\chi}Q_{\gamma,V,2}^{(\lambda)}(x')\right]\frac{j_\ell(x')}{x'},
\eeq
if $m=-\lambda$, and zero else. Integrating over redshift via \eqref{eq: redshift-binned-kernels} gives
\beq
    \left.{}_{+2}\gamma_{\ell m,a}(k\hz,\lambda)\right|_{V,\rm ext} &=& \frac{\sqrt{4\pi(2\ell+1)}}{2\sqrt{2}}\sqrt{\frac{(\ell-2)!(\ell+1)!}{(\ell+2)!(\ell-1)!}}i^{\ell-\lambda}\\\nonumber
    &&\qquad\qquad\,\times\,\int_0^{\chi_H}\frac{d\chi}{\chi} B_\lambda(\vk,\chi)\left[m_a(\chi)\hat Q_{\gamma,V,1}^{(\lambda)}(x)+(m_a(\chi)-q_a(\chi))Q_{\gamma,V,2}^{(\lambda)}(x)\right]\frac{j_\ell(x)}{x},
\eeq
exchanging the order of integration and introducing $q_a$ and $m_a$ kernels. This yields spectra of the general form \eqref{eq: general-spectra} with the kernels given in \eqref{eq: ext-vector-tensor-shear-spectra}.

A similar form may be obtained for the shear spectrum from tensor modes, starting from \eqref{eq: lensing-schmidt-paper}. Here, we implement the $h_{\pm}(\bf 0,0)$ term by subtracting the $k\to0$ limit of the scalar shear kernel (which is non-zero only for $\ell=2$), and absorb the $-\frac{1}{2}h_{\pm}(\chi\hn,\chi)$ term into the intrinsic tensor shear contribution, redefining $\mathcal{I}_Th_{ij}\to\left(\mathcal{I}_T-1/2 b_T(\chi)\right)h_{ij}$. For the remaining terms, a procedure analogous to the above above (described in detail in \citep{2012PhRvD..86h3513S}) leads to the kernels given in \eqref{eq: ext-vector-tensor-shear-spectra}.

\section{Bessel Function Operators}\label{appen: kernels}
\noindent Below, we list the various $\hat Q(x)$ Bessel function operators appearing in the full-sky power spectra of \S\ref{subsec: full-spectra}. Firstly, the relevant operators for shear sourced by intrinsic scalar, vector, and tensor perturbations (\S\ref{subsec: intrinsic-spectra}) are given by:
\beq\label{eq: Q-gamma}
    \hat{Q}_{\gamma,S}^{(\lambda)}(x) &=& x^2\left[4+12\partial_x^2+8x\partial_x(1+\partial_x^2)+x^2(1+\partial_x^2)^2\right]\\\nonumber
    \hat{Q}_{\gamma,V}^{(\lambda)}(x) &=& x\left[4x+(12+x^2)\partial_x + 8x\partial_x^2 + x^2\partial_x^3\right]+i\lambda x^2\left[x+4\partial_x + x\partial_x^2\right]\\\nonumber
    \hat{Q}_{\gamma,T}^{(\lambda)}(x) &=& \left[12-x^2+8x\partial_x + x^2\partial_x^2\right] + 2i\lambda x\left[4+x\partial_x\right],
\eeq
whilst those for the spin-$\pm1$ flexion are
\beq\label{eq: Q-F}
    \hat Q_{\F,S}^{(\lambda)}(x) &=& 3x^3\left[x+4\partial_x+x\partial_x^2\right](1+\partial_x^2)\\\nonumber
    \hat Q_{\F,V}^{(\lambda)}(x) &=& x^2\left[4+12\partial_x^2+3x\partial_x(1+\partial_x^2) + i\lambda x(1+\partial_x^2)\right]\\\nonumber
    \hat Q_{\F,T}^{(\lambda)}(x) &=& x\left[x+12\partial_x + 3x\partial_x^2+2i\lambda x\partial_x\right],
\eeq
and for the spin-$\pm3$ flexion:
\beq\label{eq: Q-G}
    \hat Q_{\G,S}^{(\lambda)}(x) &=& -x^3\left[x(x^2+18)+18(x^2+4)\partial_x+3x(x^2+36)\partial_x^2+12(3x^2+10)\partial_x^3+3x(x^2+30)\partial_x^4+18x^2\partial_x^5+x^3\partial_x^6\right]\nonumber\\
    \hat Q_{\G,V}^{(\lambda)}(x) &=& -x^2\left[6(x^2+4)+x(x^2+66)\partial_x+24(x^2+5)\partial_x^2+2x(x^2+45)\partial_x^3+18x^2\partial_x^4+x^3\partial_x^5\right.\\\nonumber
    &&\,\left.\qquad+i\lambda x\left(6+x^2+12x\partial_x+2(x^2+15)\partial_x^2+12x\partial_x^3+x^2\partial_x^4\right)\right]\\\nonumber
    \hat Q_{\G,T}^{(\lambda)}(x) &=& x\left[x(x^2-30)-6(x^2+20)\partial_x - 90x\partial_x^2-18x^2\partial_x^3-x^3\partial_x^4-2ix\lambda(6x+(x^2+30)\partial_x+12x\partial_x^2+x^2\partial_x^3)\right].
\eeq
The scalar, vector, and tensor operators act on $j_\ell(x)$, $j_\ell(x)/x$, and $j_\ell(x)/x^2$ respectively.

The kernels relevant to the extrinsic power spectra (\S\ref{subsec: extrinsic-spectra}) are given by
\beq\label{eq: Q-gamma-ext}
    \hat Q_{\gamma,V,1}^{(\lambda)}(x) &=& -x\left[x(x^2+8) + (11x^2+12)\partial_x + 2x(x^2+14)\partial_x^2+11x^2\partial_x^3+x^3\partial_x^4-i\lambda x\left(x +4 \partial_x+x\partial_x^2\right)\right]\\\nonumber
    \hat Q_{\gamma,V,2}^{(\lambda)}(x) &=& x\left[x(x^2+12) + 12(x^2+2)\partial_x + 2x(x^2+18)\partial_x^2 + 12 x^2\partial_x^3 + x^3\partial_x^4\right],\\\nonumber
    \hat Q_{\gamma,T,1}^{(\lambda)}(x) &=& -\frac{x}{2}\left[x(x^2+14)+2(7x^2+20)\partial_x+2x(x^2+25)\partial_x^2+14x^2\partial_x^3+x^3\partial_x^4\right.\\\nonumber
    &&\,\left.-2i\lambda\left(4 + x^2 + 6 x\partial_x + x^2\partial_x^2\right)\right]\\\nonumber
    \hat Q_{\gamma,T,2}^{(\lambda)}(x) &=& \frac{1}{2}\left[24(x^2+1)+x^4+16x(x^2+6)\partial_x+2x^2(x^2+36)\partial_x^2+16x^3\partial_x^3+x^4\partial_x^4\right].
\eeq

In the large-$\ell$ limit, the action of the $\hat Q$ operators on the spherical Bessel functions simplifies considerably. The intrinsic functions have the following asymptotic forms:
\begin{alignat}{6}
    \hat{Q}^{(\lambda)}_{\gamma,S}(x)j_\ell(x) &\approx \ell^4 j_\ell(x), \quad  &&\hat{Q}^{(\lambda)}_{\gamma,V}(x)\frac{j_\ell(x)}{x} & &\approx \ell^2j_\ell'(x) + i\lambda\ell^2j_\ell(x), \quad &&&\hat{Q}^{(\lambda)}_{\gamma,T}(x)\frac{j_\ell(x)}{x^2} && &\approx \left(\frac{\ell^2}{x^2}-2\right)j_\ell(x) + 2i\lambda j_\ell'(x),\nonumber\\
    \hat{Q}^{(\lambda)}_{\F,S}(x)j_\ell(x) &\approx 3\ell^4 j_\ell(x), \quad  &&\hat{Q}^{(\lambda)}_{\F,V}(x)\frac{j_\ell(x)}{x} & &\approx 3\ell^2j_\ell'(x) + i\lambda\ell^2j_\ell(x), \quad &&& \hat{Q}^{(\lambda)}_{\F,T}(x)\frac{j_\ell(x)}{x^2} && &\approx \left(3\frac{\ell^2}{x^2}-2\right)j_\ell(x) + 2i\lambda j_\ell'(x),\\\nonumber
    \hat{Q}^{(\lambda)}_{\G,S}(x)j_\ell(x) &\approx -\ell^6 j_\ell(x), 
    \quad  &&\hat{Q}^{(\lambda)}_{\G,V}(x)\frac{j_\ell(x)}{x} & &\approx -\ell^4j_\ell'(x) - i\lambda\ell^4j_\ell(x), \quad
    &&&\hat{Q}^{(\lambda)}_{\G,T}(x)\frac{j_\ell(x)}{x^2} && &\approx -\ell^2\left(\frac{\ell^2}{x^2}-2\right)j_\ell(x) - 2i\lambda\ell^2 j_\ell'(x),
\end{alignat}
where $j_\ell'(x) \equiv \partial_x j_\ell(x)$. A similar form can be obtained for the extrinsic kernels:
\begin{alignat}{4}
    \hat{Q}^{(\lambda)}_{\gamma,V,1}(x)\frac{j_\ell(x)}{x} &\approx -\frac{\ell^4}{x}j_\ell(x)+i\lambda\ell^2j_\ell(x), \qquad &&\hat{Q}^{(\lambda)}_{\gamma,V,2}(x) & &\approx \frac{\ell^4}{x}j_\ell(x)\\
    \hat{Q}^{(\lambda)}_{\gamma,T,1}(x)\frac{j_\ell(x)}{x^2} &\approx -\frac{\ell^4}{2x^2}j_\ell(x)+i\lambda\frac{\ell^2}{x}j_\ell(x), \qquad &&\hat{Q}^{(\lambda)}_{\gamma,T,2}(x) & &\approx \frac{\ell^4}{2x^2}j_\ell(x).
\end{alignat}

\bibliographystyle{JHEP}
\bibliography{refs}%

\end{document}